\def\insfig#1{#1}
\def\endinsfig{\end{document}}

\def\manuscript{
	\documentstyle[12pt,aasms4]{article}
	\def\insfig##1{}
	\def\endinsfig{}
	\def\manufig##1{}   
	\def\capfig##1{##1}  
	}

\def\typeset{
	\documentstyle[12pt,aasms4]{article}
	\def\insfig##1{}
	\def\endinsfig{}
	\def\manufig##1{##1} 
	\def\capfig##1{}    
	}

  \documentstyle[11pt,aas2pp4]{article}  

\font\smallrm=cmr8

\def\PSF{\hbox{\smallrm PSF}}
\def\PD{\hbox{\smallrm PD}}

\def\hst{HST}
\def\kms{\hbox{$\,$km$\,$s$^{-1}$}}
\def\kmsMpc{\hbox{$\,$km$\,$s$^{-1}\,$Mpc$^{-1}$}}

\def\Mpc{\hbox{$\,$Mpc}}
\def\etal{{et~al.}}
\def\solar{\ifmmode_{\mathord\odot}\else$_{\mathord\odot}$\fi}
\def\mi{\ifmmode\overline{m}_I\else$\overline{m}_I$\fi}
\def\mv{\ifmmode\overline{m}_V\else$\overline{m}_V$\fi}
\def\Mi{\ifmmode\overline{M}_I\else$\overline{M}_I$\fi}
\def\Miz{\ifmmode\overline{M}_I^0\else$\overline{M}_I^0$\fi}
\def\mim{\ifmmode\overline{m}_I^0\else$\overline{m}_I^0$\fi}
\def\dmod{\ifmmode(m{-}M)\else$(m{-}M)$\fi}
\def\vi{\ifmmode(V{-}I)\else$(V{-}I)$\fi}
\def\viz{\ifmmode(V{-}I)_0\else$(V{-}I)_0$\fi}
\def\dn{\ifmmode D_n{-}\sigma\else$ D_n{-}\sigma$\fi}
\def\avemi{\ifmmode\langle\overline{m}_I^0\rangle\else$\langle\overline{m}_I^0\rangle$\fi}

\lefthead{Tonry, Blakeslee, Ajhar and Dressler}
\righthead{SBF Large Scale Flows}

\begin{document}
\
\
\
\
\
\
\
\
\
\

\
\
\
\
\title{The SBF Survey of Galaxy Distances. II. \\
     Local and Large-Scale Flows\altaffilmark{1,2}}

\author{John L. Tonry\altaffilmark{3}}
\affil{Institute for Astronomy, University of Hawaii, Honolulu, HI 96822}
\affil{Electronic mail: jt@ifa.hawaii.edu}
\authoremail{jt@ifa.hawaii.edu}

\author{John P. Blakeslee}
\affil{Department of Astronomy, MS 105-24, California Institute of Technology, 
Pasadena, CA 91125}
\authoremail{john@arneb.mit.edu}
 
\author{Edward A. Ajhar\altaffilmark{3}}
\affil{Kitt Peak National Observatory, National Optical Astronomy
Observatories, P. O. Box 26732,}
\affil{Tucson, AZ 85726}
\authoremail{ajhar@noao.edu}
 
\author{Alan Dressler}
\affil{Carnegie Observatories, 813 Santa Barbara St., Pasadena, CA 91101}
\authoremail{dressler@omega.ociw.edu}
 
\altaffiltext{1}{Observations in part from the Michigan-Dartmouth-MIT 
(MDM) Observatory.}
\altaffiltext{2}{Based on observations with the NASA/ESA {\it Hubble
Space Telescope,} obtained at the Space Telescope Science Institute,
which is operated by the Association of Universities for Research in
Astronomy (AURA), Inc., under NASA contract NAS5-26555. These 
observations are associated with proposals ID~5910 and ID~6579.}
\altaffiltext{3}{Guest observers at the Cerro Tololo Inter-American
Observatory and the Kitt Peak National Observatory, National Optical
Astronomy Observatories, which are operated by AURA, Inc., under
cooperative agreement with the National Science Foundation.}

\begin{abstract}

We present results from the Surface Brightness Fluctuation (SBF)
Survey for the distances to 300 early-type galaxies, of which
approximately half are ellipticals.  A modest change in the zero point
of the SBF relation, derived by using Cepheid distances to spirals
with SBF measurements, yields a Hubble constant 
$H_0 = 77\pm4\pm7$~\kmsMpc, 
somewhat larger than the HST Key Project result.
We discuss how this difference arises from a different choice of zero
point, a larger sample of galaxies, and a different model for large
scale flows.  The zero point of the SBF relation is the largest source
of uncertainty, and the SBF value for $H_0$ is subject to all the 
systematic uncertainties of the KP zero point, including a 5\%
decrease if a metallicity correction is adopted.

To analyze local and large-scale flows --- departures from smooth
Hubble flow --- we use a parametric model for the distribution
function of mean velocity and velocity dispersion at each point in
space.  These models include a uniform thermal velocity dispersion and
spherical attractors whose position, amplitude, and radial shape are
free to vary.  Our modeling procedure performs a maximum likelihood
fit of the model to the observations.

Our models rule out a uniform Hubble flow as an acceptable fit to the
data.  Inclusion of two attractors, one of which having a best fit
location coincident with the Virgo cluster and the other having a fit
location slightly beyond the Centaurus clusters (which we refer to by
convention as the Great Attractor) reduces $\chi^{2}/N$ from 2.1
to 1.1.  The fits to these attractors both have radial profiles such
that $v\approx r^{-1}$ (i.e. isothermal) over a range of overdensity
between about 10 and 1, but fall off more steeply at larger radius.
The best fit value for the small scale, cosmic thermal velocity is
$180 \pm 14$\kms.

The quality of the fit can be further improved by the addition of
a quadrupole correction to the Hubble flow.  
The dipole velocity offset from the CMB frame for the volume we survey
(amplitude $\sim 150$\kms) and the quadrupole may be genuine
(though weak) manifestations of more distant density fluctuations, but we
find evidence that they are more likely due to the inadequacy of spherical
models to describe the density profile of the attractors.
Our models can account almost perfectly for the CMB motion of the
Local Group as arising from the attractors within our survey volume
($R < 3000 \kms$); in other words, our sampled volume is, in a mass
averaged sense, essentially at rest with respect to the CMB. This
contradicts claims of large amplitude flows in much
larger volumes that include our sample. 

Our best-fitting model, which uses attenuated isothermal mass
distributions for the two attractors, has enclosed mass overdensities
at the Local Group of $7\times 10^{14}$~M\solar\ for the Virgo
Attractor and $8\times10^{15}$~M\solar\ for the Great Attractor.
Without recourse to information about the overdensities of these
attractors with respect to the cosmic mean we cannot provide a good
constraint on $\Omega_M$, but our data do give us accurate
measurements of $\delta\,\Omega_M^{2/3} = 0.33$ for the Virgo
Attractor and $\delta\,\Omega_M^{2/3} = 0.27$ for the Great
Attractor.

\end{abstract}

\section{Introduction}

There is now convincing evidence that the evolution of large-scale
structure is driven by intergalactic dark matter.
Understanding the nature of this dark matter remains
one of the major unsolved problems in astronomy.  It is for this
reason that mapping the overall mass distribution, independent of
the distribution of luminous matter in galaxies, must be considered a
fundamental endeavor in cosmological research.  Unfortunately, few
tools are available for the purpose: weak gravitational lensing can
trace, statistically, the presence of intervening dark matter at large
distances, and X-ray emitting gas can be used to map the gravitational
potential of the dark matter that binds rich clusters.  However, for a
point-by-point comparison of the density of dark and luminous matter,
measuring peculiar velocities---departures from pure Hubble flow
caused by an uneven distribution of dark matter---is the only
effective method available.

Using galaxies as test particles to sample the local gravitational
field requires accurate knowledge of their positions in space, because
galaxy motions are dominated by the expansion of the universe for
all but the closest objects.  Since most methods deliver an accuracy
that (at best) scales proportionately with the distance, there is a
premium on more accurate distance measurements driven by the desire
to measure peculiar velocities over a large volume of space.
The peculiar velocity that is known best, by far, is the
370~\kms\ motion of our Sun with respect to the
cosmic microwave background (CMB), known with exquisite accuracy from
the COBE measurement of the CMB dipole anisotropy, which is then
converted with less certainty to the $600$~\kms\ motion of our
Local Group of galaxies with respect to the CMB.
A study of peculiar velocities for other galaxies in the local
universe can be thought of as an exercise with two complementary aims:
(1) to map the dark matter distribution in the local universe and
compare it with the local galaxy distribution (in the process,
measuring a representative value of the cosmic density parameter
$\Omega_M$) and (2) to account for our Galaxy's motion as the
consequence of anisotropies of the dark matter distribution.  Such
knowledge has application in interpreting large-scale structure
inferred from the galaxy distribution alone, and to measuring the
degree and nature of ``bias'' that may exist between the dark and
luminous matter distributions, so important for testing theoretical
models of structure formation.  In addition, it is likely that a
detailed comparison of the dark-matter/baryon/galaxy distributions on
the scale of the correlation length for galaxies will shed light on
the processes of galaxy formation and the role of dark
matter in that process.

The observation of Rubin \etal\ (1976) first raised the possibility of
a large-scale deformation in the local Hubble flow.  Unfortunately,
the method of distance used in this pioneering work was too crude to
fight its way through the Malmquist-like systematic errors that are
endemic to the field.  The first solid detections of large-scale flows
and inferred dark structures began around 1980, including Tonry
(1980), Schechter (1980), Yahil \etal\ (1980), Aaronson \etal\ (1980),
Tonry \& Davis (1981), and Aaronson \etal\ (1982), using spiral and
elliptical galaxies to map the Local Supercluster.  All studies
detected the infall pattern that was anticipated for a sizeable
overdensity roughly centered on the Virgo cluster, but with an amplitude
at the position of the Local Group that ranged from 125~\kms\ (Yahil
\etal\ 1980) to 480~\kms\ (Aaronson \etal\ 1980).  As we shall see, 
the flow within the Virgo supercluster is much more complicated than
these early models permitted, and this range of ``infall velocity'' is
not surprising.  On thing that was clear, however, was that in both
amplitude and direction the pull of the Local
Supercluster was insufficient to be the sole cause of the CMB dipole
anisotropy, {\it i.\,e.,} the motion of the Milky Way or Local Group
with respect to the CMB.  An additional pull, in the direction of the
Centaurus supercluster, implicated as well by the high galaxy density
in this direction, was suggested by Shaya (1984), Tammann \& Sandage
(1985), and Aaronson \etal\ (1986).  Of particular relevance for the
discussion of our own results is the suggestion by Lilje, Yahil, \&
Jones (1986) that the effect of this more distant mass can be seen
as a quadrupole term in the peculiar velocity field of the Local
Supercluster.

Based on a refined distance estimator for elliptical galaxies, actually
a projection of the fundamental plane (Dressler \etal\ 1987, Djorgovski
\& Davis 1987), Lynden-Bell \etal\ (1988) used a sample of over 400
elliptical galaxies to map a much large volume of space.  This
study confirmed the infall pattern toward Virgo, but credited a
considerable fraction of the peculiar velocity measured for local
galaxies to the quadrupole of a more distant mass concentration.
Surprisingly, however, the Lynden-Bell \etal\ study did not find the
Centaurus cluster to be the center of a more distant infall pattern,
but put the center at a distance approximately 50\% greater, with
Centaurus and its associated groups themselves falling into the more
distant attractor at a velocity of order 1000~\kms.  The pull on the
Local Group from this ``behind Centaurus'' attractor was reckoned to
be of order 500~\kms, perhaps twice as big as the Virgo pull but from a
velocity distance of around 4000~\kms, over three times as distant.  The
implied order-of-magnitude greater mass earned it the nickname
``Great Attractor'' (GA).

However, doubts have remained about whether the GA is indeed a well
defined overdensity with such a large gravitational influence.
While Dressler \& Faber's (1990a, 1990b) distance measurements of
additional elliptical and spiral galaxies appeared to show the
distinctive {\sf S}-shaped pattern of infall, including ``backside infall,''
the larger spiral sample of Mathewson, Ford, \& Buckhorn (1992) did not.
Instead, these authors argued for a continuing high amplitude flow
beyond the distance identified as the GA center, perhaps the result of
a more distant gravitational pull, such as the ``Giant Attractor''
suggested by Scaramella \etal\ (1989) associated with the Shapley
concentration of rich clusters, more than three times as distant as the GA.

Further evidence that the GA's role in producing the local flow
pattern might have been overestimated is discussed by Courteau \etal\
(1993).  Adding a more complete sample of spirals, particularly in the
Perseus-Pisces region, led to this study's finding of a considerable
amplitude ``bulk flow''---that is, a non-converging flow whose source,
if gravitational in origin, would be on a scale considerably larger
than 6000~\kms, a region encompassing the GA.  However, a recent study
of spiral galaxy distances by Giovanelli \etal\ (1999) contradicts
this result by claiming that all of the Local Group's motion with
respect to the CMB can be accounted for by sources within the $V =
6000 \kms$ sphere.  Likewise, the Type Ia supernovae observations of
Riess, Press, \& Kirshner (1995) suggest a more distant frame that is
at rest with the CMB.

This lingering uncertainty in the sources of the local flow
pattern---that is, the contribution of the Local Supercluster, the GA,
Perseus-Pisces, the Shapley concentration, and structures
beyond---have been made all the more important by studies of Lauer \&
Postman (1994), Willick (1999), and Hudson \etal\ (1999) who find
large, but not always consistent bulk flows, over even larger volumes
of the local universe.  Indeed, these flows have amplitudes large
enough, over such large volumes, that they appear inconsistent
with the small scale anisotropy $\Delta T / T \sim 10^{-5}$ measured
for CMB fluctuations.

The surface brightness fluctuation (SBF) method (Tonry \& Schneider 1988)
of measuring early-type galaxy distances is perhaps the most
promising means for understanding these issues and resolving the disagreements.
A recent comprehensive review of the SBF method is given by
Blakeslee, Ajhar, \& Tonry (1999). SBF offers an accuracy several times
greater than either Tully-Fisher or Fundamental Plane distances; as a
consequence, its susceptibility to Malmquist-like biases is reduced by
an order-of-magnitude.  SBF also includes an implicit correction for
variations in age and metal abundance, in the form of color term, which
previous methods for measuring distances to elliptical galaxies did not
include. The data we present here will have bearing on all the issues
raised above, but they are limited to relatively nearby galaxies ($V <
4000 \kms$) for all but a handful of observations with the {\it Hubble
Space Telescope} (\hst).  Nevertheless, we hope to demonstrate in this
paper the power and promise of the method for eventually producing a
highly accurate map of the local distribution of dark matter, out to
velocity distances of 10,000~\kms\ or more, thereby answering the
question ``over what scale does the motion of the Local Group with
respect to the CMB arise?''

In this paper we will discuss the characteristics of our SBF data set
and use the data to construct parametric models of the local flow
pattern.  We will show not only that the convergent flows into the
Local Supercluster and Great Attractor dominate the departures from
smooth Hubble flow but also that the data are sufficiently accurate to
provide other constraints on bulk flow, local voids, and other
possible attractors.  The combination of these two attractors and a
moderate thermal component accounts for 90\% of the variance in our
sample.  Specifically, evidence for the GA comes not only from the
large peculiar velocities in the Centaurus direction but also from the
clear quadrupole signature in the Local Supercluster flow, whose
amplitude and direction are consistent with an order-of-magnitude
larger mass concentration in the Centaurus direction.  
The amplitude and scale of these two flows are consistent with
low-$\Omega_M$ cosmologies and with the measured small-scale anisotropy
of the CMB. 

We extensively explore the covariances between various model parameters,
such as the attractor distances and amplitudes, the Hubble parameter,
and the overall bulk flow.  The uncertainties will greatly diminish when
additional \hst\ SBF measurements spanning the GA region are available. 
A future paper will discuss more detailed models for the GA and compare
our results to previous works, including a point-by-point comparison of
SBF and Lynden-Bell \etal\ distances.  In addition, while we have chosen
on this first examination of our data to use {\it only\/} the peculiar
velocity information inherent in the SBF distances and to completely
ignore the distribution of galaxies, we intend in future papers to
compare the SBF peculiar velocities in detail with the galaxy density
field.
\section{The Data}

\subsection{The SBF Distance Survey}

The first paper of this series (Tonry \etal\ 1997, hereafter SBF-I)
described the galaxy sample and observations of the SBF survey in detail
and gave a new calibration of the method.  That calibration was derived
by comparing averaged SBF measurements within 7 galaxy groups to
Cepheid distances to spirals purportedly within the same groups.  
It gave an absolute magnitude $\Miz = -1.74$ at our fiducial early-type
galaxy color of $\viz = 1.15$.  The alternative, direct calibration
from SBF measurements for 5 Cepheid-bearing galaxies gave a brighter
zero point of $\Miz = -1.82$.  As these two zero points agreed within
the uncertainty, and because of the potential for systematic differences 
arising from the relative difficulty of SBF measurements in spiral
bulges, we felt more secure in adopting the group calibration.

For a number of reasons it has become necessary to revise the
calibration yet again.  First of all, we wish to incorporate all the
latest \hst\ Cepheid distances, and we adopt the values as tabulated by
Ferrarese \etal\ (1999). These authors have made a number of small revisions
to the earlier results for purposes of homogeneity.
A bigger change is that we
have switched to the new extinctions of Schlegel, Finkbeiner, \& Davis
(1998, hereafter SFD), which are determined from the COBE/DIRBE and
IRAS/ISSA dust maps.  These have better spatial resolution, uniformity,
and accuracy than the old estimates derived from H{\,\sc i} maps
(Burstein \& Heiles 1984, hereafter BH).  While there is a good
correlation between the two sets of reddening estimates, there is also a
significant zero-point offset.  SFD find a mean extinction that is
greater by 0.02~mag in $E(B{-}V)$ than that of Burstein and Heiles.
Therefore when we correct our SBF magnitudes and colors according to
these new extinctions, our \Mi\ zero point should change.

Increasing the extinction estimate to any given galaxy (while keeping
a fixed calibration) causes the SBF distance to that galaxy to increase.
This is due to the sensitivity of \Mi\ to \vi\ color;
if the extinction is changed by $\delta E(B{-}V)$, the distance modulus
changes by
\begin{equation}
  \delta\dmod \,=\, [\,4.5\,(A_V{-}A_I) - A_I] \times \delta E(B{-}V)\,,
 \label{eq:dMm}
\end{equation}
where $A_V$ and $A_I$ are the ratios of the total $V$ and $I$ band
extinctions to the selective $E(B{-}V)$ extinction.  For the average
increase of $\delta E(B{-}V) \sim +0.02$ in going from BH to SFD
extinctions, we find a mean $\delta\dmod \sim +0.1$ mag.  However, if all
the calibrator galaxies underwent this same average shift in extinction,
then the zero point we would derive through comparison with the Cepheids
would change by $\delta \Mi \sim +0.1\,$mag, and our overall distance
scale would not change.  This is basically the case for the
group-derived calibration, which now gives $\Miz = -1.61\pm0.03$ from a
total of 6 groups.  However, the direct Cepheid-galaxy calibration did
not change, since the mean BH and SFD extinctions towards these spirals
are the same.  For M31, which is excluded from the SFD map, we follow
Bianchi \etal\ (1996) and Ferrarese \etal\ (1999) in adopting the BH
value of $E(B{-}V) = 0.08$, and then we use the SFD extinction ratios
for consistency.  Thus, the galaxy calibration still gives a weighted
average of $\Miz = -1.80\pm0.08$, from 6 individual comparisons
now, and a median of $\Miz = -1.74$. The group and galaxy 
calibrations have diverged from a 0.08~mag difference in SBF-I to a
0.13--0.19~mag difference with the SFD extinctions.

In contrast to the approach of SBF-I, we have decided to adopt the
direct galaxy calibration rather than the group calibration.  This
fairly major shift is inspired by new worries that the \hst\
Cepheid-bearing spirals may be systematically in the foreground for
Virgo and other groups (see Ferrarese \etal\ 1999).  Since we are not
confident about the formal error estimates for either the
spiral SBF or Cepheid distances, we adopt the median rather than the
weighted average of the spiral zero points.  This new calibration
is within 0.07~mag of the theoretical calibration
from stellar population models of Worthey (1994) (see discussions by
SBF-I and Blakeslee \etal\ 1999).  The new SBF calibration derived with
all the latest measurements and the SFD extinctions and used in this
paper is
\begin{equation}
\Mi \;=\; -1.74 + 4.5\,[\viz - 1.15] \,.
\label{eq:newcal}
\end{equation}
It is slightly fainter than the SBF calibration
of Ferrarese \etal\ for two reasons: they used the extinction
ratios from Cardelli, Clayton, \& Mathis (1989) whereas we have used the
ones recommended by SFD, and we have used a median rather than an
average.  Despite the numerical coincidence of our new calibration
with SBF-I, the change due to the 
increase in average extinction [Eq.~(\ref{eq:dMm})] means that
our distance moduli increase on average by $\sim\,$0.1 mag.
We provide complete details about our choice of zero point in Appendix B.

Finally, the survey data reductions were not yet complete at 
the time of SBF-I.  In particular, we have added a large body
of new measurements from the 2.4\,m Las Campanas telescope that
were generally taken under superior conditions and
have enabled us to improve some distances.
From detailed comparisons of these new data to older, poorer
quality data, we discovered that bad data sometimes yield
spurious SBF signals.  We have found two simple quantities
useful in comparing data qualities:
$\PD = \PSF\times v_{\rm CMB}$, where \PSF\ is the full-width at
half-maximum of the point spread function in arcseconds and 
$v_{\rm CMB}$ is the CMB velocity in units of 1000 \kms, and
$q = \log_2(\#e/\overline{m} / \PD^2)$, where $\#e/\overline{m}$
is the number of electrons that would be detected in the image
from an object of magnitude $\overline{m}$ (estimated from the CMB
velocity), and $\PD^2$ is proportional to the metric area within a
resolution element, hence the number of stars for fixed surface
brightness. 
When $\PD > 2.7$ or $q < 0$, the observation is marked as
unreliable and ignored (see \S\ref{ssec:bias}).
For multiple observations of a given galaxy, we adopt a weighted
average if the difference in \PD\ is less than 0.3 or if each 
individual observation has $\PD < 1.3$; otherwise, we reject the
observation with the larger \PD.

The entire SBF Survey data set will be published in 
a forthcoming supplementary paper (Tonry \etal\ 1999,
in preparation).

\subsection{HST Data}

We incorporate eight distances from \hst\ SBF observations:
four from Lauer \etal\ (1998), NGC~4373 from Pahre \etal\ (1999),
and three from our Cycle 6 program (Dressler \etal\ 1999).
We have adjusted these distances slightly from the published ones to
agree with the SFD extinctions and have revised the zero point from
Ajhar \etal\ (1997) to agree with our new $I$-band zero point. 
Although the results presented here do not change 
significantly when the \hst\ data are included, they do become more robust
to covariances between, for example, $H_0$ and infall amplitudes.

\subsection{Biases and Uncertainties}
\label{ssec:bias}

When we compare these \hst\ distances with the difficult ground-based
efforts beyond 3000\kms\ or so, we find that there may be a bias in the
ground-based data that correlates with \PD.  Typical ground-based data for
more nearby galaxies agree well with \hst\ measurements (Ajhar \etal\
1997; see also the discussion by Blakeslee \etal\ 1999).  However, at
$\PD=3$ we find that $d_{\rm g\mbox{-}b}/d_{HST}$ ranges between 1 and 0.8,
but we do not have enough \hst\ observations to be confident that this is
the onset of a bias in the ground-based data.  If it truly is a bias, we
believe the source could have origins including unresolved dust and
structure, unresolved color gradients, and instrumental effects, and we
cannot predict reliably when it may be present.  We therefore restrict
ourselves to $\PD < 2.7$.  When we fit our models to subsets of the data
with different \PD\ cuts, we find no change in the parameters to within
the errors for $2.0<\PD<3.5$, which gives us confidence that we are not
affected by a distance-dependent bias.

As in SBF-I, we try to estimate the accuracy of our error estimates by
looking at the scatter within groups.  Since we do not know the radial
extent of groups, we ask how much cosmic scatter in $\overline M$ is
necessary if the groups have no radial depth at all, and find the answer
is 0.10 mag.  Conversely, if we assume that groups have the same depth
as breadth then the SBF cosmic scatter is 0.06 mag.  Including this 0.06
mag of cosmic scatter in \Mi, the quartiles of the distribution in
error in SBF distance modulus are 0.17, 0.21, and 0.29 mag, so this is a
fairly sensitive test that the SBF error estimates are accurate.
(Note that these errors are dominated by the observational compromises
inherent in doing such a large survey on relatively small telescopes;
in almost all cases the distances could be significantly improved.)

Because the differential volume
increases with distance, there will be more galaxies scattered
down in distance by the errors than are scattered up.
The level of this Malmquist-like bias in our sample is difficult to quantify.
We have the usual problem of not knowing what the true distribution
of galaxies is, and we may also be subject to possible biases and selection
effects which depend on distance.  We try to minimize these effects
by tailoring our exposure times and seeing conditions
to the distance of the galaxy we are observing.  
In addition, we are not actually sampling a population that
increases rapidly with distance (mostly we tend to be seeing 
early-type galaxy in groups).
Nevertheless, if we calculate the $e^{3.5 (\delta d / d)^2}$ 
correction (Lynden-Bell \etal\ 1988) for the median error,
we get a bias factor of 1.038.  
Including this correction in the models described below
does not change the results, and in fact it slightly increases the
$\chi^2$ values.  

The level of bias due to the irreducible cosmic scatter in \Mi\ is
very small. We preferentially measure galaxies to be too close because 
of their cosmic scatter at the 0.3\% level (assuming 0.06 mag of cosmic
scatter).  Moreover, unlike the case for other methods, such as the
``forward'' Tully-Fisher relation, $D_n$-$\sigma$/fundamental plane, 
or SN\,Ia, that use some of the same observables 
(magnitude, effective radius, surface brightness) for the distance
measurement as were used in the selection, the cosmic scatter in
the \Mi\--\vi\ relation is not known to correlate with any observables
measuring galaxy size or luminosity.  The bottom line is that we
can safely ignore bias correction for cosmic scatter, and because
the dubiously applicable Malmquist bias factor of $\sim\,$1.04
yields no measurable improvement, we ignore it.

The galaxy velocities come from ZCAT (Huchra \etal\ 1992, version
dated 1998 November 23), and are converted to the CMB frame 
according to Lineweaver \etal\ (1996), and to the Local Group
frame as defined by Yahil \etal\ (1977).  (A recent study by 
Courteau \& van den Bergh 1999 finds almost exactly the same 
Local Group reference frame.)

There are 336 galaxies in our sample, which is reduced to 295 after
applying the $\PD<2.7$ cut.  We restrict another 10 of these from
contributing to the model fits, three
because we think the SBF distance may be in error despite passing the
\PD\ cut, and seven because their velocities are quite discordant with
our model velocity distribution function.  The former cases include a
messy galaxy at fairly low galactic latitude with marginal seeing
(NGC~6305), and galaxies with extremely blue color where we
do not believe our calibration is reliable (NGC~5253 and IC~4182).  The
latter cases include three galaxies in Cen-45 (NGC~4709, NGC~4616, and
D~45), NGC~1400 (the high velocity companion of NGC~1407 in Eridanus),
NGC~4150 (a Coma~I galaxy with a very low velocity), and NGC~4578 and
NGC~4419, which we think have just passed through Virgo at high speed
and therefore have extremely unusual velocities for their location.

\section{Models of Large Scale Flows}

\subsection{Overview}

In this paper
we shall limit ourselves to a simple, parametric model for the
velocity field of galaxies. 
In addition, we try to avoid {\it all} group information.  The
assignment of galaxies to groups has had a checkered history, and we
prefer to deal with the virial velocities directly, rather than trying
to average them away.

There are two methods commonly used for constructing a merit function
for a peculiar velocity model $V(r)$.  The first is to assemble 
\begin{equation}
\chi^2 = \sum {(v_{\rm obs} - V(r))^2 \over \epsilon_{\rm v}^2}\,,
\label{eq:chi2}
\end{equation}
where $\epsilon_{\rm v}^2 = \delta r^2 + \delta v^2 + \sigma_{\rm v}^2$, and
$\delta r$ is the uncertainty in position expressed in units of \kms,
$\delta v$ is the (small) uncertainty in the measurement of velocity,
and $\sigma_{\rm v}$ is an allowance for thermal or virial velocity
dispersion (cf. Hudson \etal\ 1997).   This
tacitly assumes that the slope $dV(r)/dr = H = 1$ everywhere, otherwise
it is not correct to use distance error $\delta r$ in the denominator.
This can be very problematic for a model which has a significant
disturbance to the Hubble flow, for example in the local supercluster.
It is possible to ``correct'' $\delta r$ by a spatially varying
$H(r)$, but then this $\chi^2$ is no longer maximum likelihood (unless
it includes the often neglected $-2\ln\sigma$ term) and will tend to
bias parameters in such a way as to maximize $H \delta r$.  The other
weakness of this approach is that it involves some arbitrariness of 
when objects should be grouped and what virial velocity $\sigma_{\rm v}$
should be allocated for them.

The second common method (e.g., Davis, Strauss, \& Yahil 1991; Strauss
\etal\ 1992; Shaya, Tully, \& Pierce 1992) is to invert the velocity
model and use the observed velocity to provide a model distance which
can be compared with the observed distance.  This has the advantage that
it compares the distances and model directly, but it has the
disadvantage that it cannot cleanly deal with virial velocities,
necessitating grouping to the point that virial velocity is reduced to a
negligible magnitude compared to the flow velocity and distance error.
It also has the disadvantage that the inversion of $V(r)$ can be
multiple valued, and is necessarily multiple valued in the regions of
space where a supercluster flow is most prominently seen.  One is then
forced to an arbitrary or probabilistic choice of which model distance
to use.

What we have chosen to do is to accept the fact that there is a
distribution function of galaxy velocity which varies from place to
place, both in mean velocity (large scale flow) and in dispersion.
Our models consist of a velocity distribution function at each
location, $P(v|r)$, which we take to be Gaussian of mean $v_0(r)$ and
dispersion $\sigma_{\rm v}(r)$.  A given distance measurement, itself a
probability distribution, is multiplied by this model distribution
function and integrated over distance, giving a velocity probability
distribution.  The merit function is then the likelihood, the product
of these velocity distributions evaluated at the observed velocities.

The components of our model for the mean velocity $v_0(r)$ include a
Hubble flow of amplitude $H_0$, an constant dipole velocity $\vec w$,
possibly a quadrupole $Q$ with zero trace which acts like an anisotropic
Hubble constant, a density parameter $\Omega_M$ for the universe, and
attractors  
which are assumed to be spherical power laws in density with a core and
cutoff radius.  The positions of these attractors can be free parameters
as well as the power-law exponent, the core and cutoff radii, and the
overall normalization.
Our model for the velocity dispersion as a function of position,
$\sigma_{\rm v}(r)$, consists of
the quadrature sum of an overall thermal velocity, a virial component
at the center of each attractor, and possibly other (non-attracting)
virial components at the locations of groups such as the Fornax cluster.
These components have a velocity dispersion which varies spatially 
as $\exp(-r^2/2 r_{\rm virial}^2)$. In principle we have enough distance
accuracy to constrain both the background thermal component and these
virial components (except for GA/Centaurus at the limit of our survey)
with a smoothing length of perhaps 4~Mpc.  
In practice we fit only for the thermal background 
and Virgo velocity dispersions and put in by hand
the Fornax (235\kms) and Centaurus (500\kms) dispersions
with $r_{\rm virial} = 2$~Mpc.
Unless otherwise stated, we use a thermal velocity dispersion
of 187\kms, which is the best-fit value in our
most refined models.

Figure \ref{fig:probs} illustrates how this works.  
\insfig{
\begin{figure}[t]
\epsscale{1.0}
\plotone{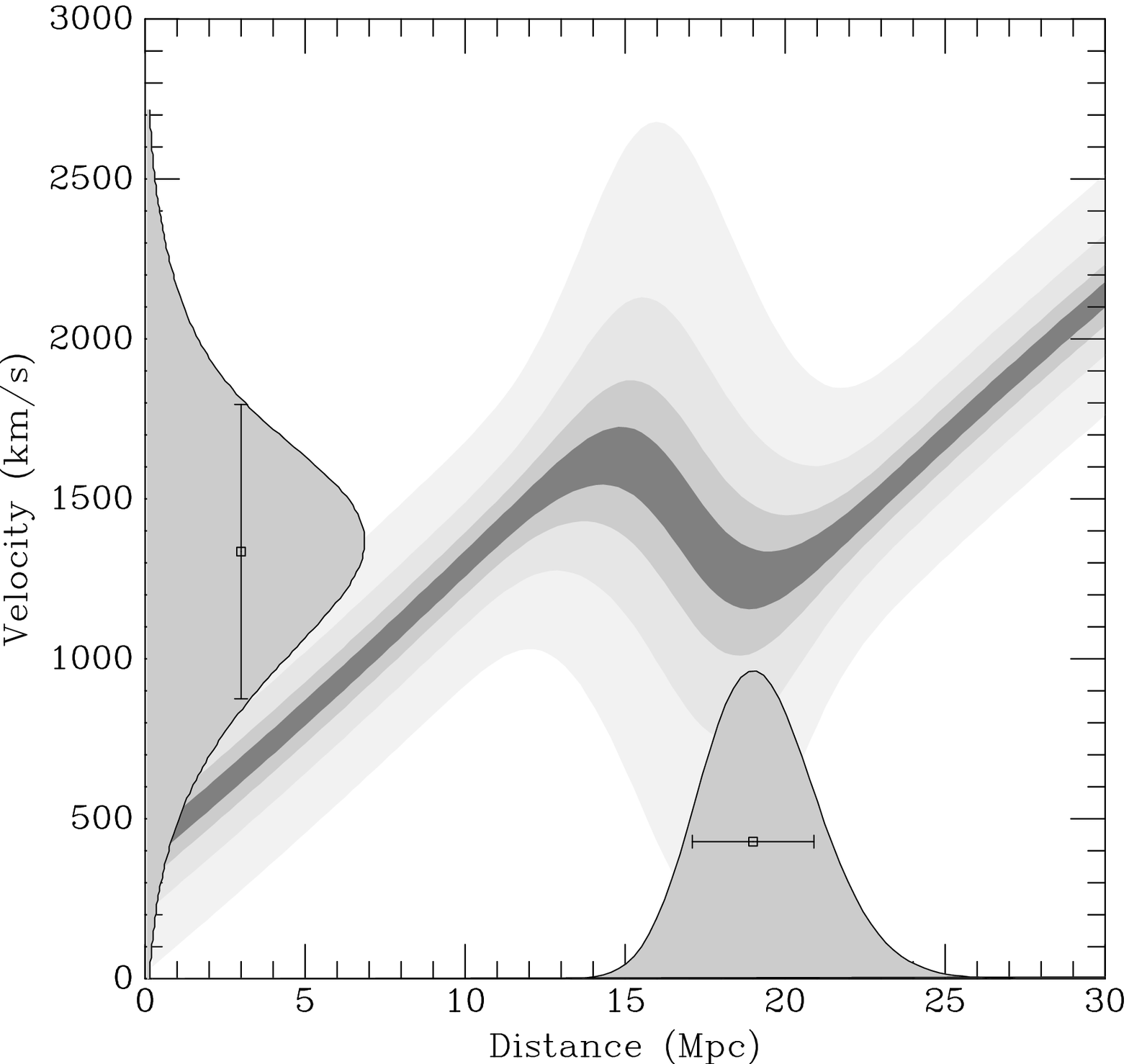}
\caption[f01.eps]{
The model velocity distribution function $P(v|r)$ is shown for a line
of sight which passes through the Virgo cluster.  The different
grayscale levels show the 2, 1, 0.5, and 0.2 sigma points on the
velocity distribution at a given distance.  A distance
observation is shown on the abscissa as a distance probability
function, and the resulting velocity probability function is shown on
the ordinate.  This is evaluated at the observed velocity and forms a
term in the likelihood product.
\label{fig:probs}}
\end{figure}
}
The model $P(v|r)$ is shown
along a line of sight near the Virgo cluster.  The mean velocity has a
Hubble flow term modified by a peculiar velocity which grows
approximately as $r^{-1}$ (for an $r^{-2}$ density distribution), but
then rolls off to zero at the center of the cluster because of the
core radius in the density distribution.  The velocity
dispersion at each location consists of the quadrature sum of the
thermal background component and a cluster velocity dispersion
with an amplitude of about 650\kms, which declines away
from the Virgo cluster as a Gaussian of width 2~Mpc.  
Again, these parameters are all variable components of
the velocity model.


With three shape parameters (power-law slope and core and cutoff
radii), our attractor models have much radial flexibility.
The models can emulate anything from a centrally concentrated mass to
an extended distribution with divergent total mass.

The question of spherical symmetry is harder to justify.  We
note, however, that the galaxy distribution in the Virgo supercluster
is not completely flat, that the potential is much rounder than a
mass distribution, and that the large scale flows we see today
arise from the time integral of the potential so that a spherical
model might represent the velocity data quite well even though the
potential is today significantly flattened.  Moreover,
we do allow for the possibility of a quadrupole correction to the flow
model, and this can start to modify a spherical flow into a more
flattened one.  As we shall see, there are indications that the Virgo
Attractor is not spherical.

We convert a mass distribution to a peculiar velocity distribution
using the usual non-linear ``nested Friedmann universe'' model, where 
each mass shell evolves from the Big Bang according to its interior mass
density.  We do not do this computation exactly, however, but rather use
the Yahil (1985) ``$\rho^{1/4}$ law'' approximation
\begin{equation}
u_{\rm infall} = {1\over3}\, H_0\, r \; \Omega_M^{0.6}\; \delta\, (1+\delta)^{-1/4}\,,
\label{eq:upec}
\end{equation}
where $r$ is the radius of the shell and $\delta$ is the mean mass
density interior to the shell in units of the background density.
Giavalisco et al. (1993) examined this approximation with N-body
simulations and found that it is remarkably accurate for 
$-0.5 < \delta < 20$, which is the range of interest here.
We do not expect our models to be valid all the way into the
virialized core of the attractor ($\delta > 200$), so
our velocity distribution function rolls over 
to zero in the mean at the centers of the attractors with a
compensating rise in virial velocity dispersion.

It is common to recast this parametrization in terms of a local
overdensity and peculiar velocity, which then permits $\Omega_M$ to be
derived from this formula.  However, because of our desire to avoid use
of the galaxy distribution, we lack any a priori information about
$\delta$, and thus have very little constraint on $\Omega_M$.  Only the
non-linear part of Eq.~(\ref{eq:upec}) provides any constraint at all, and
clearly that is not going to be very reliable.  Instead, we choose a
single value for $\Omega_M$ and derive the overdensities for the model
attractors from the fitted infall velocities.

\subsection{Details of Models}

The input to our models includes for each galaxy the position on the
sky, distance and error, and velocity.  The fitting program reads these
data and also accepts a variety of model components, including cosmology
($H_0$, $\Omega_M$, and dipole velocity $\vec w$), quadrupole (five
components plus origin and cutoff), extended attractors, thermal
components, and compact attractors.  The program computes a model
velocity distribution function for each observed data point, evaluates
the probability of the observed velocity given the velocity probability
distribution, forms the likelihood sum from the probabilities $P$:
${\cal L} = \sum \ln P + \hbox{constant}$, and searches parameter space
for a maximum in $\cal L$.

The model velocity field consists of a mean velocity at each point and
associated Gaussian velocity dispersion.  The Hubble constant and
dipole velocity give a contribution to the model velocity 
$\vec v_{\rm model}$ at $\vec r$ of $\vec w + H_0 \vec r$.
The quadrupole contributes a velocity at $(x,y,z)$ of
\begin{equation}
      e^{-r^2 / 2r_{\rm quad}^2}
      \left(\matrix{Q_{xx}&Q_{xy}&Q_{xz}\cr Q_{xy}&Q_{yy}&Q_{yz}\cr 
                  Q_{xz}&Q_{yz}&Q_{zz} \cr}\right)
      \left(\matrix{x-x_Q\cr y-y_Q\cr z-z_Q \cr}\right) \,.
 \label{eq:matrix}
\end{equation}
The quadrupole matrix is forced to zero trace by insisting that 
$Q_{yy} = -Q_{xx} - Q_{zz}$, which ensures that the monopole $H_0$
parameter carries the net Hubble expansion proportional to distance.
Unless otherwise specified, the origin of the quadrupole, $\vec r_Q$,
is taken to be the Local Group, and the cutoff radius, $r_{\rm quad}$
is taken to be infinity.

Extended attractors are modeled by starting with spherical density
distributions of the form
\begin{equation}
\rho(r) = {\rho_0\over (1+r^3/r_c^3)^{\gamma/3}} \,,
\label{eq:rhoofr}
\end{equation}
where $\gamma$ is the power-law exponent and $r_c$ is a core radius.
This integrates nicely to give an enclosed mass function to which we
append an exponential cutoff and divide by volume to get a mean
enclosed overdensity:
\begin{equation}
\overline\rho(r) = {\rho_0\,e^{-r/r_{\rm cut}}\over1-\gamma/3} 
\left( r\over r_c\right)^{-3} \,
\left[(1+{r^3\over r_c^3})^{1-\gamma/3} - 1\right].
\label{eq:rhobar}
\end{equation}
Since the central density is finite, the flow velocity rolls over at
approximately $r\approx r_c$ and approaches zero as $r\rightarrow0$.
We ordinarily normalize these attractors in terms of a velocity
amplitude at our location by using the inverse of Yahil's $\rho^{1/4}$
law to convert such a velocity into an expression for $\rho_0$.  The
cutoff radius $r_{\rm cut}$ ensures that these models have zero net
mass, hence do not bias $H_0$ high by making a net local modification
to $\Omega_M$.  For typical values of $\gamma$, $r_{\rm cut}$ is
approximately where the density becomes negative and
$\overline\rho(r)$ starts to decrease back to zero.

The covariance between the parameters $\gamma$, $r_c$, and $r_{\rm
cut}$ makes it difficult to interpret values for any one parameter in
isolation.  We will strive to indicate what is well constrained by the
models (e.g., mass of attractor, or run of peculiar velocity with
radius) and what is not.

As mentioned above, the model velocity dispersion consists of the
quadrature sums of background and extended thermal components.  Each
component has a velocity dispersion with a central value that falls off
spatially according to a Gaussian of width $r_{\rm virial}$.  These
thermal components can either be specified explicitly (the cosmic
dispersion is taken to be centered on the origin with an infinite
Gaussian radius; Fornax is centered on the location of the Fornax
cluster and has a core radius of 2~Mpc), or as part of an extended
attractor.  The thermal component of an extended attractor is centered
on the attractor, has a parameter describing its central virial velocity
dispersion, and shares the core parameter $r_{\rm virial}$ with the core
radius $r_c$ of the mean infall.  (We expect $r_{\rm virial}$ to be
close to $r_c$, and in our models we set the two radii to be the same.)

It is also possible to specify ``Compact Attractors'',
density distributions which are Gaussians.  These carry five
parameters, their location, their Gaussian radius, and their
amplitude.  The amplitude is given in terms of the total mass and a
fiducial radius, $R_{\rm fid} \equiv 50$~Mpc:
\begin{equation}
v_{\rm amp} = {G M_{\rm tot} \over R_{\rm fid}^2}\, {1 \over H_0} \,.
\label{eq:vamp}
\end{equation}
For a Gaussian with $\rho = \rho_0 \exp(-r^2/2\tau^2)$, the
enclosed mass is
\begin{equation}
M(r) = 4\pi\rho_0 \left[ \sqrt{\pi\over2} \tau^3
         \hbox{erf}\left(r\over\sqrt{2}\tau\right) - 
         \tau^2 r \exp\left(-{r^2\over2\tau^2}\right)\right] \,,
\end{equation}
and $M_{\rm tot} = 4\pi\rho_0 \sqrt{\pi\over2} \tau^3$.
The expression for $M(r)$ is converted to a $\delta$ and then used in
Eq.~(\ref{eq:upec}).
The velocity amplitude can be made negative, which corresponds to a
negative density void.  Of course, the void cannot become deeper than
the mean cosmological density.  These compact attractors do not carry a
thermal component.


We make no attempt to create an overall self-consistent density and
velocity distribution---these components do not interact with one
another.  Having specified the various components, the mean velocity
is just the sum of the contributions of all the components, and the
velocity dispersion is the quadrature sum of the various contributors.

Once we have a model, we then go about evaluating how well it matches
our observations.
For Gaussian statistics, we would have that the probability for
making an observation $v_i$ at a location $r_i$ where our model
predicts a mean velocity $V(r_i)$ and Gaussian dispersion
$\sigma_{\rm v}(r_i)$ is 
\begin{equation}
P_i = {1\over\sqrt{2\pi}\sigma_{\rm v}(r_i)} \,
 \exp\left[-{1\over2}\left(v_i-V(r_i)\over\sigma_{\rm v}(r_i)\right)^2\right]
 \; \Delta v \,.
\end{equation}
The choice of $\Delta v$ is arbitrary but represents how closely 
the model adheres to observation.  
We define the likelihood ${\cal L} = \ln\prod P_i$ 
\begin{eqnarray}
{\cal L}  &=& -{1\over2}\sum \left(v_i-V(r_i)\over\sigma_{\rm v}(r_i)\right)^2   \\
 & & + \sum -\ln{\sqrt{2\pi} \sigma_{\rm v}(r_i)} \,+\, \ln\Delta v \,.
	\nonumber 
\end{eqnarray}
The first term in $-2{\cal L}$ is just
the usual definition of $\chi^2$, and we can evaluate the goodness of
fit according to the $\chi^2$ per degree of freedom. 
If the dispersions $\sigma_{\rm v}(r_i)$ do
not depend on the parameters, then the second two terms can be ignored,
since they are constant for a given set of observations and do not
affect the choice of parameters which maximizes the joint probability.

Our situation is not quite so simple, because we believe we have errors
which are normally distributed in {\it log} distance, and although our
model distribution function has a Gaussian distribution of velocity at
any point, the mean velocity and velocity dispersion vary as a function
of position, and hence are not a constant over the range where a given
galaxy might lie.  We deal with this for a galaxy of observed modulus
$\mu \pm d\mu$ by forming a distance probability distribution that
consists of 11 points between $\mu-2d\mu$ to $\mu+2d\mu$ weighted
according to Gaussian statistics.  The model is evaluated at each of
these 11 points, providing a Gaussian velocity distribution with some
mean and dispersion at each point.  We sum these 11 Gaussians in
velocity, weighted by their distance probabilities, to form a net
probability distribution for the velocity we expect to see for this
galaxy.

Our likelihood function is then simply formed by evaluating this
probability at the observed velocity and summing a negative,
normalized likelihood
\begin{equation}
{\cal N} = -2\sum\left[\ln P_i(v_i) + \ln(300\,\sqrt{2\pi})\right],
\end{equation}
where the constant term is introduced for $\Delta v$ in order to shift
the zero point of ${\cal N}$ into approximate agreement with $\chi^2$.
Deviations of ${\cal N}$ about its minimum, 
${\cal N} - {\cal N}_{min} = 2({\cal L}_{max} - {\cal L})$ 
are described approximately by a $\chi^2$ distribution, so we use
${\cal N}$ to evaluate goodness of fit and confidence intervals on
parameters.  We also calculate a variance for the velocity
distribution function we derive for each galaxy and use them to form
a traditional $\chi^2$.  This $\chi^2$ is useful for evaluating a
goodness of fit, but does {\it not} correspond to maximum likelihood
and is not minimized for the best fitting model.
\section{Fits of Model to Data}

The model described in the previous section carries a lot of
parameters, and there can be significant covariance between them, for
example $\gamma$ and $r_{\rm cut}$, or $w_x$ and $u_{\rm GA}$.
We will first present a sequence of models which motivate our choices
of mass components, starting with a Hubble flow, and ending with two
attractors contributing peculiar velocities.

We will subsequently look more carefully at these covariances and try
to show what is well constrained by our data and what is not.  For
example, we cannot independently choose a unique $\gamma$ and
$r_{\rm cut}$, but all our models give nearly the same run of peculiar
velocity between 6 and 20 Mpc from the center of the Virgo cluster.
Distances will often be given in terms of their SGX, SGY, and SGZ 
components in the supergalactic coordinate system, defined
in the RC3 (de Vaucouleurs \etal\ 1991).

It is extremely important not to force the models to conform to any
particular velocity reference frame.  For example, insisting on the
CMB frame and using an $r^{-2}$ density model for the Great Attractor
leads to a very large infall velocity because of its covariance with
$w_x$.  We use the CMB frame for velocity data, but we always fit for
an arbitrary dipole vector $\vec w$ as part of our models.

\subsection{A First Look}

Figure \ref{fig:hubble} shows all of the galaxies for which we
have SBF distances 
plotted in a Hubble diagram --- redshift in the CMB frame as a function
of distance.  The expansion of the universe is apparent, and a naive
linear fit, without regard for error bars or the fact that the abscissa
carries greater errors, yields a Hubble ratio of 73~\kmsMpc.
\insfig{
\begin{figure}[t]
\epsscale{1.0}
\plotone{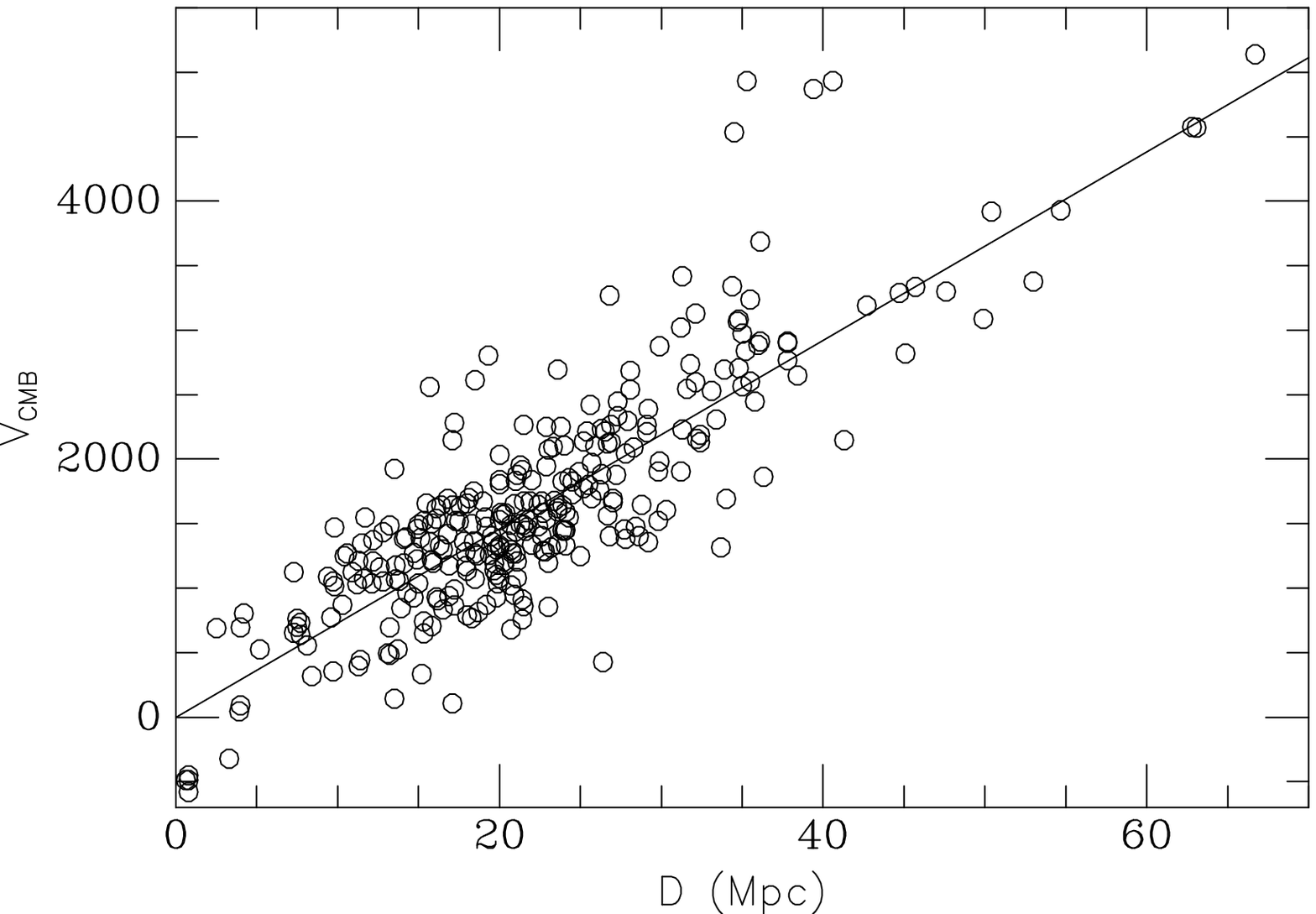}
\caption[f02.eps]{
Recession velocities in the CMB frame are plotted as a function of distance.
The line has a slope of 73\kmsMpc; 
the very deviant high points near 40~Mpc are Cen-45 galaxies.
\label{fig:hubble}}
\end{figure}
}
\noindent Figure \ref{fig:h0} focusses in on the more distant galaxies
with smaller errors and plots Hubble ratio as a function of distance.
\insfig{
\begin{figure}[t]
\epsscale{1.0}
\plotone{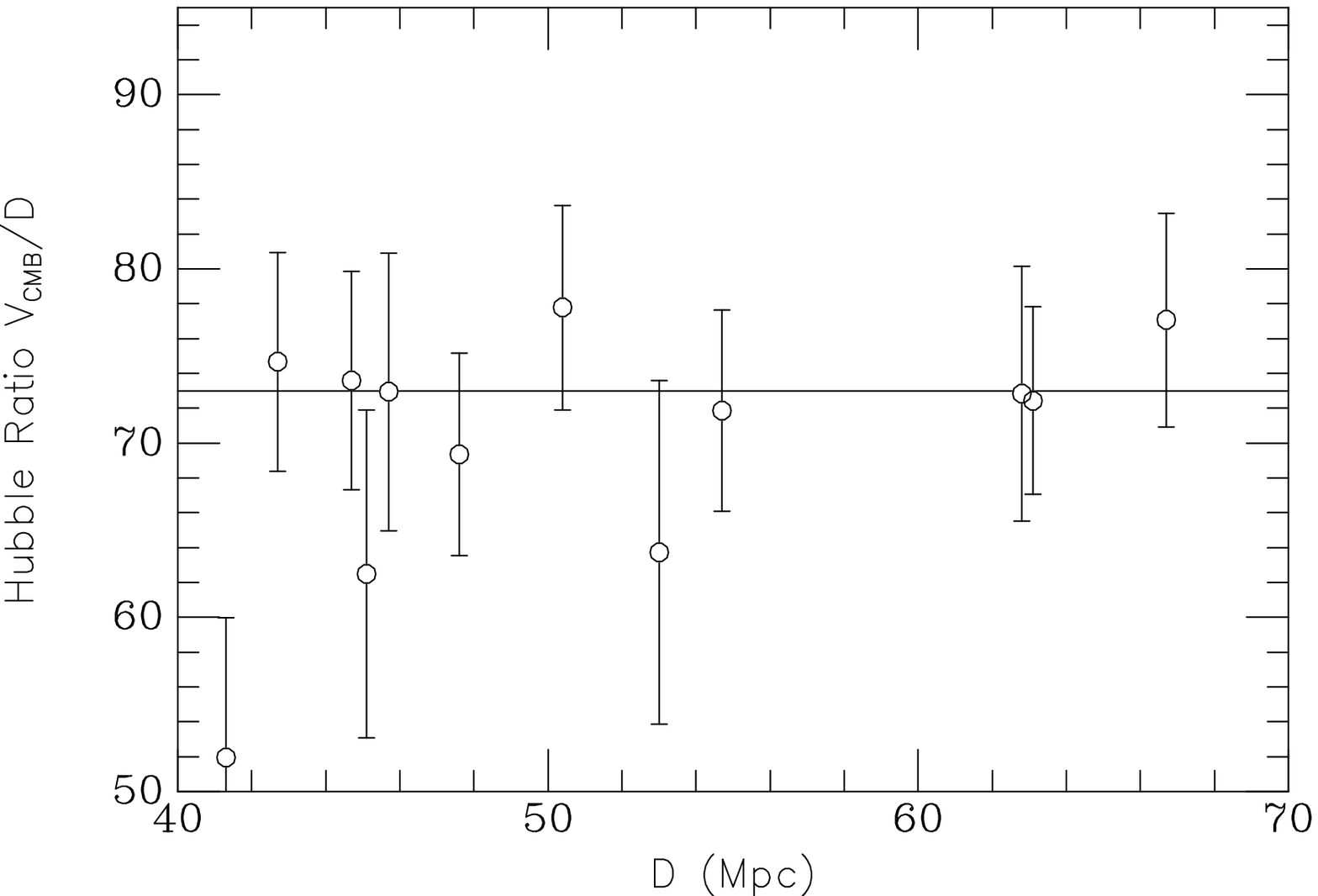}
\caption[figs/fig_h0.ps]{
Hubble ratios (in CMB frame) are plotted as a function of distance for
those points with small enough errors that the error in the Hubble
ratio is less than 10~\kmsMpc.
\label{fig:h0}}
\end{figure}
}
Local peculiar velocities cause scatter in the Hubble ratio at
distances of 40\Mpc, but it seems to settle down fairly well to an
asymptotic value of 73\kmsMpc.  However, as is seen in Figure
\ref{fig:hubble}, 
there is a lot of scatter in the Hubble plot, more than can be
explained by distance error.

The near field Hubble flow is shown in Figure \ref{fig:aniso} in
two different 
directions.  Velocities in the Local Group frame are used to avoid the
constant velocity offset incurred by using the CMB frame locally.
\insfig{
\begin{figure}[t]
\epsscale{0.8}
\plotone{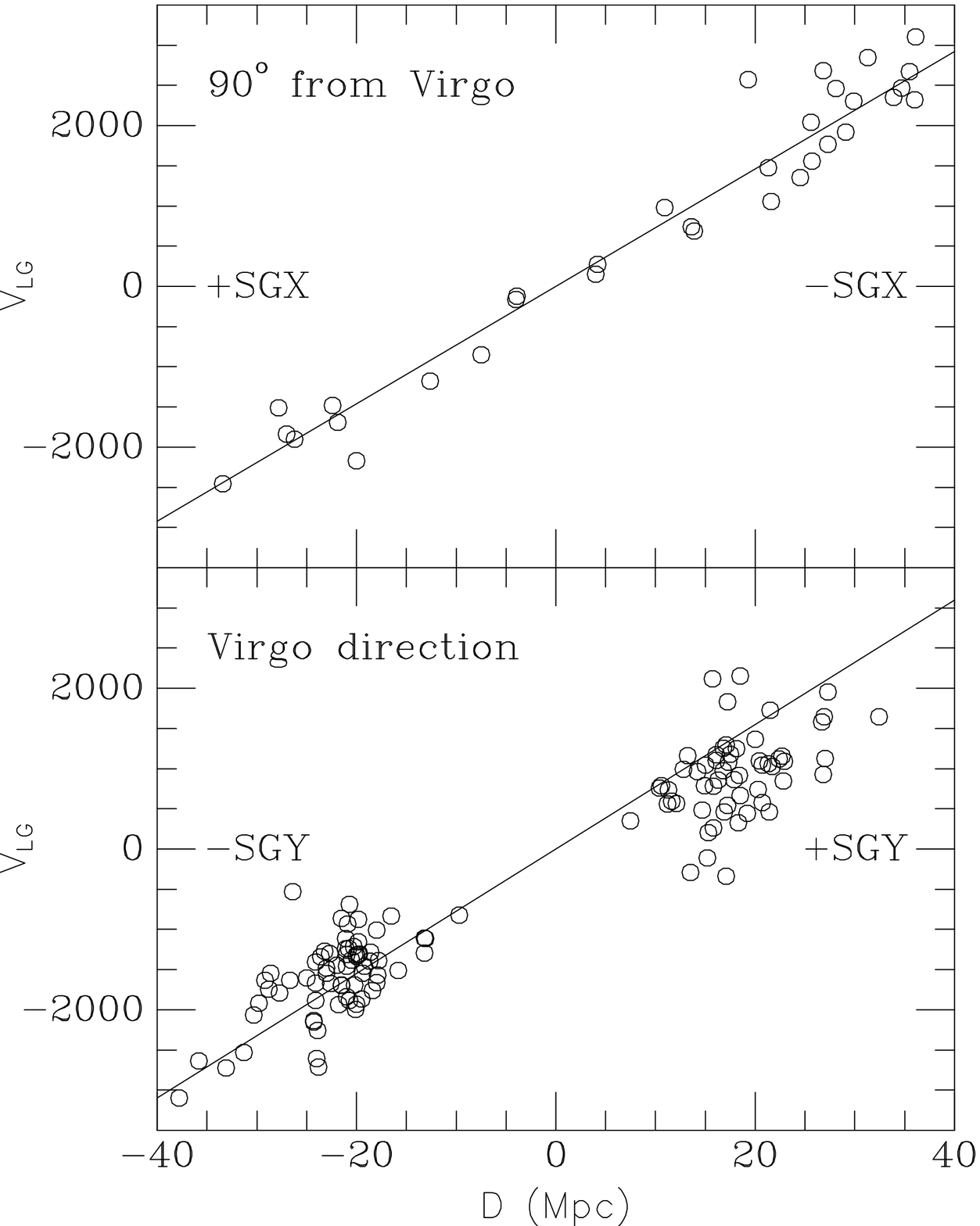}
\caption[f04.eps]{
Recession velocities (in Local Group frame) are plotted as a function of
distance in two directions.  The upper panel shows galaxies which lie
approximately in a $30^\circ$ (half angle) cone in the $\pm$SGX direction,
$90^\circ$ away from the direction of Virgo.  The lower panel shows
galaxies which lie within a $15^\circ$ cone towards Virgo (+SGY) and a
$45^\circ$ cone away from Virgo (which includes the Fornax cluster).  A
Hubble ratio of 73\kmsMpc\ is also drawn.
\label{fig:aniso}}
\end{figure}
} 
It is possible to discern the important features of the local 
large-scale flow.  The Virgo cluster in the lower pane at
$\hbox{SGY} = +20\,$Mpc lies several hundred \kms\ below the Hubble flow
--- this is often referred to as our ``infall velocity'' towards Virgo.
However, this is not the entire story, since the Fornax cluster at
$-20$\Mpc\ {\it also} has a smaller velocity than the Hubble flow,
but the quadrupole from a pure Virgo infall should cause it to have
a larger velocity (although it is quite near the quadrupole null at
45$^\circ$).  Likewise, the quadrupole from a pure flow towards Virgo
should cause the Hubble ratio to be smaller in a direction
perpendicular to Virgo, but in fact we find a ratio in the upper panel
which is slightly
{\it larger} than the nominal Hubble flow.  What we are seeing in these
local galaxies are the effects of two flows, one towards Virgo and
the other towards Centaurus in the $-$SGX direction.

\subsection{$H_0$ only}

We start by fitting the distances and velocities using a model which
includes $H_0$ and a peculiar velocity $\vec w$, for a total of
4 free parameters.  The sample consists of 285 galaxies.
Allowing 187~\kms\ of thermal velocity
but no additional virial velocities, we find
$H_0 = 70$ \kmsMpc\ and an overall dipole peculiar velocity of
$\vec w = (-330,+180,-80)\,$\kms\ in the CMB frame. 
The ${\cal N}$ for this model is 513, and
$\chi^2$ per degree of freedom (DOF) for this model
is $\chi^2_N = 2.08$.

An obvious shortcoming of this model is that there is no allowance for
virial velocities other than 187\kms\ of background thermal velocity.
If we add in additional thermal components for Fornax and
Centaurus (235 and 500 \kms, which will be used throughout)
and Virgo (650 \kms, which is about what we find when we fit for it),
${\cal N}$ drops considerably to ${\cal N} = 387$
and $\chi^2_N = 1.41$.
In the CMB frame we again get $H_0 = 71$ with a negligibly
different peculiar velocity of $\vec w = (-330,+200,-90)$.

Figure \ref{fig:h0fit} illustrates residuals of velocities after
removal of this model in the CMB frame.  In all these vector field
plots, the black points (falling stones) are blueshifted residuals
which are greater than 1-$\sigma$, the white points (rising balloons)
are redshifted residuals greater than 1-$\sigma$, and the gray points
are residuals which are less than 1-$\sigma$ in absolute value.  The
arrows indicate the magnitude of the residual according to 
the Hubble flow, i.e. an arrow of length 5\Mpc\ represents a
velocity residual of $5\Mpc\times H_0$\kms.  All the points
are projected onto the supergalactic $x{-}y$ plane, and the region where
the galactic plane cuts through is darker gray.  When galaxies are
members of a group we plot only the group residual, so as to prevent
the plot from becoming too overcrowded.
The main failing of this model is obvious.
The residuals near $(-3,+16)$ are the Virgo ``s-wave'', positive
peculiar velocities on the near side and negative on the far side of
Virgo.

\insfig{
\begin{figure}[t]
\epsscale{1.0}
\plotone{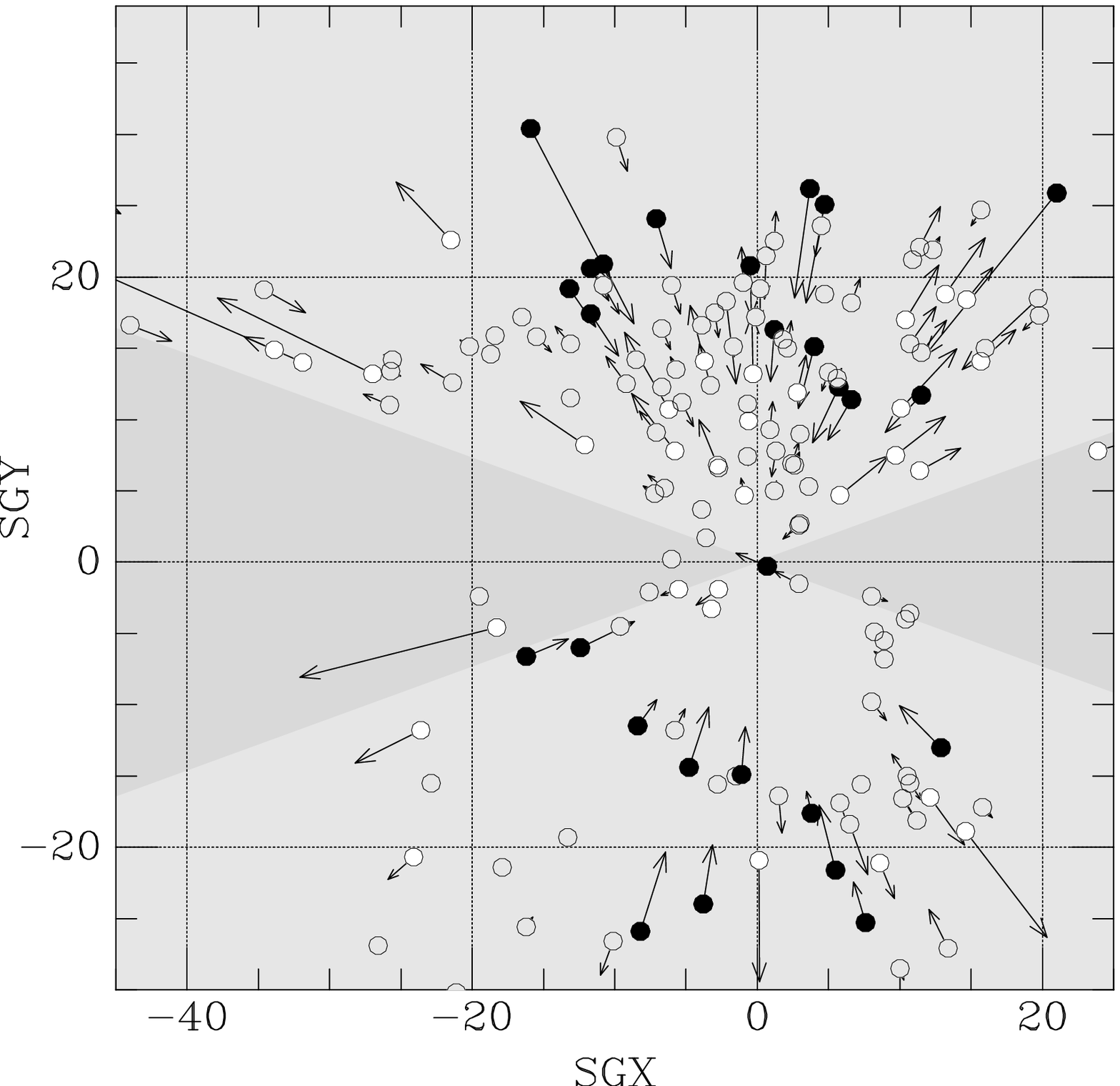}
\caption[f05.eps]{
Residual velocities after removal of a model Hubble
flow.  Residual velocities which are less than $1{-}\sigma$ are shown as
gray; greater than $1{-}\sigma$ are shown as black (approaching) or white
(receding).  For clarity only the group residual is shown for galaxies
in groups.
\label{fig:h0fit}}
\end{figure}
}

Another way of looking at the quality of the fit is to plot observed
peculiar velocity as a function of model peculiar velocity.
\insfig{
\begin{figure}[t]
\epsscale{1.0}
\plotone{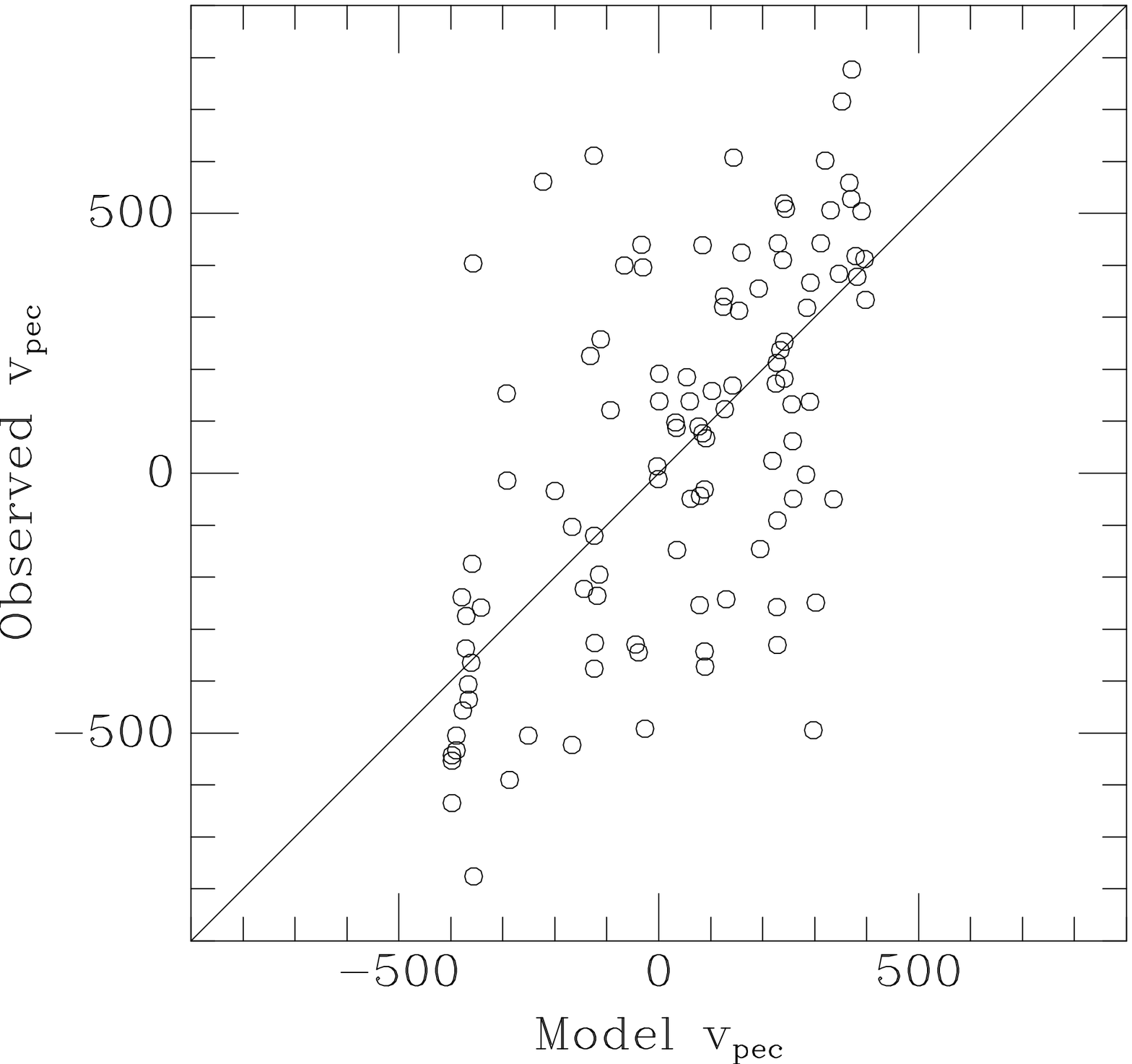}
\caption[f06.eps]{
Observed peculiar velocities are compared to model peculiar velocities.
Only galaxies with model velocity uncertainty less than 250\kms\
(essentially $[(H_0\,\delta r)^2+\delta v_{\rm virial}^2]^{1/2}$) are
shown, thereby excluding galaxies where a significant virial velocity is
expected and showing the quality of match between the model
and observed large scale flow field.
\label{fig:h0resi}}
\end{figure}
}
In order not to be overwhelmed by galaxies which have a large virial
velocity (which does not appear in the mean model velocity), we
plot only those galaxies where the model velocity uncertainty is less
than 250\kms.  There appears to be a correlation in Figure
\ref{fig:h0resi} between 
model and observed, but the agreement is not good.

\subsection{Virgo Attractor Infall}

Adding an extended ``Virgo Attractor'' (VA) with $\gamma{\,=\,}2$
near the location of the Virgo cluster at $(-3.1, 16.6, -1.6)$\Mpc,
we find an infall velocity at the Local Group of
$u_{\rm VA} = 101$\kms, a Virgo thermal velocity dispersion
of $\sigma_{\rm VA} = 676$\kms, and a cutoff radius 
$r_{\rm VA} = 9.3$\Mpc.  The Hubble constant rises slightly to 
$H_0 = 72$, and $\vec w = (-320,195,-82)$ in the CMB frame.
This position for the VA is the best fit location of a somewhat more
elaborate model which also includes a Great Attractor.  By itself the
Virgo Attractor settles to a location of $(-3.7,15.4,-1.7)$.
The likelihood is ${\cal N}=345.0$, which is a significant improvement
for just three additional parameters.
Figure \ref{fig:virgofit} shows the residuals from this
model.  The ``s-wave'' of Figures \ref{fig:hubble} and \ref{fig:h0fit} 
has been removed; the black
and white points are evenly distributed around the center of the model
attractor at $(-3,+16)$.  However, a quadrupole
signature remains evident, by which residuals in the $\pm$SGY direction
tend to be 
blueshifted and residuals in the $\pm$SGX direction tend to be redshifted.
\insfig{
\begin{figure}[t]
\epsscale{1.0}
\plotone{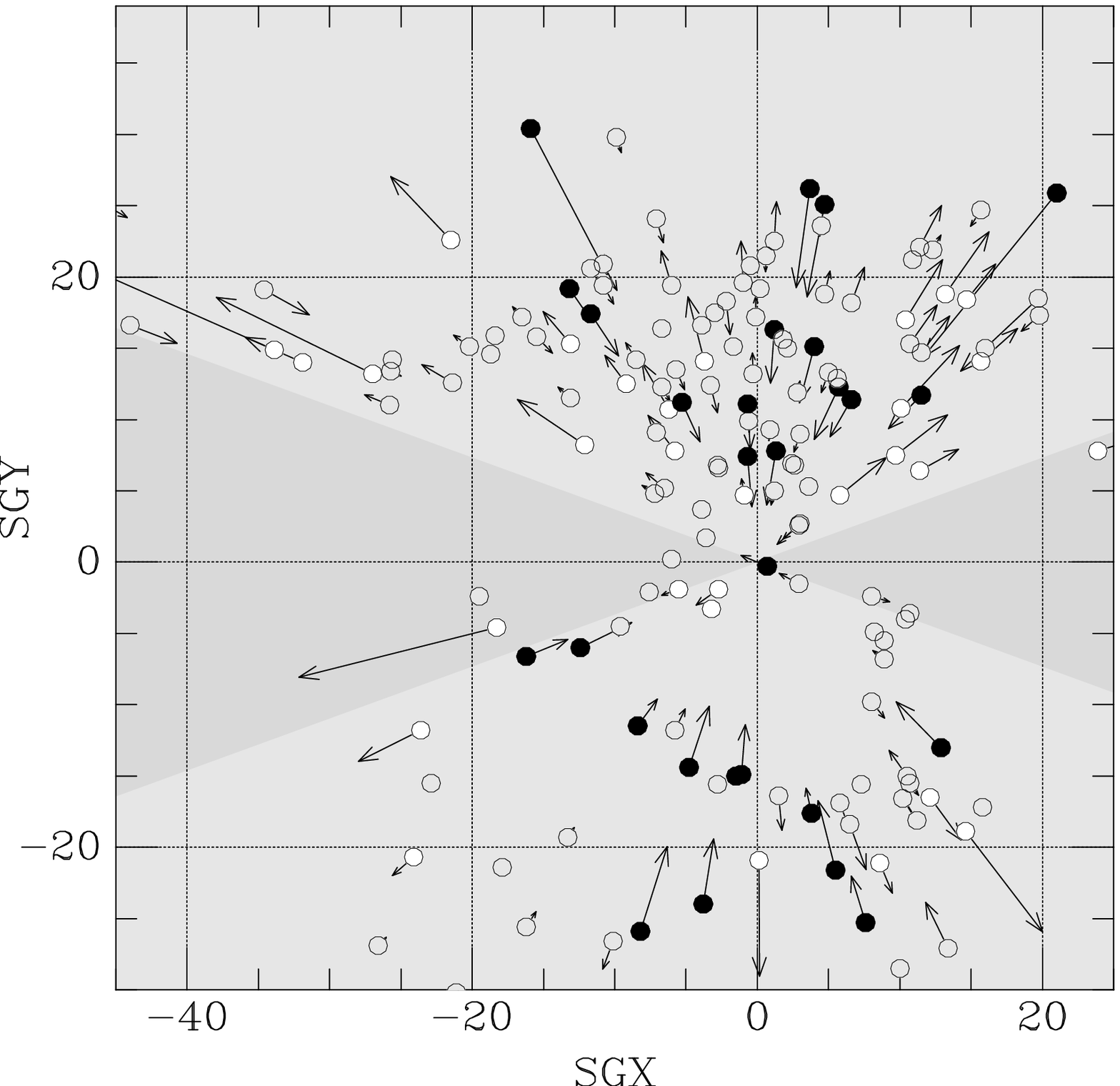}
\caption[f07.eps]{
Residual velocities after removal of a model
consisting of Hubble flow, Virgo Attractor, and constant velocity vector.
\label{fig:virgofit}}
\end{figure}
}
Figure \ref{fig:virgoresi} shows the galaxy by galaxy residuals from this
model.
\insfig{
\begin{figure}[t]
\epsscale{1.0}
\plotone{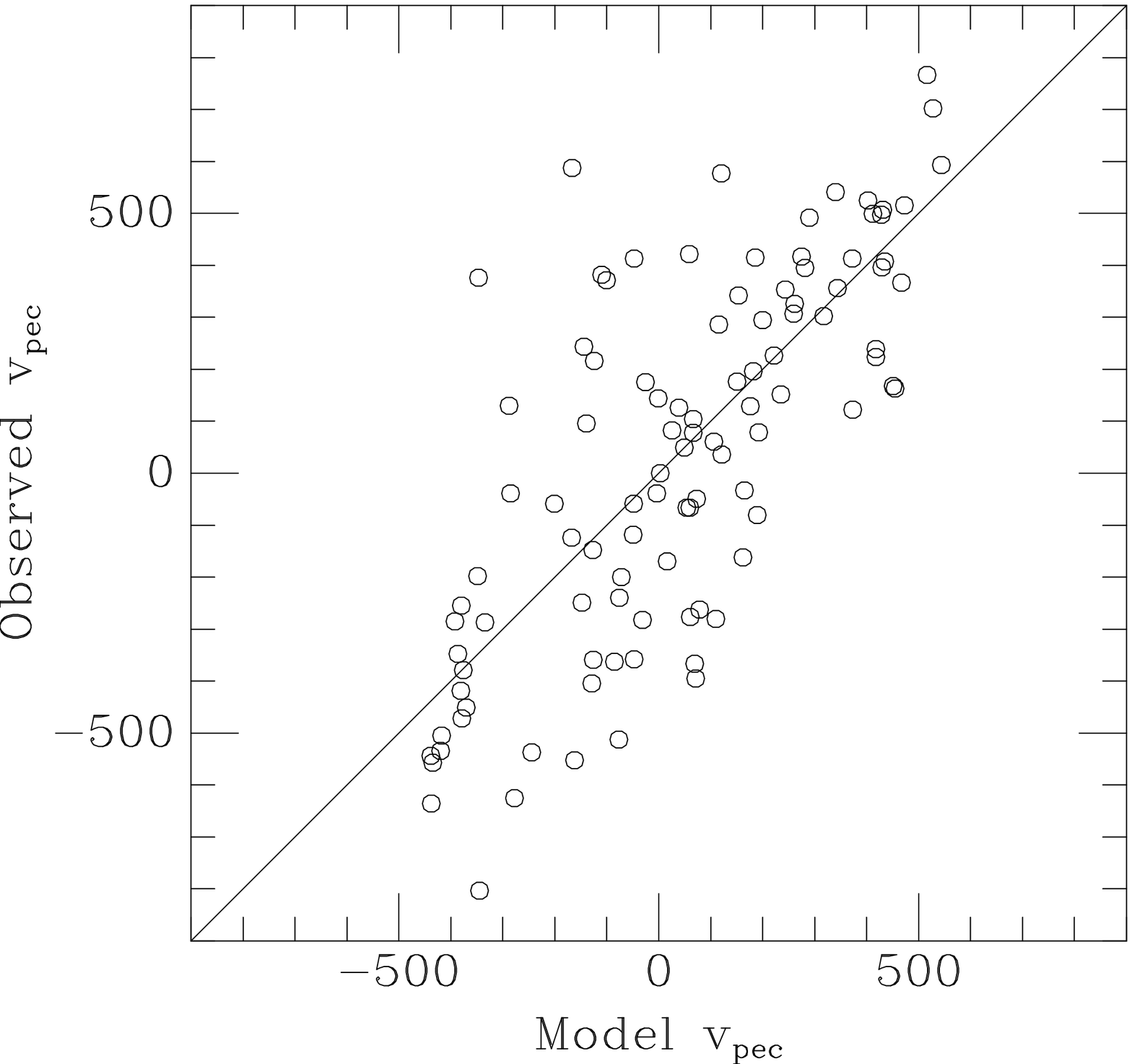}
\caption[f08.eps]{
Comparison between model and observed residual velocities
consisting of Hubble flow, Virgo Attractor, and constant velocity vector.
\label{fig:virgoresi}}
\end{figure}
}
While they are somewhat better than Figure \ref{fig:h0resi}, the
residual quadrupole causes substantial scatter.

The $\chi^2_N = 1.26$ value for this fit demonstrates that the model is
a better match of the velocity field than a pure Hubble flow,
but the evident quadrupole and the peculiar velocity
in the CMB frame which misses pointing at the Centaurus cluster
by only 10 degrees, indicates a new
flow in that direction.  Following Lilje et al. (1986)
we can add a quadrupole component to this fit, centered on the origin
and with no cutoff.
We find that ${\cal N}$ improves substantially to 320.4, and
the rest of the parameters change slightly to
$H_0 = 75$, 
$u_{\rm VA} = 132$, $\sigma_{\rm VA} = 649$, 
$r_{\rm VA} = 14.4$,
and $\vec w = (-265,200,-97)$.
The quadrupole has a
stretch of $6.5$\kmsMpc\ in roughly the +X/$-$Z direction, a stretch
of $4.1$\kmsMpc\ in the +X/+Z direction,
and a compression of $-10.6$\kmsMpc\ in the $\pm$SGY direction.

\subsection{Virgo and Great Attractor Infalls}
\label{ssec:infalls}

The above suggests that it might be profitable to try a fit with a second
attractor.  This is our ``Great Attractor'' (GA) component, and we will
allow it to vary in both amplitude and position.
We fit $\vec w$+VA+GA in the CMB frame and find
$H_0 = 73.5$, VA parameters of $u_{\rm VA} = 127$,
$\sigma_{\rm VA} = 667$, $r_{\rm VA} = 14.2$, GA parameters of
$u_{\rm GA} = 199$ and $r_{\rm GA} = 28.3$, centered on
$(-37.7, 13.3, -18.2) \pm (1.7,2.6,1.6)$~Mpc, and
$\vec w = (-146, 143, -15) \pm (79, 34, 51)$ \kms.
The likelihood has improved to ${\cal N} = 294.9$ for 273 DOF.
Figure \ref{fig:gafit} shows the residuals when this
model is removed from the observed velocities, and 
figure \ref{fig:garesi} shows the residual comparison.
\insfig{
\begin{figure}[t]
\epsscale{1.0}
\plotone{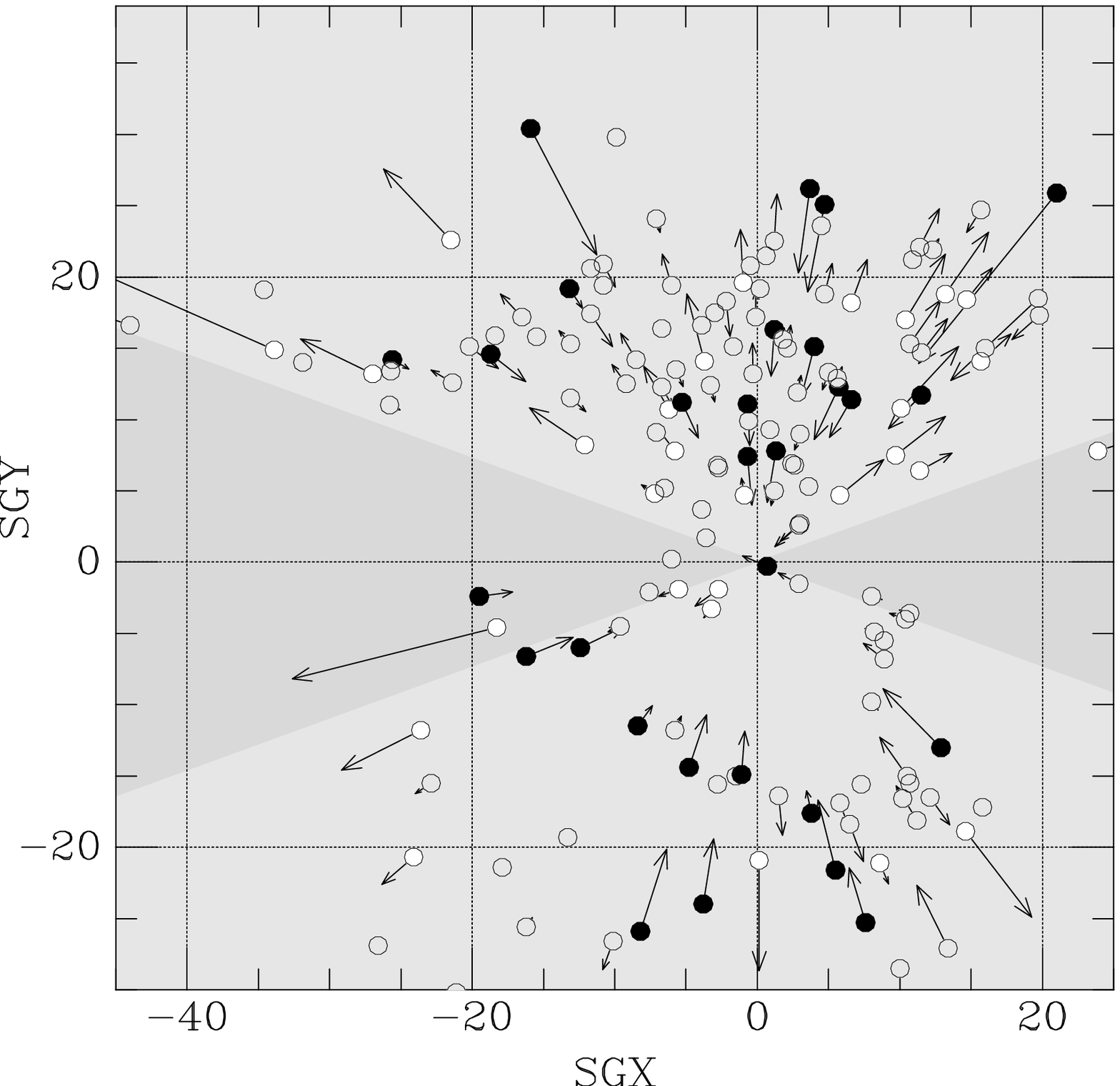}
\caption[f09.eps]{
Residual velocities after removal of a model
consisting of Hubble flow, Virgo and Great Attractors,
and constant velocity vector.
\label{fig:gafit}}
\end{figure}
}
\insfig{
\begin{figure}[t]
\epsscale{1.0}
\plotone{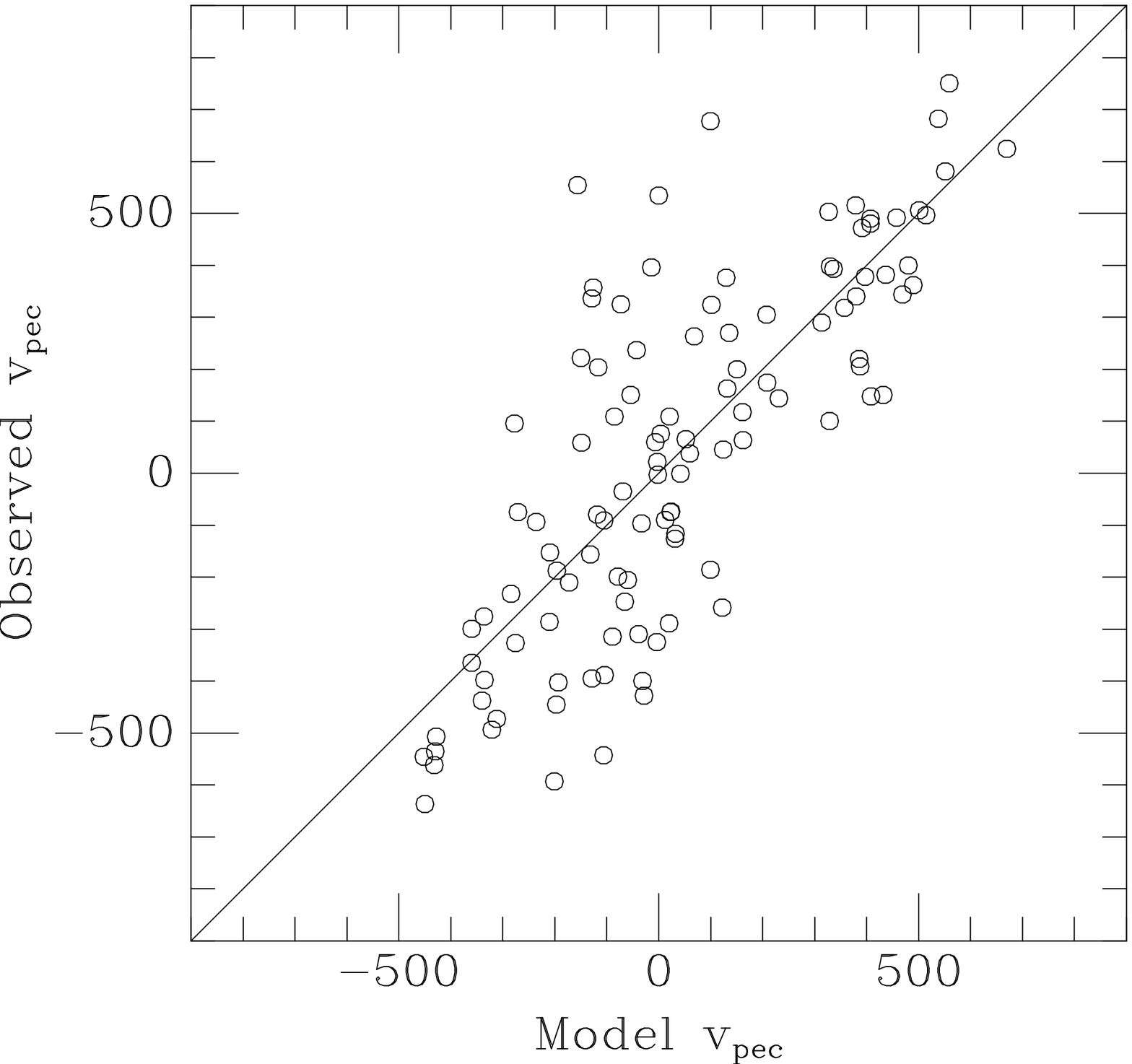}
\caption[f10.eps]{
Comparison of velocity residuals after removal of a model
consisting of Hubble flow, Virgo Attractor, GA,
and constant velocity vector.
\label{fig:garesi}}
\end{figure}
}
Like the previous one, this model uses power law exponents of
$\gamma = 2$ and core radii $r_c = 2$.

This two attractor plus dipole model appears to be a good
description of the observed velocity field.  It constitutes
a significant improvement over a simple Virgo infall model.
If we force the overall dipole to zero in the CMB frame
($\vec w = 0$), we find that the cutoff radii want to go to infinity,
and that ${\cal N}$ rises to 305.3 for 278 DOF.
The other parameters change quite a bit:
$H_0 = 79.9$, $u_{\rm VA} = 207$,
$\sigma_{\rm VA} = 675$, $r_{\rm VA} = \infty$,
$u_{\rm GA} = 397$ and $r_{\rm GA} = \infty$, centered on
$(-37.0, 16.8, -16.7)$~Mpc.
This illustrates how sensitive some of the parameters (such as GA
infall velocity) are to an assumed velocity reference frame.

\subsection{VA, GA, and a Quadrupole}

Finally, we return to the notion that influences from outside the
volume where we have data will manifest themselves to lowest order as
a dipole velocity and a quadrupole, and that our spherical flow fields
are overly simplistic.  We see a persistent $w_y$ velocity and Figure
\ref{fig:gafit} still seems to have a dipole pattern.  We now fit a
model which includes a velocity dipole, the Virgo and Great Attractors
with free amplitudes and $r_{\rm cut}$, and a 5 component quadrupole
which is centered on the origin and has a cutoff radius of $r_{\rm
quad} = 50$\Mpc.  We use a power law exponent of $-1.5$ for the VA and
$-2.0$ for the GA, as suggested by the confidence contours illustrated
below, and $\Omega_M = 0.2$.  The positions of the VA and GA are fixed
at $(-3.1, 16.6, -1.6)$ and $(-36.7, 14.6, -17.1)$ respectively (their
best-fit locations).
We find that 
$H_0 = 78\pm3$, $u_{\rm VA} = 139\pm48$,
$r_{\rm VA} = 12\pm6$,
$u_{\rm GA} = 289\pm137$, $r_{\rm GA} = 50\pm44$, and
$\vec w = (-55, 143, -8) \pm (102,41,62)$\kms.
The likelihood has improved to ${\cal N} = 269.2$ for 272 DOF.
Figure \ref{fig:qfit} shows the residuals when this
model is removed from the observed velocities, and 
Figure \ref{fig:qresi} shows the residual comparison.
\insfig{
\begin{figure}[t]
\epsscale{1.0}
\plotone{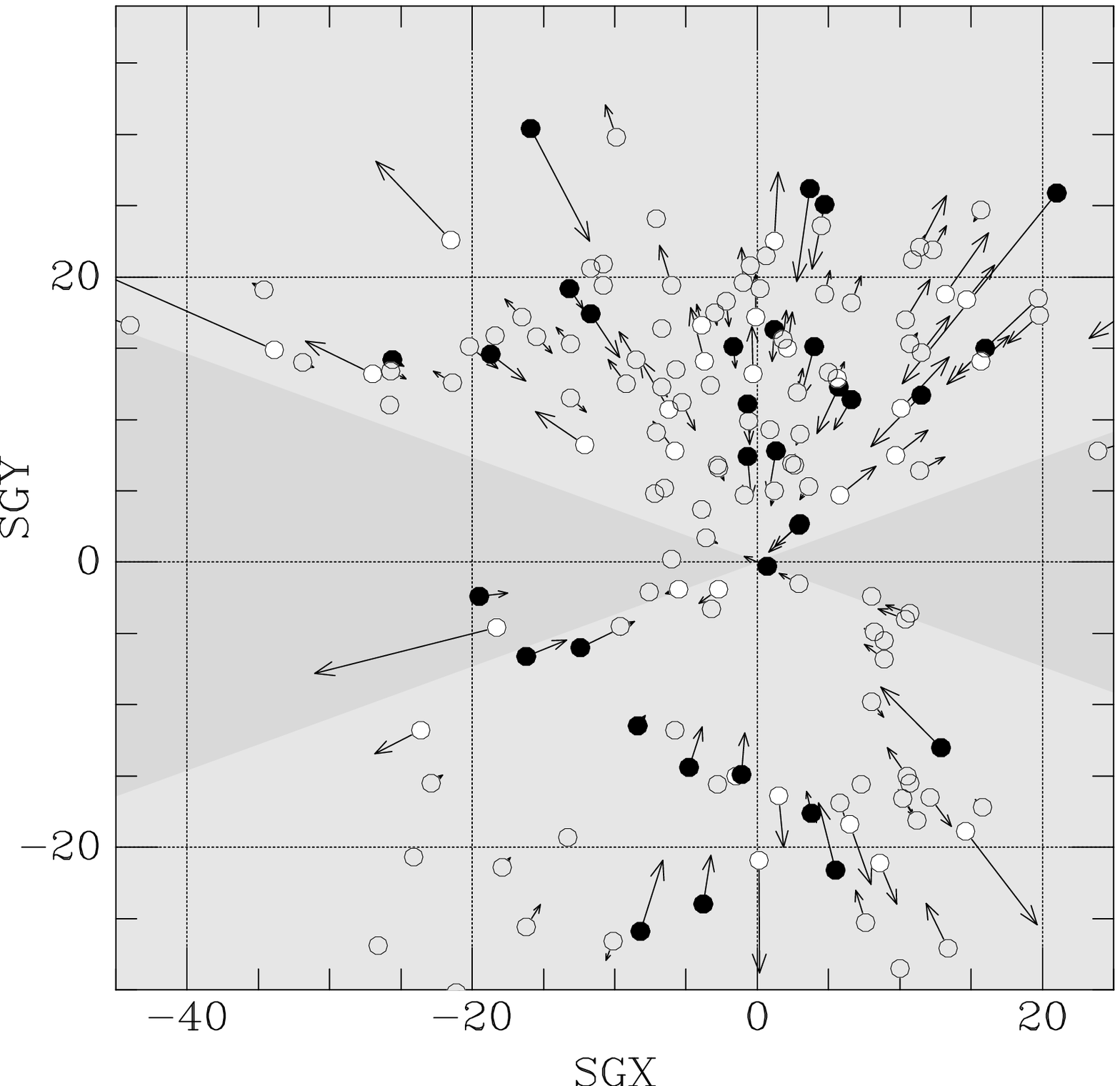}
\caption[f11.eps]{
Residual velocities after removal of a model
consisting of Hubble flow, Virgo and Great Attractors,
constant velocity vector, and quadrupole.
\label{fig:qfit}}
\end{figure}
}
\insfig{
\begin{figure}[t]
\epsscale{1.0}
\plotone{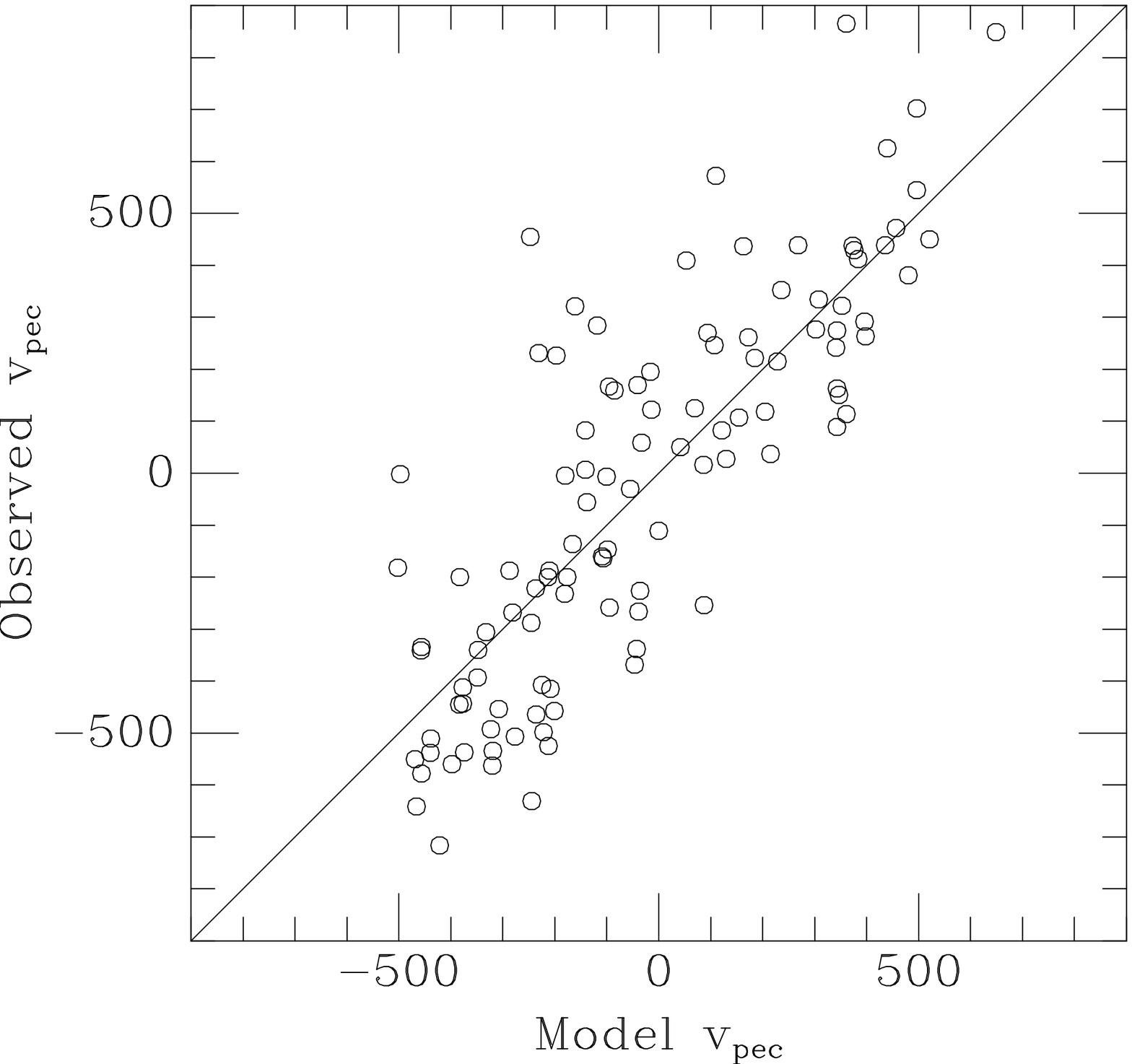}
\caption[f12.eps]{
Comparison of velocity residuals after removal of a model
consisting of Hubble flow, Virgo and Great Attractors,
constant velocity vector, and quadrupole.
\label{fig:qresi}}
\end{figure}
}
The quadru\-pole found by this model has an expansion of $14.6$\kmsMpc\
in the $(-0.56,-0.16,+0.82)$ direction, a compression of $-11.8$\kmsMpc\
in the $(-0.06,+0.99,+0.15)$ direction, and a small compression of
$-2.8$\kmsMpc\ in the $(+0.83,-0.04,+0.56)$ direction.  It and the peculiar
dipole $w_y = +143$\kms\ are significant at more than 3-sigma.

The reason that introducing a traceless quadru\-pole changes the model 
$H_0$ is that the sample galaxies are not isotropically distributed. 
In particular,
the major compression axis of the quadru\-pole is in the direction of Virgo,
which is obviously rich in galaxies.  Thus, without the quadru\-pole 
component, the model is forced to compensate for this extra compression
by lowering its value for $H_0$.
It is possible that the quadrupole arises locally
because discrete, spherical attractors are being used to
model flows in what is actually an anisotropic potential. 
We discuss this idea further in \S\ref{ssec:cmbframe}.
Alternatively the dipole and quadrupole could come
from unmodelled attractors outside our volume.  We investigate
below whether some of the well-known galaxy superclusters such as
Perseus, Coma, and Shapley could be the origin, but we will defer a
more complete investigation for a subsequent paper.

\subsection{Virial Motions}

If we allow the cosmic thermal velocity dispersion to
vary from the fixed value of 187\,\kms\ that we have
been assuming, then it now settles in at
$\sigma_{\rm cosmic} = 187\pm14$\,\kms\ for
our $\vec w$+VA+GA+Q model. The model likelihood does not change in any
significant way, of course.  This result is completely insensitive to
the motions of galaxies within the Virgo cluster, as these are handled
by the fitted cluster virial dispersion.  Actually, this best-fit
``thermal'' dispersion also includes a component from the velocity
measurement errors, but these are relatively small and correcting for
them puts $\sigma_{\rm cosmic}$ around 180\kms.

Our fitted value for the Virgo velocity dispersion of 
$\sigma_{\rm VA} \sim 650$\kms\ is identical to those found in
kinematical studies.  The agreement is expected, since the velocity
data are largely the same.  For instance, Girardi \etal\ (1996)
concluded $\sigma = 640^{+85}_{-65}$\kms\ for galaxies within 2\Mpc\
of the Virgo core.
If we delete the 12 galaxies within 2~Mpc of the center of the
Virgo cluster, the fit parameters for the
($\gamma_{\rm VA}=1.5$) model become
$H_0 = 78$, $u_{\rm VA} = 139$,
$u_{\rm GA} = 290$
and $\vec w = (-53, 142, -8)$.  This is virtually identical
to the model that includes the Virgo core, and demonstrates that our
models are able to accept large virial velocities and still reliably
follow large-scale flows.

\subsection{Uncertainties, Covariances, and Constraints}

We have deliberately tried to provide a great deal of flexibility in
the modelling (for example by always fitting for a velocity reference
frame).  However, there are so many parameters making up this model
that it can be deceptive to quote a formal error on a parameter
without disclosing the covariances it has with other parameters.  The
greatest covariances come from the $r_{\rm cut}$ parameters,
interacting with $\gamma$ and infall amplitude especially.  The next
most troublesome covariances come from the SGX component of the
reference frame velocity vector $\vec w$ mixing with the GA infall
amplitude and $H_0$, and there is also a little bit of direct
covariance between $H_0$ and $u_{\rm GA}$.  Apart from these, the
formal covariances between parameters are less than 0.6 in magnitude.

The contour diagrams of this section illustrate these covariances 
as well as graphically showing the uncertainties in other interesting
parameters. The shaded regions of these plots enclose the 68, 90, and 95
percent confidence regions for two variables, and the bars on the axes 
show the $\pm1\sigma$ limits for each variable considered separately.
Figure~\ref{fig:locvir} shows the joint confidence contours on the
distance, infall velocities, and location on the sky of the Virgo
Attractor, and Figure \ref{fig:locga} shows the same thing for the
Great Attractor.
\insfig{
\begin{figure}[t]
\epsscale{1.0}
\plottwo{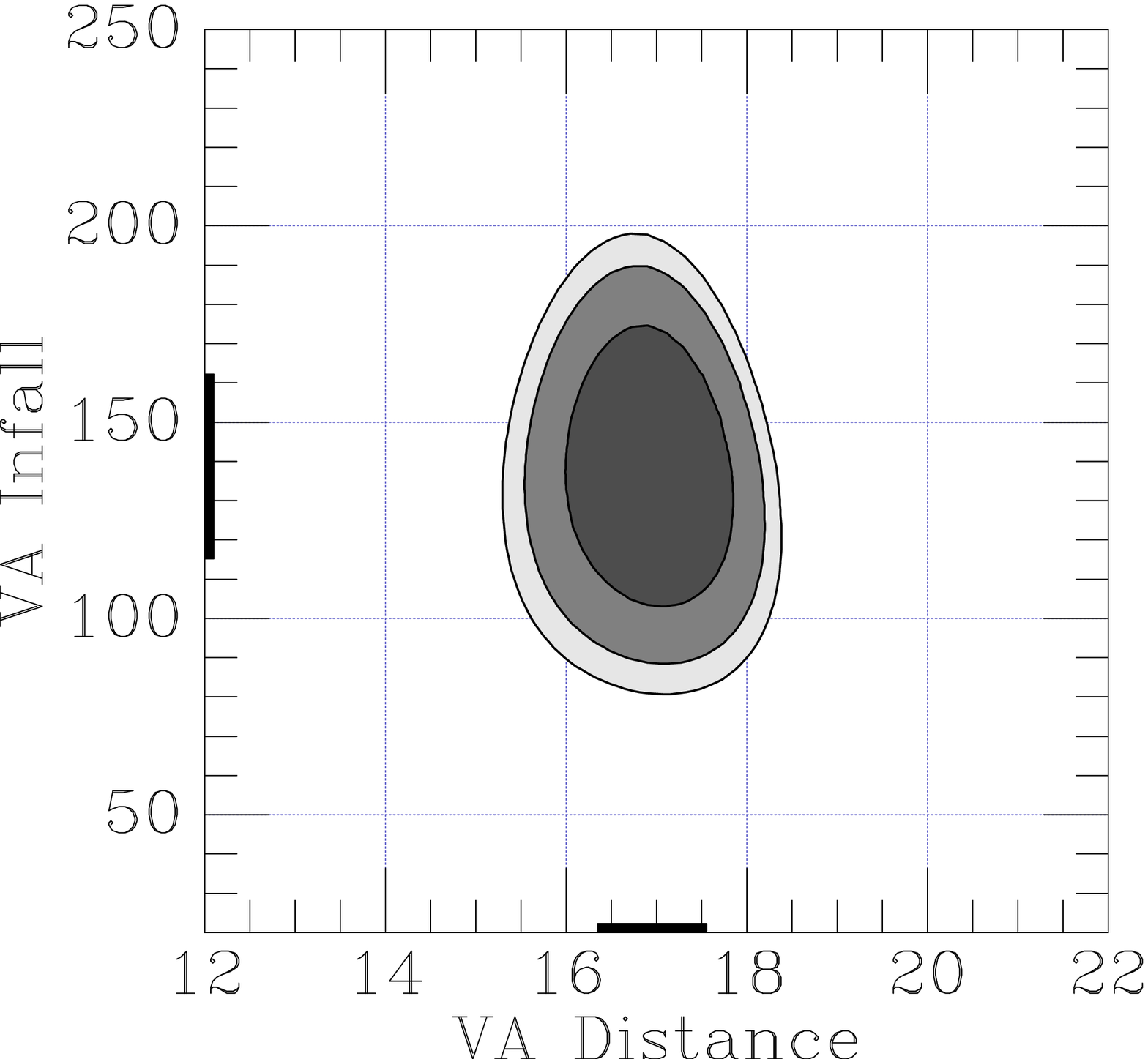}{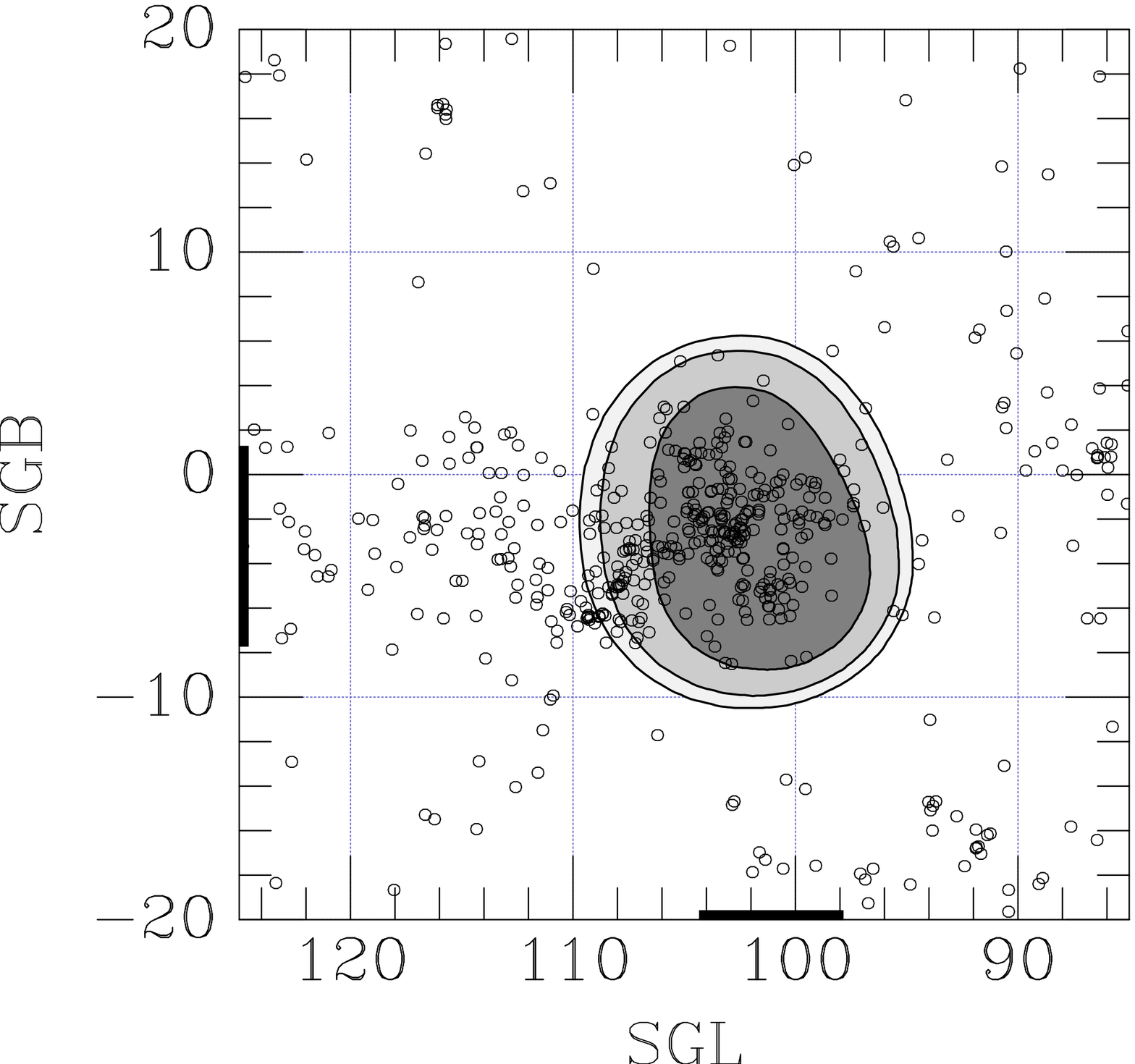}
\caption{
Joint confidence contours of the fitted infall velocity,
distance, and location of the sky of the Virgo Attractor.
The points show the location of the galaxies in the RC3 with $v_{h} < 2800$.
\label{fig:locvir}}
\end{figure}
}
\insfig{
\begin{figure}[t]
\epsscale{1.0}
\plottwo{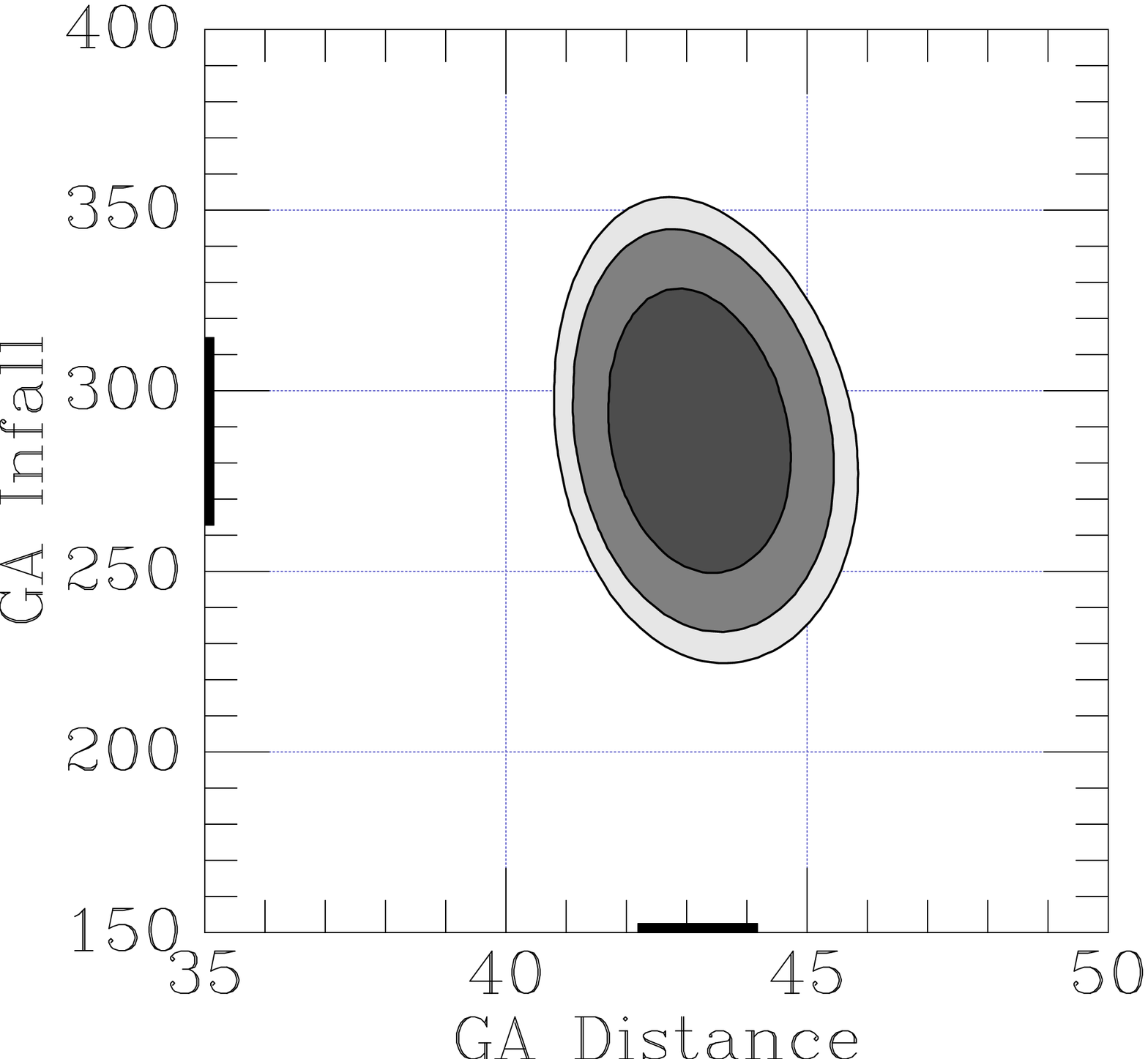}{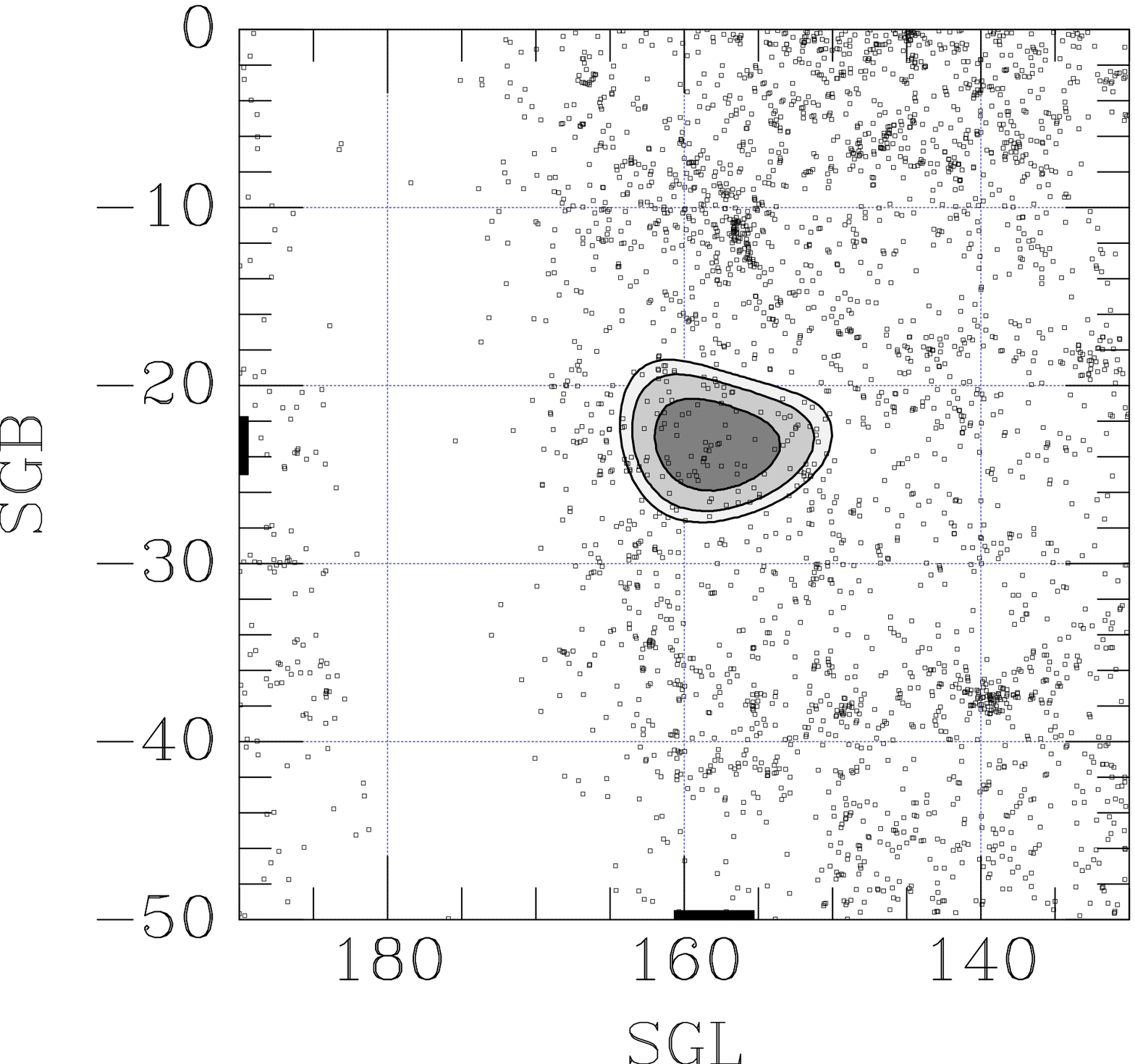}
\caption{
Joint confidence contours of the fitted infall velocity,
distance, and location of the sky of the GA attractor.
The points are from the SPS survey; Centaurus lies at $(156,-12)$,
Abell~1060 is the cluster at $(139,-37)$, and the Galactic plane is
evident on the left side.
\label{fig:locga}}
\end{figure}
} 
The locations of these attractors are evidently quite well
constrained purely by the velocity data.  The best-fit distance for
the Virgo Attractor of 17.1\Mpc\ is coincident with both the location
of NGC~4486 and the median of the galaxies in the core of the Virgo
cluster to within the errors.  The location of the GA is similar
in distance to the most distant galaxies attributed to the Cen-30
cluster but lies about 15 degrees from 
Cen-30.  As figure \ref{fig:locga}
illustrates, this is a complicated region.

Figure \ref{fig:wcov} shows how the SGX component $w_x$ of the dipole
velocity affects both the GA infall amplitude and the Hubble constant.
\insfig{
\begin{figure}[t]
\epsscale{1.0}
\plottwo{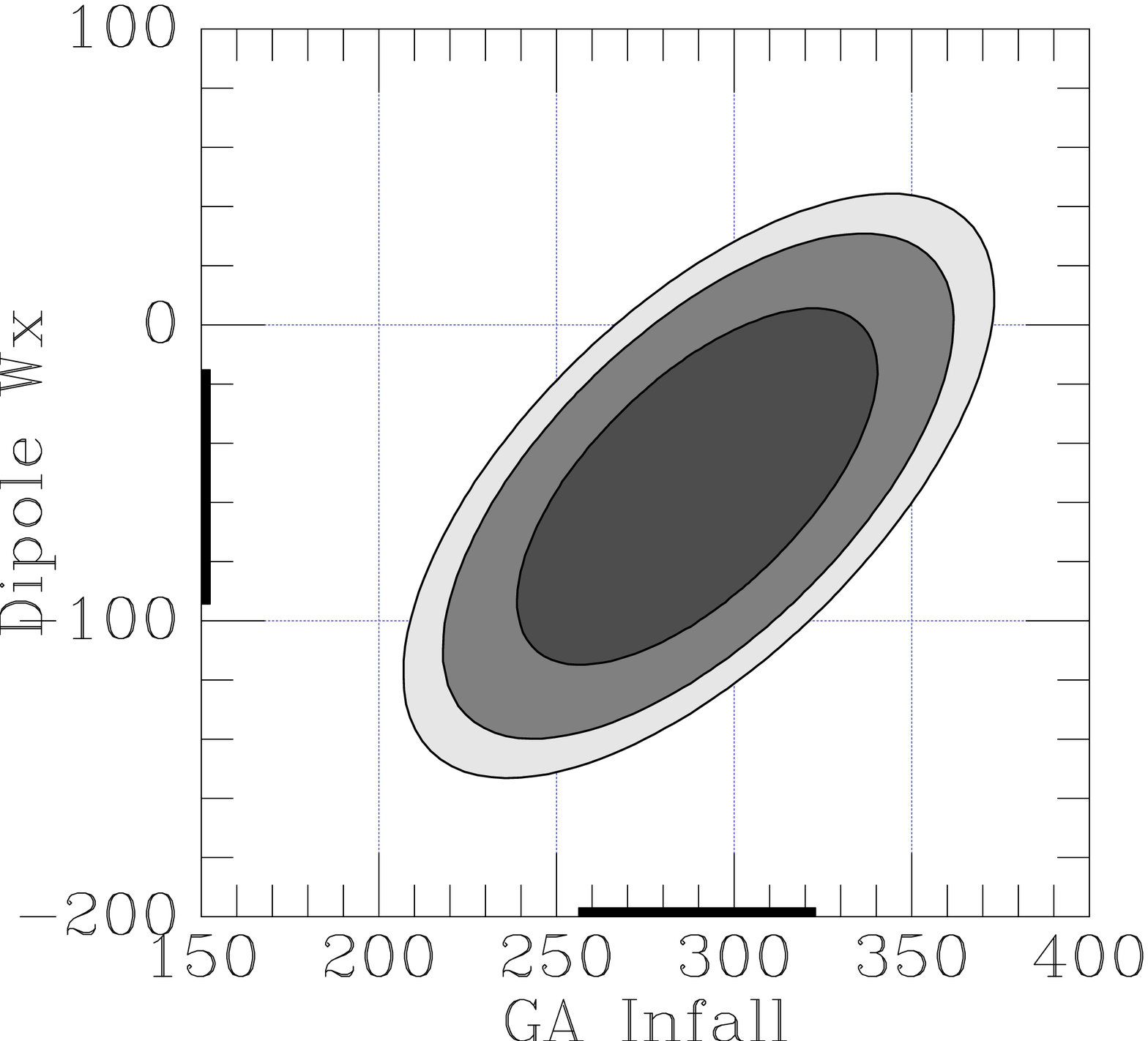}{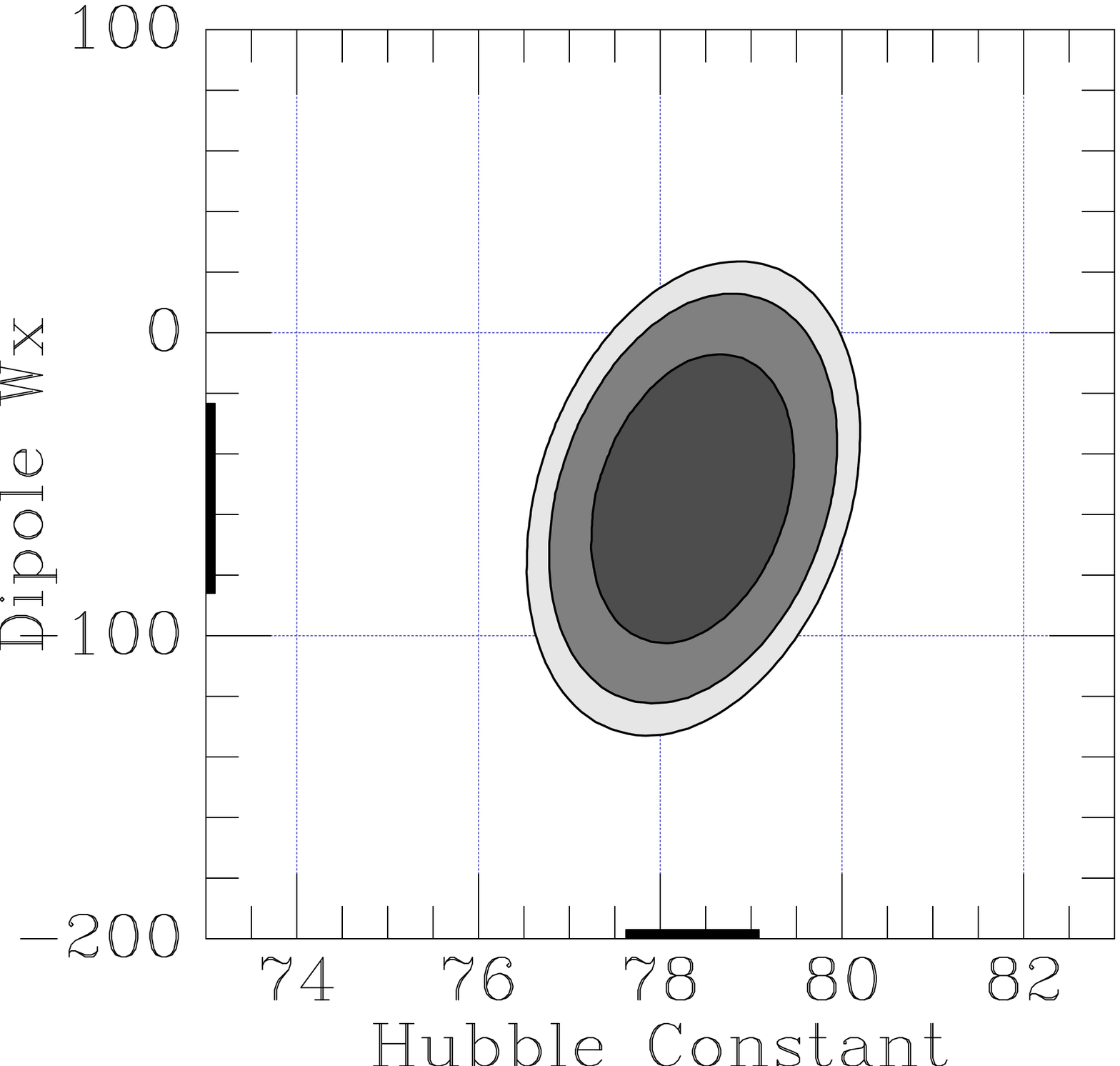}
\caption{
Joint confidence contours of the SGX component $w_x$ of the dipole
velocity with GA infall amplitude (left) and $H_0$ (right).
\label{fig:wcov}}
\end{figure}
}
The cosmic thermal velocity is
little affected by any other parameter, and is quite well constrained
at 187\kms.
The Hubble constant $H_0$ has a formal covariance of 0.6
with the GA infall amplitude, but such a small covariance scarcely
affects the determination of each independently.

Figures \ref{fig:rcutvir} and \ref{fig:rcutga} show the confidence contours
for the $r_{\rm cut}$ parameter on the power law exponents and the
infall velocities of the Virgo and Great Attractors.
\insfig{
\begin{figure}[t]
\epsscale{1.0}
\plottwo{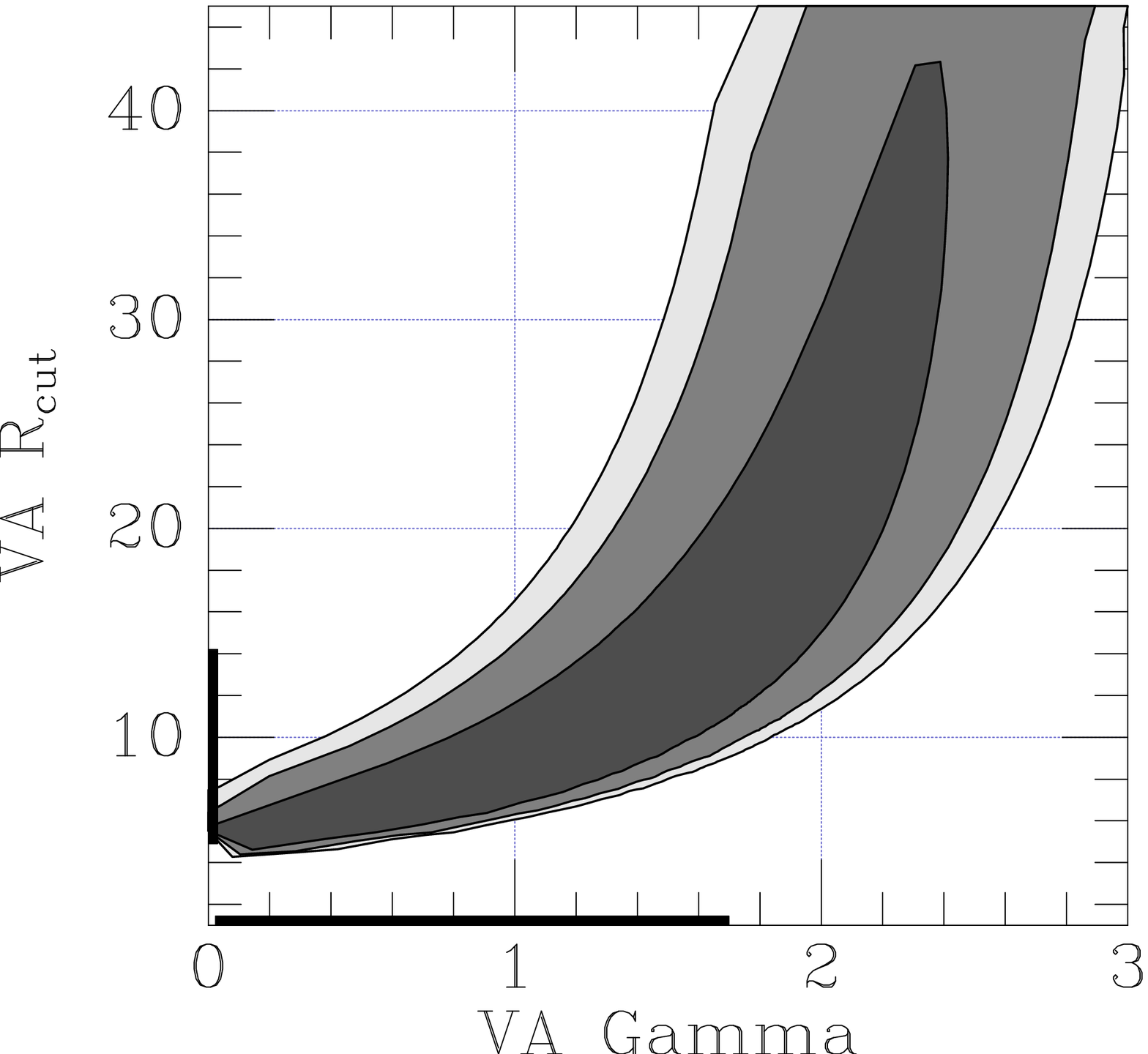}{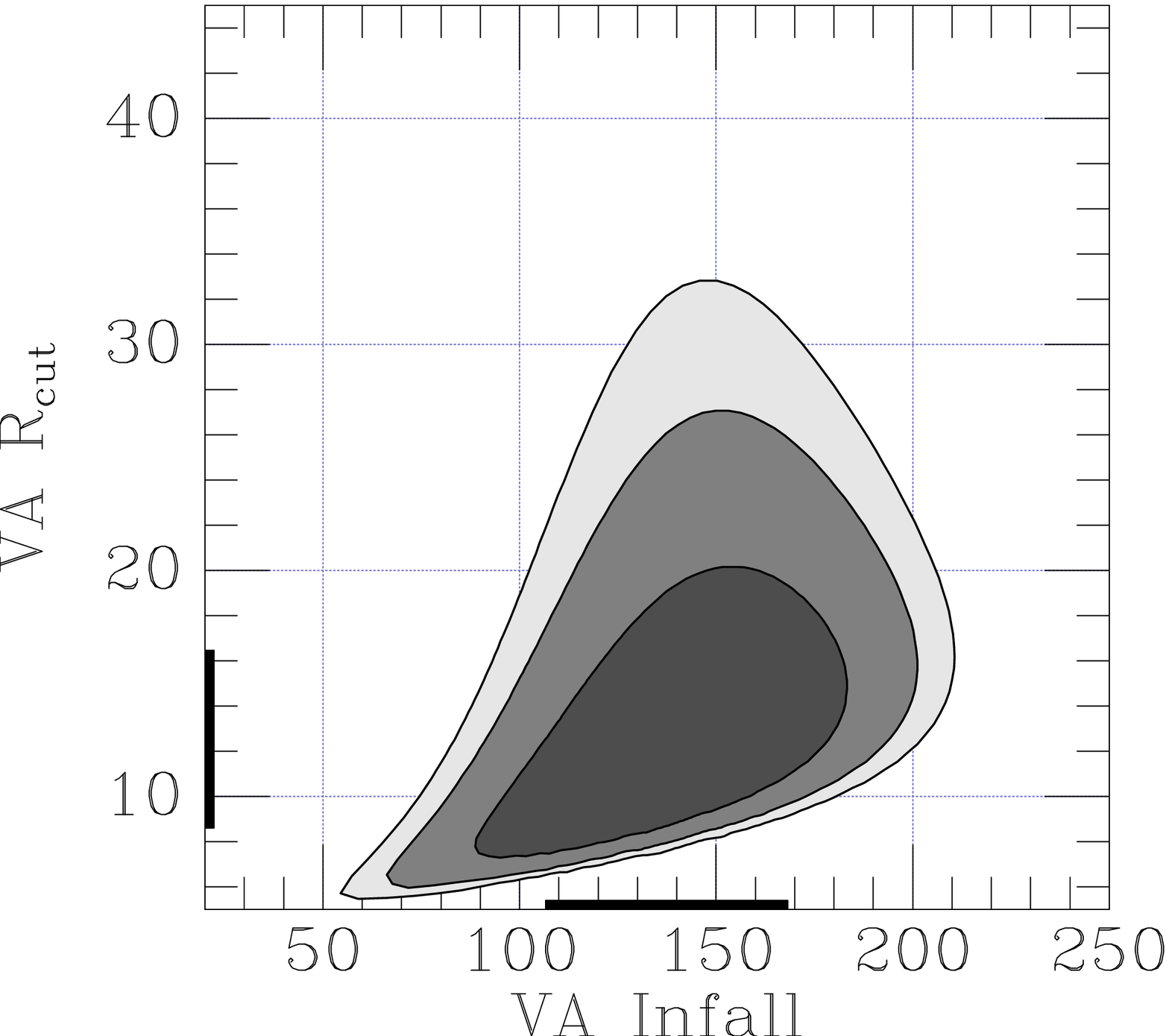}
\caption{
Joint confidence contours of the cutoff radius $r_{\rm cut}$ and power
law slope $\gamma$ (left) and infall amplitude (right) for
the Virgo Attractor.
\label{fig:rcutvir}}
\end{figure}
}
\insfig{
\begin{figure}[t]
\epsscale{1.0}
\plottwo{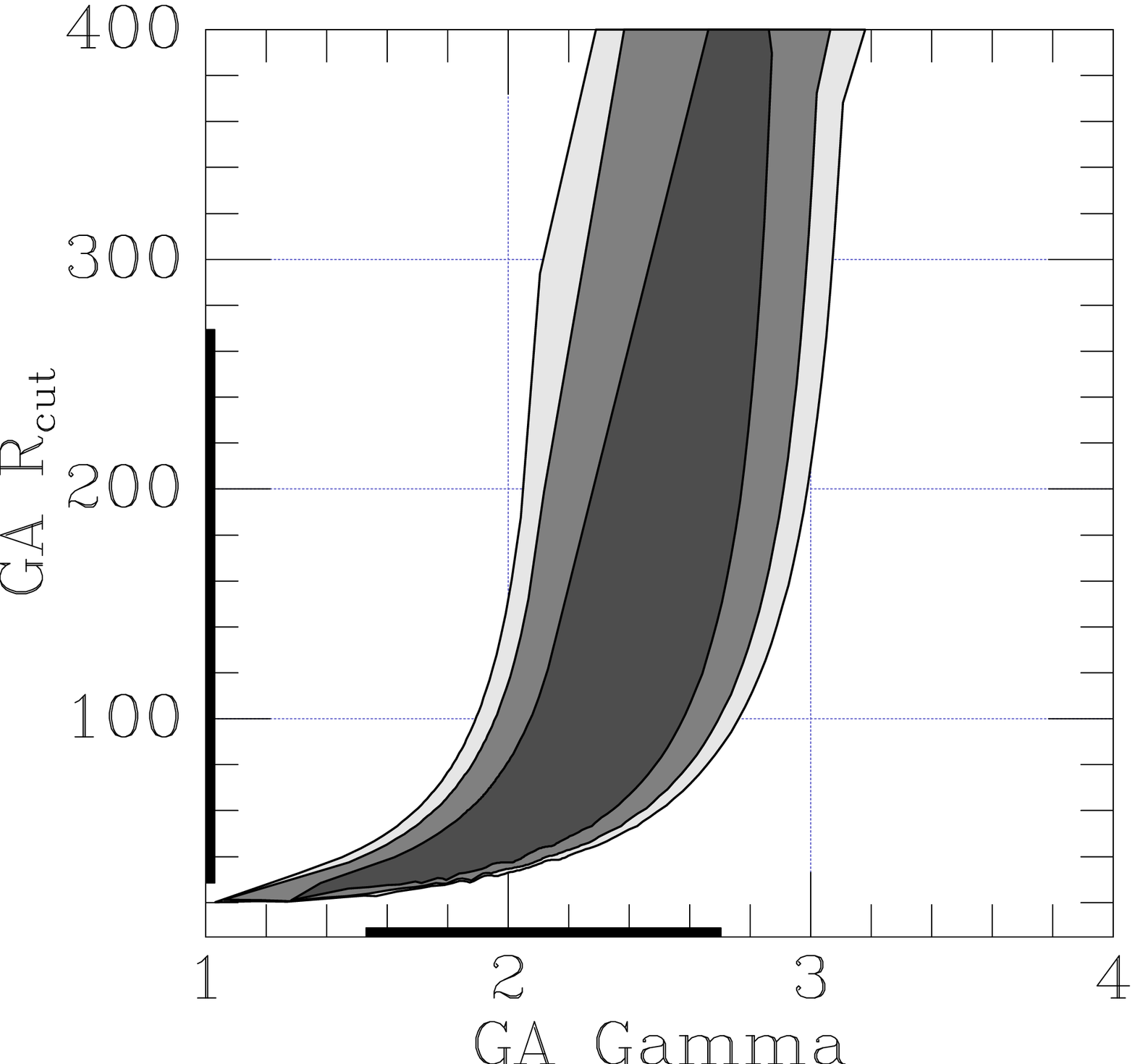}{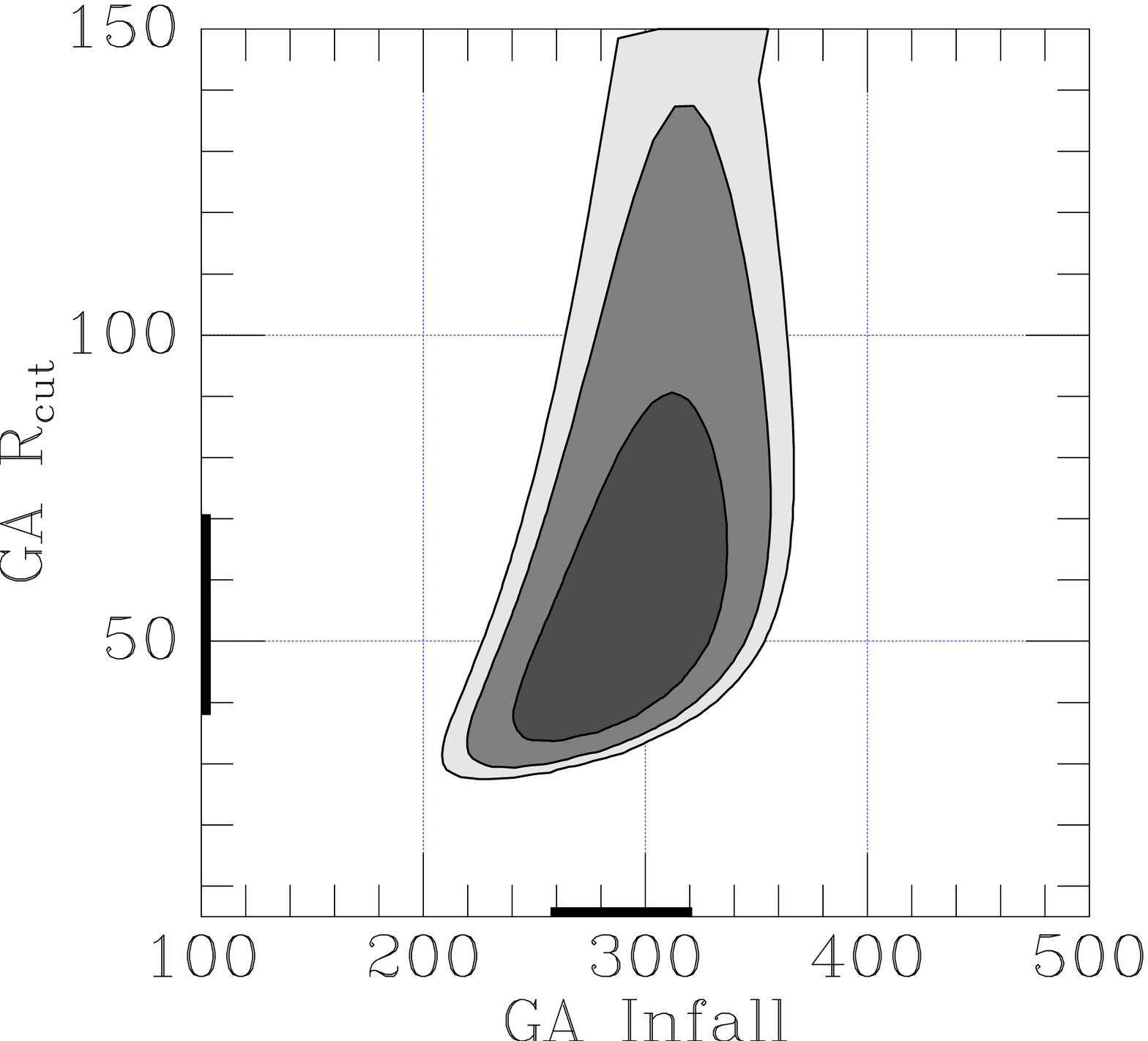}
\caption{
Joint confidence contours of the cutoff radius $r_{\rm cut}$ and power
law slope $\gamma$ (left) and infall amplitude (right) for
the GA attractor.
\label{fig:rcutga}}
\end{figure}
} 
Evidently these parameters are very poorly constrained
individually, and values of $0 < \gamma < 2$ for VA and
$1 < \gamma < 3$ for GA can be offset by appropriate changes in
$r_{\rm cut}$ to yield acceptable models.  Nevertheless,
overall run of peculiar velocity with position
varies little among these models.

Figure \ref{fig:flows} shows the run of net model velocity as a function
of position along vectors towards the Virgo and Great Attractors.  The
four curves are $\gamma = 0.1, 1, 2, 3$, and 
$r_{\rm cut} = 5, 9, 19, \infty$ (VA) and 
$\gamma = 1, 2, 2.5, 3$, and $r_{\rm cut} = 15, 50, \infty, \infty$ (GA)
for which $\cal N$ is virtually constant around 269.  (The enclosed density
profile of the Virgo Attractor can be described quite accurately as an
exponential with scale length 5~Mpc.)  Except in the
cores of the attractors, the curves differ very little.  We also plot
the galaxies which lie within $25^\circ$ of the vectors ($|\cos\theta|
> 0.9$) to give a sense of the number of constraining points we have 
near the attractors (although the attractor models are affected by
galaxies all over the sky).
\insfig{
\begin{figure}[t]
\epsscale{1.0}
\plotone{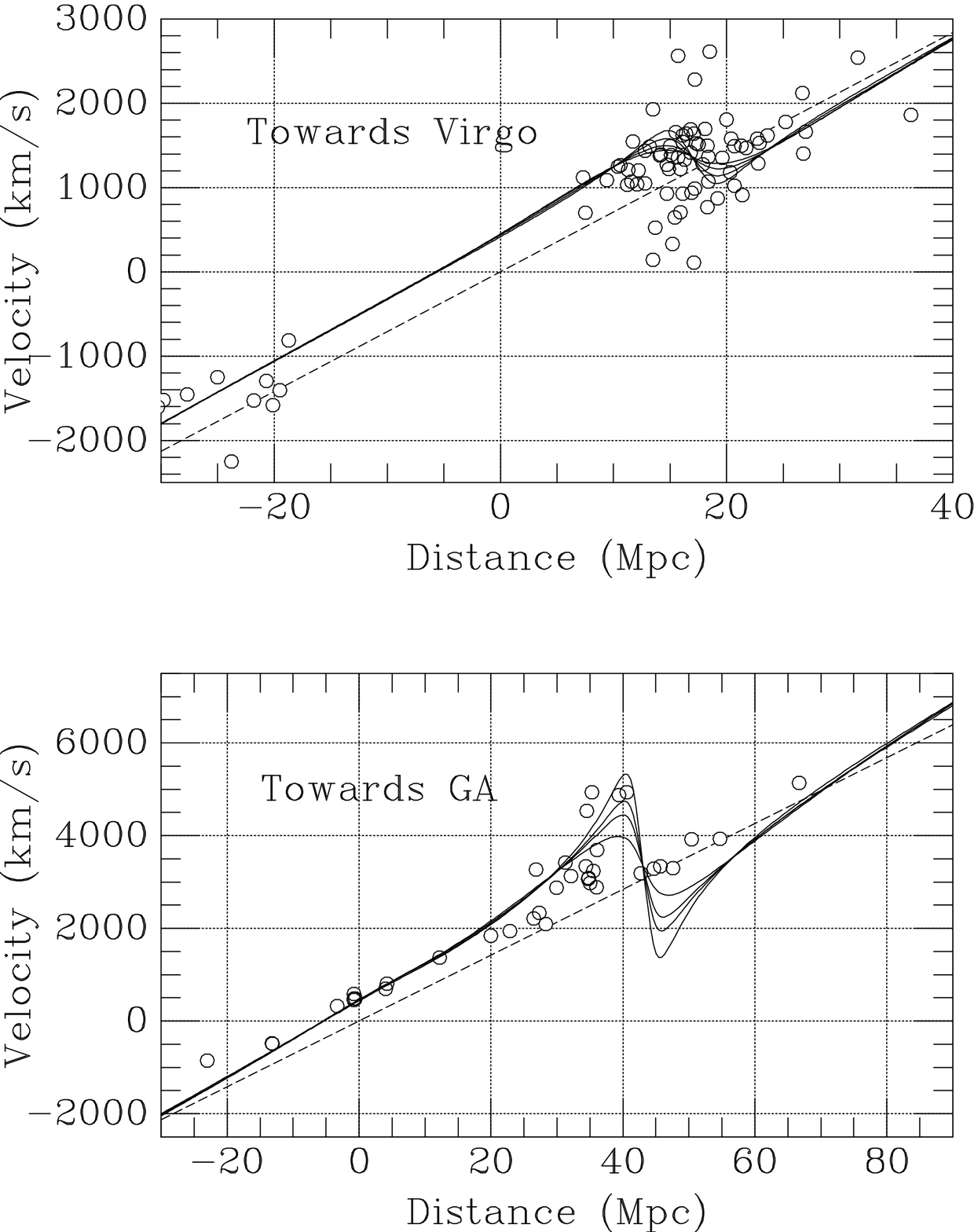}
\caption{
The net model velocity is shown as a function of position along vectors
running through the Virgo and Great Attractors using various values of
$\gamma$ and $r_{\rm cut}$ which leave $\cal N$ constant.  The dashed
line is just $H_0\,r$.  The points are those galaxies within
$25^\circ$ of the vectors ($|\cos\theta| > 0.9$).
\label{fig:flows}}
\end{figure}
}
The agreement between the galaxy and model velocities in this plot should
only be very approximate close to the attractors because the angle
between the line of sight and the vector from the galaxy to the
attractor becomes much greater than $25^\circ$.

Figure \ref{fig:infalls} illustrates this slightly differently.
These panels show how $u_{\rm infall}$ for each of the two attractors
individually varies 
as a function of distance from the attractor, given these four models
with very different $\gamma$\,'s.  
\insfig{
\begin{figure}[t]
\epsscale{1.0}
\plotone{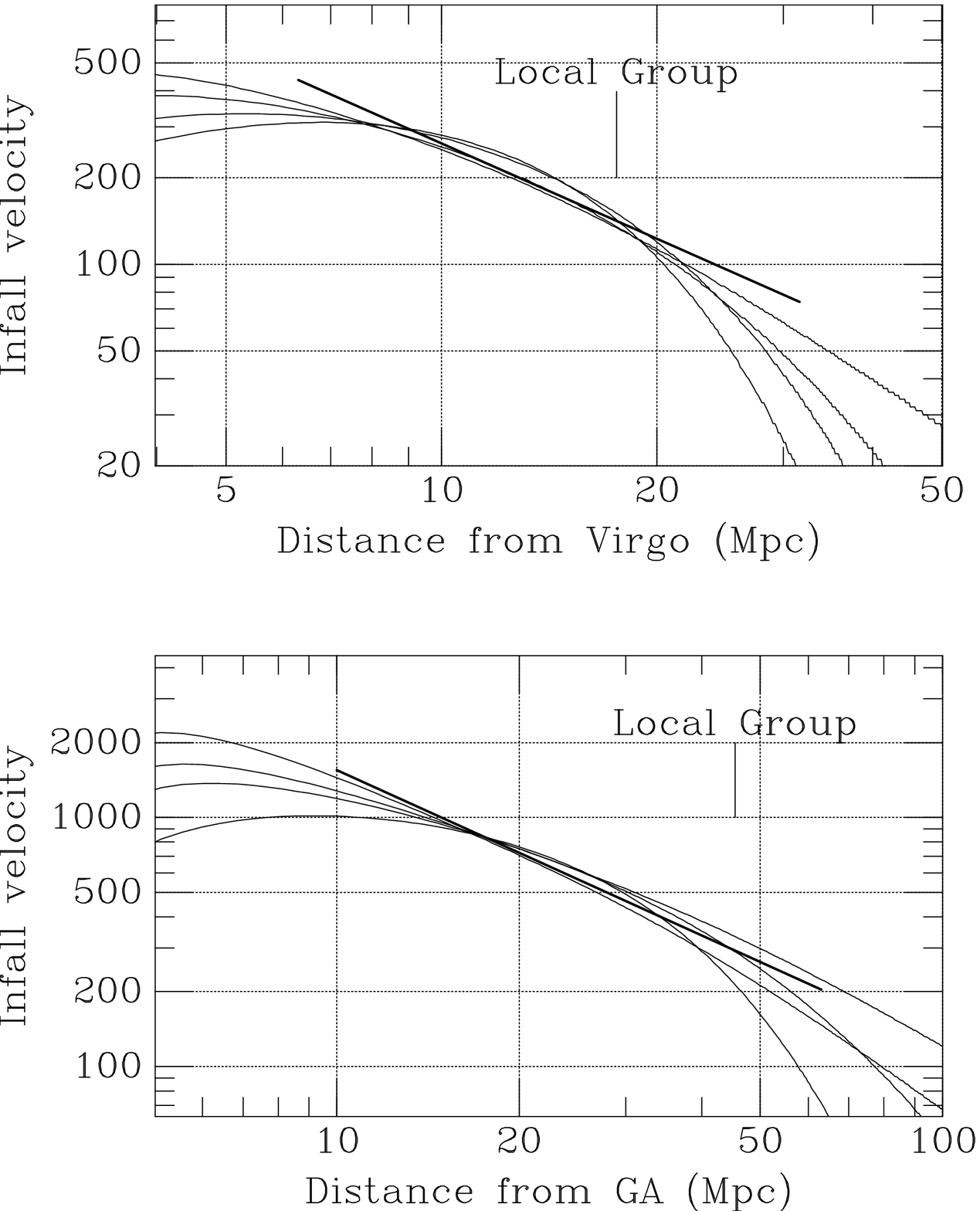}
\caption{
The model infall velocity is shown as a function of distance from the
Virgo and Great Attractors for the four sets of $\gamma$ values.
A solid line delineates the range where there is substantial agreement
and where we believe the models are well constrained by the data.
\label{fig:infalls}}
\end{figure}
}
The models diverge close to and far from the attractors but there is a
broad range where the models give substantially the same values for
infall velocity ($6 < r < 25$~Mpc for the VA and
$10 < r < 50$~Mpc for the GA).  We draw power-law approximations
for these two attractors on the plots whose slopes are 
$-1.1$, although it is apparent that all models for both attractors
curve away from a pure power law over these radial ranges.
The ratio of attractor infall velocity to Hubble flow is
%
for the Virgo Attractor
\begin{equation}
{u_{\rm VA} \over H_0\,r} = 0.11\; \left(r\over r_{\rm VA}\right)^{-2.1}
\label{eq:vflow}
\end{equation}
and for the Great Attractor
\begin{equation}
{u_{\rm GA} \over H_0\,r} = 0.09\; \left(r\over r_{\rm GA}\right)^{-2.1},
\label{eq:gflow}
\end{equation}
where $r_{\rm VA}$ and $r_{\rm GA}$ are the distances of the Local
Group from the attractors.

Not surprisingly, for a given value of $\Omega_M$ 
we find quite consistent values for $\delta$, the
overdensity within our radius, for these models.
For the Virgo Attractor, values of $0.1 < \gamma < 3$ lead to values of
$\delta = 1.0\pm0.07$, or $M_{\rm VA} = 7\times10^{14}$~M\solar.
However, if the quadrupole is caused by a non-spherical Virgo (as we
suggest below), these estimates for $\delta$ and $M_{\rm VA}$ are
not meaningful, although Equation~(\ref{eq:vflow}) is.
For the Great Attractor, models with $1 < \gamma < 3$  lead to values of
$\delta = 0.73\pm0.14$, or $M_{\rm GA} = 8\times10^{15}$~M\solar.
(These masses, of course, are the masses in excess of background
density within spheres centered on the attractors with radius reaching
the Local Group.)
From this it is easy to solve for the radius and angle subtended by
the zero velocity surface (sphere where infall cancels outward Hubble
flow) around these attractors.  For the Virgo Attractor this is 6\Mpc,
or $20^\circ$; for the GA it is 14\Mpc, or again about $18^\circ$.
We see from Figure \ref{fig:infalls} that by $\delta \approx 1$
($u/H_0\,r \approx 0.1$) the models do not fit well
unless they are falling at least as steeply as $r^{-3}$.
In particular, pure isothermal models
($\gamma{\,=\,2}$, unmitigated by any cutoff) would provide
poor representations of the mass distributions.

The GA distance is well constrained in these models
at $d_{\rm GA} = 43\pm3$\Mpc.
This is a bit behind most of the galaxies comprising the
Cen-30 and Cen-45 clusters.  In fact, the Cen-30 ``cluster''
appears to be galaxies distributed over 5-10\Mpc\ on the near side of
the GA and the ``s-wave'' distortion of the Hubble flow causes
them to share similar redshifts.  This is illustrated in Figures
\ref{fig:vcont} and \ref{fig:vcontgxy}, which show contours of our
flow model with and without survey galaxies overlaid.
\insfig{
\begin{figure}[t]
\epsscale{1.0}
\plotone{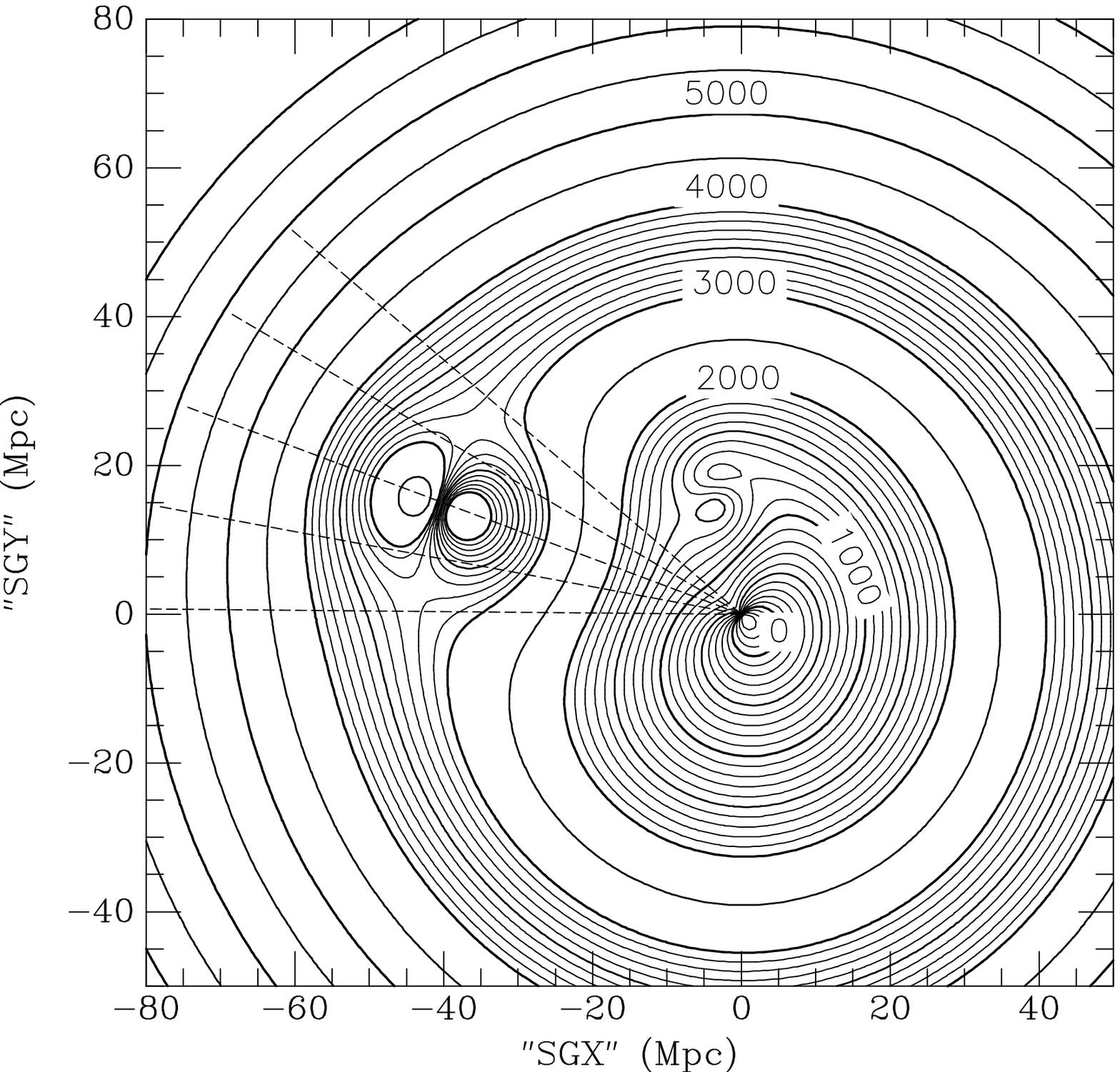}
\caption{
Contours are shown of our model of the radial component of the
flow field in a plane which cuts through the Local Group, Virgo,
and Great Attractors.  The model velocity of the Local Group with
respect to the CMB frame is apparent in the discontinuity at the
origin.  The Virgo Attractor is found at ``SGX,SGY'' = $(-3,+17)$
(quotes because the plane doesn't correspond perfectly to
supergalactic coordinates); the Great Attractor is at $(-42,+16)$,
and radial lines are drawn at $\pm10^\circ$ and $\pm20^\circ$ from it.
\label{fig:vcont}}
\end{figure}
}
\insfig{
\begin{figure}[t]
\epsscale{1.0}
\plotone{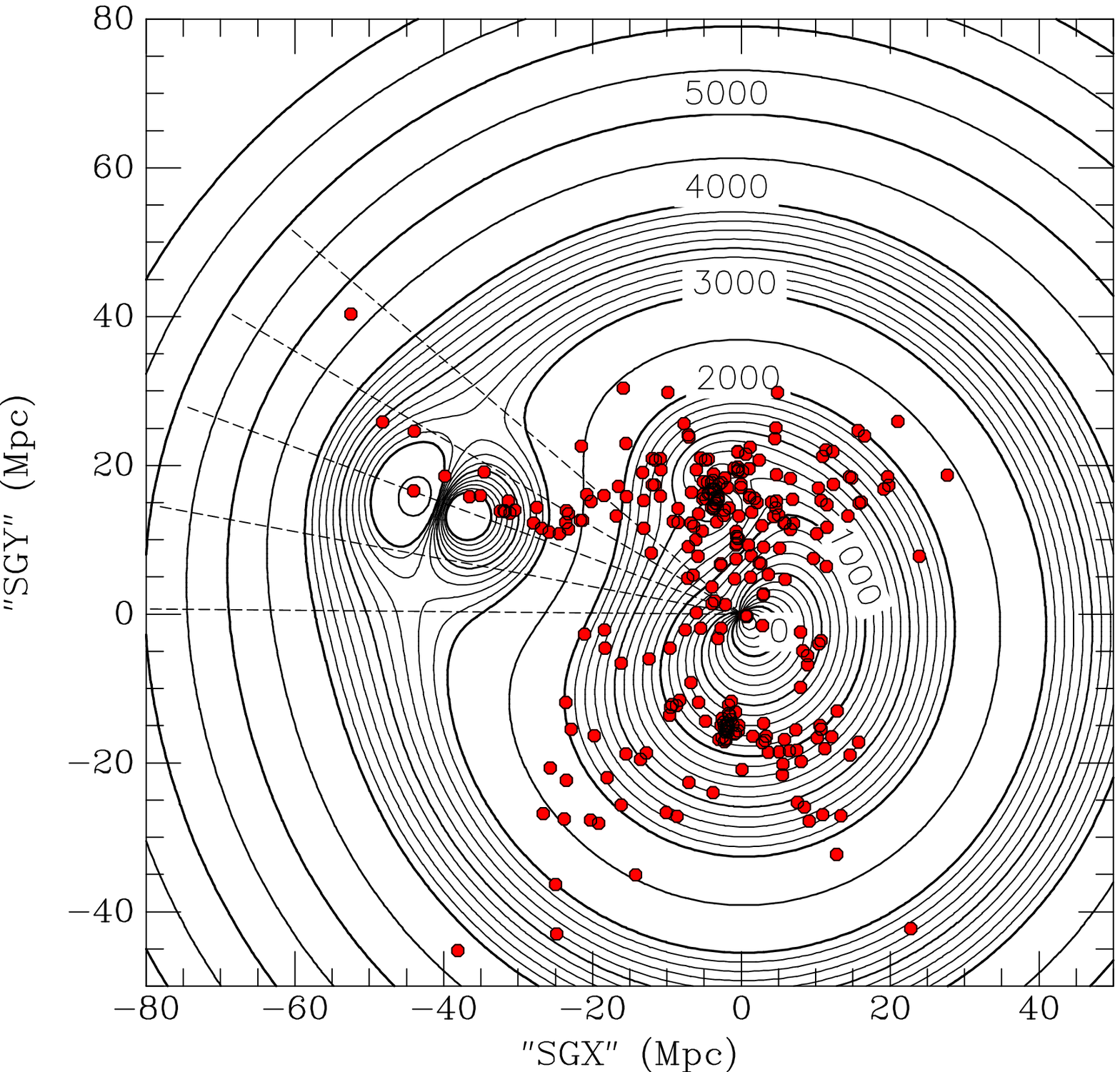}
\caption{
This is the same figure as the previous one except that our survey
galaxies are overplotted.  Note that the Centaurus galaxies which
appear to be going through the Great Attractor actually pass above it
by about $15^\circ$, hence lie in the stall zone.
\label{fig:vcontgxy}}
\end{figure}
}
We do not have a good constraint on the GA amplitude closer than 10\Mpc,
so the very dramatic peculiar velocities inside that radius
may well be overestimates. In this projection, the Centaurus galaxies
fall above the GA, but they lie in the stall region near
$\pm15^\circ$ angular separation, and the model and data agree on a
typical CMB velocity of 3150\kms.  Similarly, the Ursa Major group
lies in the stall region near the VA with a CMB velocity of
1150\kms, and the Virgo Southern extension lies in a stall zone with
CMB velocity of 1500\kms. 
The general trend of the model for $v_{\rm CMB}$ to
range from 1100\kms\ to 1500\kms\ as SGL swings from $85^\circ$ to
$125^\circ$ is apparent in redshifts surveys of the region, as is
illustrated in Figure~\ref{fig:gauma}.
\insfig{
\begin{figure}[t]
\epsscale{1.0}
\plottwo{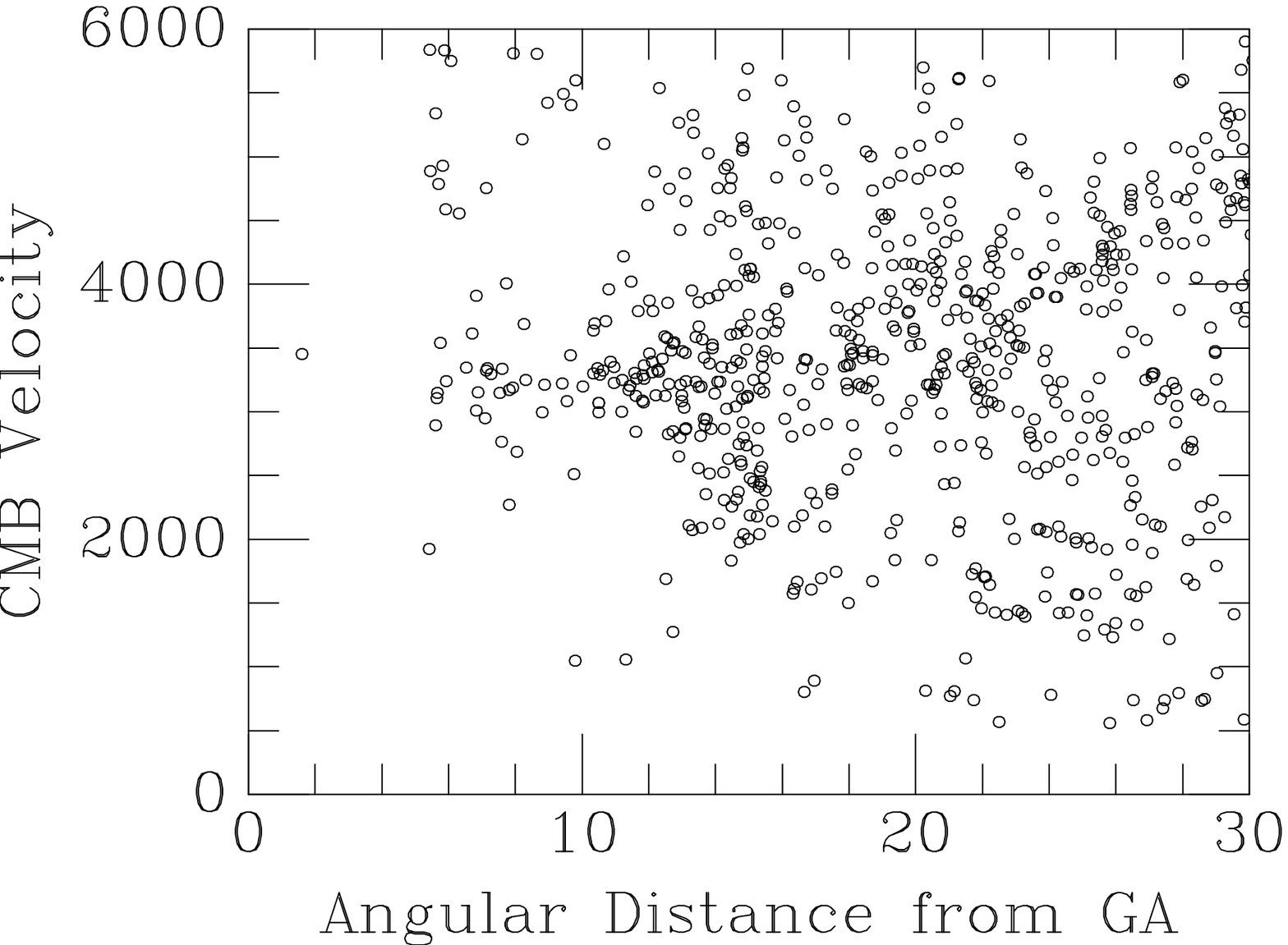}{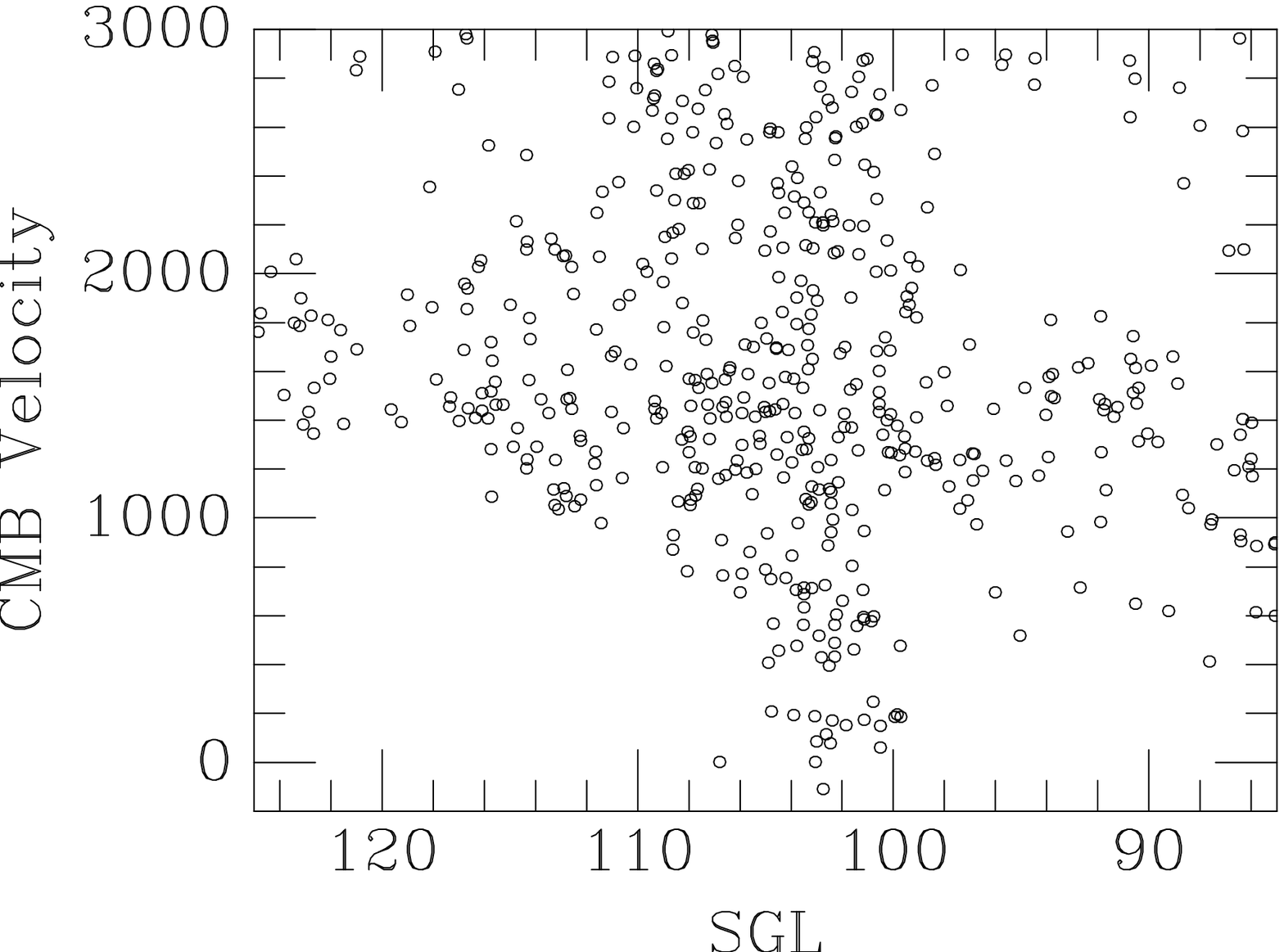}
\caption{
The CMB velocities of
galaxies from the RC3 are plotted as a function of angular separation
from the GA (left), and as a function of supergalactic longitude SGL
(right).  The concentration at $v_{\rm CMB} = 3150$
(left) and the trend for $v_{\rm CMB}$ to change from 1100 to
1500\kms\ (right) inherent in our model is apparent in these redshift
data.
\label{fig:gauma}}
\end{figure}
}

We note that the GA distance increases by about 10\% when the 3 
high velocity Cen-45 galaxies (with good quality data) are included 
in the modeling. However, the likelihood of the model suffers badly
in trying to accomodate just these 3 additional DOF, rising from
${\cal N} = 269$ to ${\cal N} = 295$.
Thus, for now we have chosen to exclude them.
A detailed discussion of the distribution and motions of galaxies in
the GA/Centaurus region, and a comparison of the SBF flow model results
with those of previous GA models will be given by Dressler \etal\ (1999).

\subsection{Influence of Perseus-Pisces and Shapley}

We now ask if the Perseus-Pisces supercluster and/or the Shapley
Concentration could be sources of the peculiar dipole $\vec w$ 
and quadrupole that we find.  To test this idea, we add
to our standard $\vec w$+VA+GA+Q model 
compact attractors representing Perseus-Pisces
at $(65,-15,-17)$ and Shapley at $(-170,120,-30)$, allowing the
masses to be free parameters.   The likelihood
$\cal N$ improves to 267.2, which is not a significant
improvement for two new parameters over the previous value of 269.2.
For Perseus-Pisces, we find an insignificant velocity amplitude 
of $v_{\rm amp} = 59\pm314$\kms; recall that this is
normalized at the fiducial radius of 50 Mpc.
For Shapley, we find an extraordinary
$v_{\rm amp} = 17926$\kms, which for the assigned distance of 210 Mpc
corresponds to a velocity of 1013\kms\ at the Local Group. 
The other parameters change a great deal to offset this,
especially $\vec w$ which adjusts itself to $(920,-530,210)$.
Equally telling however, is the fact
that the quadrupole {\it grows} to have eigenvectors of $22$,
$-20$, and $-2$ \kmsMpc, indicating that the inclusion of Perseus and
Shapley has made this model less plausible. 

There is a great deal of covariance between the distant
Shapley attractor and the $\vec w$ term, so we ran another model 
including the Shapley attractor but with $\vec w$ set to zero.
This model gave a velocity amplitude $v_{\rm amp} = 2533\pm717$\kms,
corresponding at the Local Group to $144\pm41$\kms.
The likelihood of this model is ${\cal N} = 274.0$, which for only
one additional parameter is a substantial improvement over the $\cal
N$ of 278.5 
found for a model with VA+GA+Q but no $\vec w$ or Shapley component.
However, the fit is worse than our standard model 
with $\vec w$ and no distant attractors, although the difference in the 
fitted quadrupole components for these two models is now insignificant.
(Adding a Perseus attractor gives no further improvement; it is in the
wrong direction to compensate for the lost flexibility of $\vec w$.)

We conclude that Perseus-Pisces exerts no significant influence on
our local volume, consistent with previous studies in this direction
(e.g., Willick 1990; Courteau \etal\ 1993; Hudson \etal\ 1997), and
that the Shapley Concentration is not the source of the peculiar
quadrupole in our modeling, although it might contribute some
part of the observed dipole.  Bearing this last point in mind,
we continue with our standard model because it provides a better
fit to the data than the model which replaces $\vec w$ with
a Shapley attractor.  In addition, we show below in \S\ref{ssec:cmbframe}
that the dipole can be removed by simply translating the quadrupole
origin to the center of Virgo.

Similarly we experimented with including nearby Abell clusters,
selected under the assumption that their galaxy counts are proportional
to their masses and ranking them by $M/r^2$.  Using a suite of the
22 most significant ones (including Coma, Perseus, A2199, etc.)
and allowing their mass-to-luminosity ratio to be a free parameter
gives a likelihood $\cal N$ = 268.8, not a significant improvement.
The model parameters are virtually unchanged, including the quadrupole
and the dipole velocity.

\subsection{$\Omega_M$}

Lacking any a priori information about the values of the $\delta$'s,
we have very little constraint on $\Omega_M$.  Only the
non-linear component to Yahil's $\rho^{1/4}$ law (Eq.~\ref{eq:upec})
provides any leverage at all.
Nevertheless, if we permit $\Omega_M$ to be a free parameter 
the best fitting model comes in with $\cal N$ = 269.2, and
$\Omega_M = 0.08\pm 0.33$.  If we refit our standard model with 
$\Omega_M = 1$, we find $\cal N$ = 269.5, again indicating how little
direct handle we have on $\Omega_M$.  The other parameters change
very little, but as expected,
the combination of $\delta\,\Omega_M^{2/3}$ for the two attractors is
relatively constant for the various values of $\Omega_M$ at 0.33 for
the Virgo Attractor and 0.27 for the Great Attractor.
We plan in a future work to compare our measured peculiar velocities
to the galaxy distribution and so obtain constraints on $\Omega_M$.
%
%
\newpage
\section{Discussion}

\subsection{Mass Distributions of the Virgo and Great Attractors}

We find good fits to the velocity field using attractors whose density
distributions go as $\rho \sim r^{-2}$ when $r$ is small enough that
$\delta>1$, but 
steepening to $\rho \sim r^{-3}$ by $\delta\approx1$.
Most previous parametric flow models have adopted isothermal $\gamma = 2$
attractors (e.g., Lynden-Bell \etal\ 1988; Han \& Mould 1990), as
$\gamma$ was not well-constrained by their data sets.
Aaronson \etal\ (1989) did allow $\gamma$ to vary in their model,
although they constrained it to be the same for both the Virgo and
Great Attractor mass concentrations, and found $\gamma\approx2$.
Faber \& Burstein (1988) also experimented with different $\gamma$'s
and found that Virgo was roughly isothermal, and the more
distant GA preferred $\gamma\sim2.7$.  This is similar to what we 
find with $r_{\rm cut}$ set to infinity and $\gamma$ 
allowed to vary, but we prefer the approach adopted in the previous
section because of the substantial improvement of the fit.

The isothermal cluster assumption was motivated 
by dynamical evidence that the enclosed mass in clusters increased
approximately linearly (e.g., Press \& Davis 1982).  Observations
continue to show flat velocity dispersion profiles for many clusters,
in support of this assumption (e.g., Fadda \etal\ 1996).
Weak lensing observations are also consistent with isothermal,
or somewhat steeper, mass distributions (Squires \etal\ 1996a, 1996b).
However, none of these studies probe beyond a few Mpc from the cluster
centers, whereas we are dealing here with the distributions over tens of Mpc.

Numerical studies by Navarro, Frenk, \& White (1996, 1997) suggest that
cluster dark matter halos in hierarchical collapse models should 
go as $r^{-3}$ at large radii.  
Earlier numerical work on the collapse of isolated halos produced
even steeper $r^{-4}$ profiles at large radii (Dubinski \& Carlberg 1991).
The total mass of course depends strongly on $\gamma$, with the 
$\gamma \leq 3$ yielding an infinite mass; thus at some point the
distribution must steepen.  Our observed mass profiles therefore are
quite reasonable in terms of flat velocity dispersion profiles and the
numerical simulations.

\subsection{Dipoles and Quadrupoles}
\label{ssec:cmbframe}

Although we find a 4-$\sigma$, 150\kms\ dipole residual pointing in
the SGY direction, we regard this result as an indication that our
survey volume, which extends to about 3000\kms\ or 40\Mpc, is
essentially at rest with respect to the CMB.  Indeed, we can clearly
create or remove any dipole velocity we like by shifting the origin of
the quadrupole contribution, as is evident from
Equation~(\ref{eq:matrix}), provided the quadrupole matrix is invertible.
Applying the inverse of the quadrupole to the dipole velocity 
$\vec w$ of our best fit model implies a quadrupole origin of
$(-17,13,-9)$, rather close to the VA, although moved somewhat towards
the GA.  If we shift the quadrupole origin 
from the Local Group to the center of the Virgo Attractor (which can
then partially mimic the effects of a flattened attractor), choose
a cutoff $r_{\rm quad} = 35$~Mpc, and fit
with $\vec w = 0$, we find $\cal N$ = 272.0, which is only 3 greater
with effectively three fewer degrees of freedom; it is 6.5 less (with
the same DOF) than the model with $\vec w = 0$ and the
quadrupole centered on the LG.   Alternatively we can
fit for $\vec w$ with the quadrupole centered on the Virgo Attractor,
and we find $\vec w = (-91, -7, -63)$ or within 1-$\sigma$ of zero
in each coordinate.  We therefore think it likely that the galaxies
in our survey volume are on average at rest with respect to the CMB,
and that the dipole and quadrupole terms of our standard model are
acting as lowest order correction terms to our spherical Virgo Attractor.

We found previously that addition of a quadrupole component to the two
attractor model caused the Hubble constant to increase
from 73 to 78 as the likelihood improved.  This is illustrated in the
left panel of 
Figure~\ref{fig:hqsig} which shows confidence contours of our standard
model as a function of $H_0$ and quadrupole amplitude relative to the
best fit value.
We find that moving the origin of the quadrupole to Virgo also has a
slight side effect on the value of the Hubble constant.  If there were
no cutoff in the quadrupole its origin would be completely degenerate
with $\vec w$, but when when the quadrupole is centered on Virgo we
find that $H_0$ drops slightly to 76, and there is some covariance 
between $H_0$ and $r_{\rm quad}$ in the sense that larger $r_{\rm quad}$
leads to larger $H_0$, as seen in the right panel of 
Figure~\ref{fig:hqsig}.
\insfig{
\begin{figure}[t]
\epsscale{1.0}
\plottwo{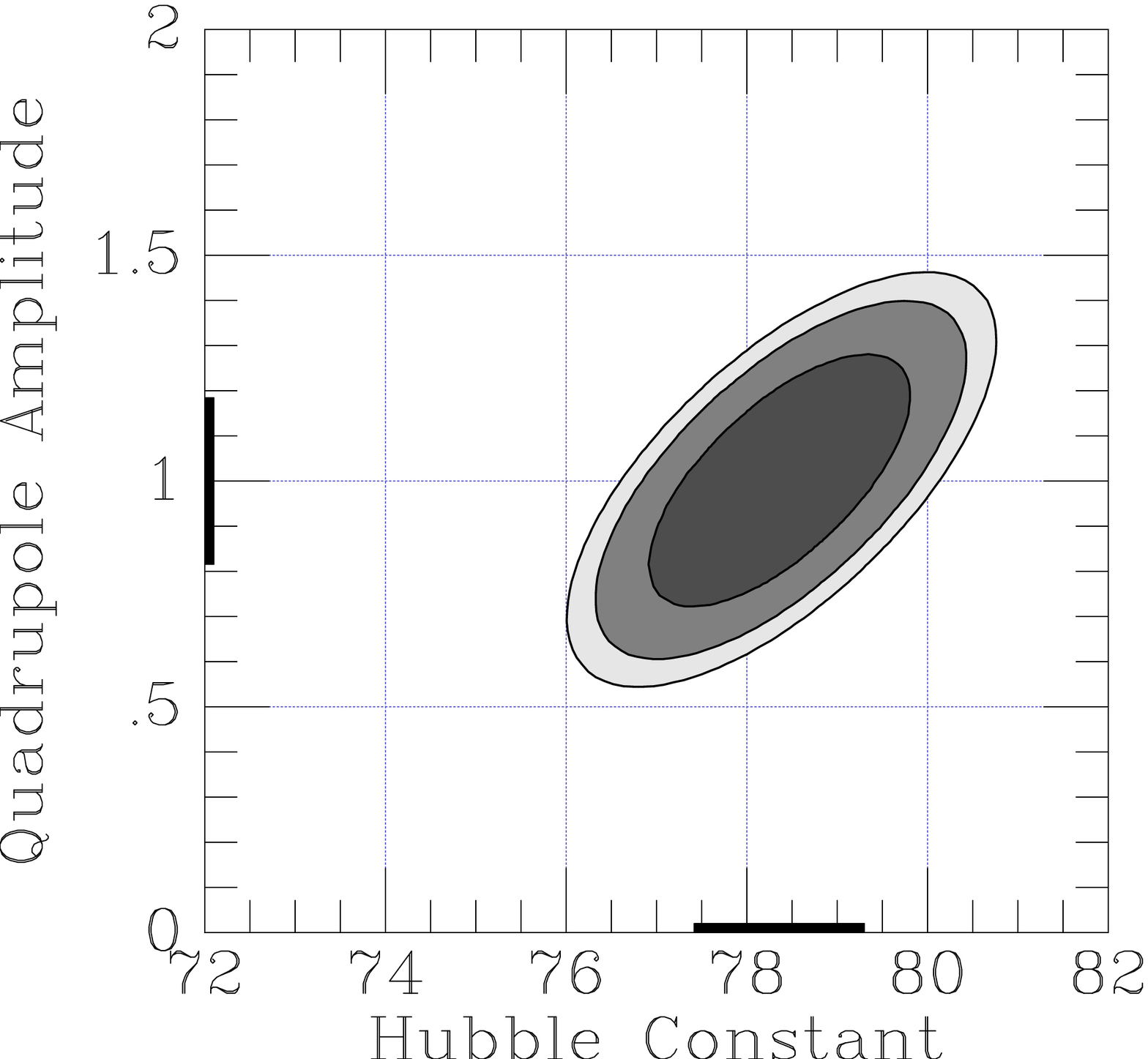}{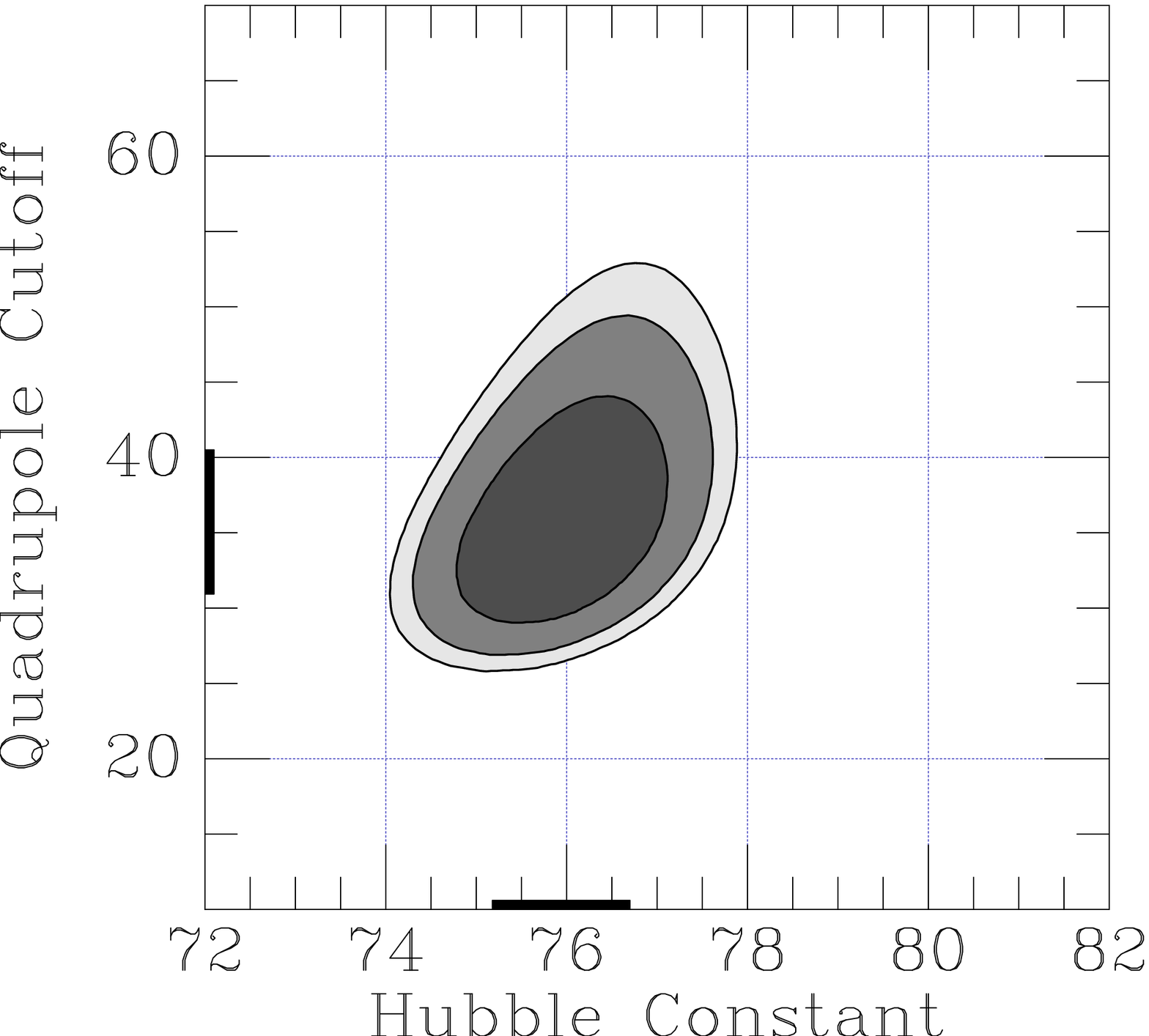}
\caption{
Joint confidence contours of the Hubble constant and relative
quadrupole amplitude when the quadrupole is centered on the Local
Group are shown on the left. Confidence contours of Hubble constant
and quadrupole cutoff radius when the quadrupole is centered on Virgo
instead are displayed on the right.
\label{fig:hqsig}}
\end{figure}
}
Because this use of a quadrupole correction to a spherical infall
model seems like a poor approximation to a flattened potential, and
because the eigenvectors of the quadrupole are tipped by $45^\circ$
from the SGZ symmetry axis of the galaxy distribution, we do not
prefer this model over the one which has the quadrupole centered on
the Local Group.  Nevertheless, we are mindful of the sensitivity of
$H_0$ to this possibility and we therefore choose
$H_0 = 77\pm4$ as the most likely range we can derive from these data
and models.

Unlike Lynden-Bell \etal\ (1988), we do not find that Centaurus
has a much larger radial peculiar velocity in the CMB frame than the Local
Group.  Instead it appears to have a significant extent falling into
the attractor at 43\Mpc\ and at least some galaxies (e.g., NGC~4767)
have no radial velocity with respect to the CMB.  Dressler et al. will
discuss this issue in greater depth, but it appears that the flow in
the Great Attractor region is fairly complex and may need many more
high accuracy distances before it can be completely untangled.

\subsection{Motion of the Local Group}

The Local Group is moving at $627$\kms, with components
$(-406, 352,-324)$\kms, with respect to the CMB frame,
but we have not used this fact at all in our modelling.
To what extent can we account for this motion of the Local
Group reference frame in the context of our model?  The three
contributors to the model velocity at the Local Group are as follows:

\newlength{\notecolwidth}
\setlength{\notecolwidth}{1in}
{\renewcommand{\arraystretch}{1.2}
\begin{tabular}{l@{ }c@{ (}r@{}r@{}r@{) }}
W         & = & $-$55,&  143,&   $-$8 \\
Virgo     & = & $-$25,&  136,&  $-$13 \\
GA        & = &$-$247,&   98,& $-$115 \\
\cline{1-5}
Net       & = &$-$327,&  377,& $-$136 \\
Obs $-$ Net  & = & $-$79,&$-$25,& $-$188 \\
\end{tabular}\\}
We can account for $w_x$ and $w_y$ quite well; the residual of
$-79$\kms\ is not statistically significant and is in any case
a reasonable peculiar velocity for the Local Group 
with respect to the nearby Hubble flow.  

The $w_z$ residual of $-188$\kms\ is much too large to be an unsuspected
motion of the Milky Way with respect to the LG frame.  The models by
Yahil et al. (1977) and Courteau \& van den Bergh (1999), among others,
require that this be a motion of the entire Local Group.
Our sample does not have any nearby galaxies which are far enough out of
the supergalactic plane to ascertain whether this motion is restricted
to the Local Group or extends farther.  The nearest galaxy in our sample
with $|\rm{SGB}|>20^\circ$ is 8~Mpc distant, and at SGB$=
-33^\circ$ we could not distinguish between $v_z = 0$ and $v_z = -188$.
We note that this velocity is similar in magnitude to our best-fit 1-d
thermal velocity.

This peculiar motion of the Local Group in the general direction of
$-$SGZ has been noted before and was named the ``Local Anomaly'' (LA)
by Tully (1988) and Faber \& Burstein (1988), who both
attributed it to inhomogeneities in the galaxy distribution within
$c{z}\sim1000$ \kms, including the ``Local Void'' (Tully \& Fisher 1987).
These authors found fairly large LA velocities: the Faber \& Burstein 
value was 360\kms\ towards $(l,b) = (199^\circ,0^\circ)$, or 
$(169,30,-317)$ \kms\ in supergalactic components. Han \& Mould (1990) 
modeled the LA in more detail and found a smaller net motion of 236\kms\
towards $(l,b) = (205^\circ,11^\circ)$, or $(88,68,-208)$\kms. 
All three of these sets of
authors used the Aaronson \etal\ (1982b) infrared Tully-Fisher data set to
measure the LA.  Aaronson \etal\ (1982a) themselves found a LG peculiar
motion of $(38, 86, -150)$ \kms\ and attributed the large $-$SGZ velocity
to a displacement of the Local Group above the supergalactic plane
and a resultant downward acceleration.

Lahav \etal\ (1993) suggested that the LA could be explained by the combined
effects of the Local Void in the $+$SGZ direction and the Puppis concentration
of galaxies towards $-$SGZ, with possible smaller contributions from Fornax
and Eridanus.  The Puppis concentration was revealed by H\,{\sc i} surveys
of {\it IRAS}-selected galaxies; it is a rich, but loose, association of
galaxies lying in the Galactic plane at 
$(l,b) = (245^\circ,0^\circ)$ with a mean velocity of about 2000\kms\
(Yamada \etal\ 1994).  Lu \& Freudling (1995) surveyed the region behind 
the Galactic plane towards the direction of the Faber \& Burstein (1988) 
and Han \& Mould (1990) LA vectors and disputed the conclusion of Lahav et~al.
Since they found no other significant excesses of nearby galaxies,
Lu \& Freudling concluded that any motion arising from a gravitional
pull of nearby galaxies and a push from the Local Void, whose center
is nearly diametrically opposite Puppis, would have to be directed almost
straight towards Puppis, nearly 45$^\circ$ from the supposed direction of
the LA vector.  

In fact, the Local Anomaly we find here has an apex at 
$(l,b) = (250^\circ,-13^\circ)$, remarkably close to the direction of the
Puppis concentration and to the anticenter of the Local Void as sketched
out by Lu \& Freudling.  However, the magnitude we find is significantly
smaller than the Faber \& Burstein value adopted by Lahav \etal\ (1993);
thus, the $\sim\,$50\kms\ contribution they estimated for the Puppis galaxies
becomes less necessary.  Likewise, Han \& Mould (1990) concluded that the
Local Void might account for their fairly modest LA motion, if the
mean position of the Void was a free parameter.

We experimented with adding a Gaussian void, as defined by
Eq.~(\ref{eq:vamp}), to our standard model along the positive SGZ axis.
We first looked at a very small Local Void with both the distance and
Gaussian size set to 10~Mpc and fitted for the velocity amplitude.
Since there is some covariance between $v_{\rm amp}$
and $w_z$, we set $w_z=0$.
We find a best-fit amplitude 
of $v_{\rm amp} = -8.5\pm5.6$ \kms, and an improvement in ${\cal N}$
from 269.2 to 267.7, marginally significant.  This void provides a
push of $-65$ \kms\ at the location of the LG.
The model parameters show little change, although $H_0$ drops from 78
to 77\kmsMpc.  Alternatively, we tried adding a Local Void using the
distance and size of 20\Mpc\ suggested by redshift surveys and fixing
the amplitude at the value of $v_{\rm amp} = -88$ \kms\ which provides
the push of $-188$ \kms\ at the Local Group that we are trying to
understand.  The central $\delta$ of this void is close to $-1$.
This time we do allow $w_z$ vary.  The fit improves still more to
${\cal N} = 265.9$ and $w_z$ is still negligible at $20$\kms, but the
amplitudes of the VA and GA shift quite a bit to 172 and 159\kms\
respectively, $w_x$ and $w_y$ also move to $-136$ and $+151$\kms, and
$H_0$ decreases to 73\kmsMpc.
Obviously fine-tuning the free parameters could produce an even better
match to the excess LG motion, but this is precisely the sort of
ad~hoc modelling which we think is not appropriate.  This void's
properties are not well-constrained in our model nor are there
well-motivated external constraints, so we choose not to make this
adjustment to our standard model, but simply note that our peculiar
velocity data do seem to support the idea that there is a 20\Mpc\ void
above the Local Group which is nearly empty at its center.

Another interesting issue is the peculiar velocity of the Local Group
with respect to the Virgo Attractor, since this was the historical
means of determining the Virgo infall amplitude.  The Local Group is
approaching Virgo at about 139\kms\ due to the Virgo Attractor itself;
the Great Attractor causes the LG and Virgo to converge at 97\kms; the
extra quadrupole component results in the LG and Virgo approaching
each other at an additional 173\kms.  The total model peculiar
velocity between the Local Group and Virgo is therefore 409\kms.


It is unclear how much ``infall velocity'' to ascribe to the Virgo
Attractor.  Using our model of the Virgo Attractor as a spherical
distribution, it only tugs on us at 139\kms.  However, if we regard
the entire quadrupole as a component of
an anisotropic Virgo Attractor, the net tug on the LG is
303\kms.  These considerations may partially account for the widely
disparate values which have been reported for the Virgo infall
amplitude.  Depending on the sample one is using and whether one's fit
is dominated by model amplitudes near our location or averaged all
around the Virgo supercluster, it is possible to get naive ``infall
velocities'' (i.e., peculiar velocity between Virgo and the LG)
anywhere between 140\kms\ and 400\kms.

\subsection{Comparison with Other Flow Studies}

Our analysis finds more modest infall velocities at the Local Group
and a smaller bulk flow than most previous studies.  Part of the
reason for our smaller infall velocities is that our modeling has been
more flexible in not requiring a single power-law profile for the
massive attractors.  Thus, our model can accommodate fairly large 
infalls close to the attractors without requiring large motions 
at the Local Group.  In addition, we have not required the spherical
infalls to account for all of the Local Group motion in the CMB frame,
but rather have allowed the reference frame to be a free parameter.
In fact, setting $\vec w = 0$, $r_{\rm cut} = \infty$,
and omitting the quadrupole term yields a model similar 
to that of Lynden-Bell \etal\ (1988), but at a significant cost 
in the likelihood (see \S\ref{ssec:infalls}).

Our best-fit bulk flow $\vec w$ of $\sim\,$150 \kms\ towards 
$(l,b) = (294^\circ, +67^\circ)\pm43^\circ$, and the strong indications
that even this modest value may be an artifact of the model,
appears inconsistent with several recent studies which find
large-scale streaming motions of amplitude
$\sim\,$700 \kms\  for volumes encompassing our survey volume 
(e.g., Lauer \& Postman 1994; Hudson \etal\ 1999; Willick 1999).
However, it is quite consistent with the bulk flows found by the $I$-band
Tully-Fisher study of Giovanelli \etal\ (1998) 
(using their ``Case b'' weighting).  These authors report bulk flows of
$151\pm120$\kms\ towards $(l,b) = (295^\circ, +28^\circ)\pm45^\circ$ for their full
sample of 24 clusters extending out to 9000 ~km\,s$^{-1}$, and
$131\pm90$\kms\ towards $(l,b) = (325^\circ, +62^\circ)\pm60^\circ$ 
for their 17 clusters within 6000~km\,s$^{-1}$.


Our best-fit 1-d thermal velocity is $\sigma_{\rm cosmic}=180\pm14$~km/s,
after making allowance for velocity measurement uncertainty.
Lynden-Bell \etal\ (1988) found a best-fit thermal dispersion of
$250 \pm 40$ in their favored model and
concluded that this result was mainly due to small-scale flows, not
virial dispersions in groups.

Strauss, Cen, \& Ostriker (1993) calculated the thermal velocity dispersions
from several peculiar motions surveys in order to compare the 
resulting Mach numbers (the ratio of the mean bulk flow to the
3-d thermal dispersion) to expectations from various cosmologies.  
Their estimates of the 1-d dispersions from the observational
surveys ranged from 250 to 420 \kms.  
More recent large-scale flow studies often simply assume a 
value $\sigma_{\rm cosmic}=250$ \kms\ for clusters and find that the resulting
$\chi^2_N \sim 1$ (e.g., Willick 1999; Hudson \etal\ 1999).
The Giovanelli \etal\ (1998) cluster study allowed this to be a 
free parameter and found $\sigma_{\rm cosmic}=300\pm80$ \kms.
However, on small enough scales there is evidence that the dispersion 
of galaxies (field spirals in particular)
about their local flow field is closer to $\sim\,$125\,\kms\
(Davis, Miller, \& White 1997; Willick \etal\ 1997). 
This value may better reflect the virial
velocities of galaxies within small groups.

\subsection{The Revised Calibration and $H_0$}
\label{sec:Hnaught}

The revised \Mi\ calibration of Eq.~(\ref{eq:newcal}) carries
a formal zero-point uncertainty of 0.10~mag from the SBF/Cepheid
galaxy comparison.  This uncertainty is larger than the 0.07~mag
quoted in SBF-I because that paper used the group comparison, 
which averaged many SBF distances for each Cepheid distance.
As stated above, the group comparison is 0.13--0.19 mag fainter than
the galaxy comparison.
By choosing the galaxy calibration over the group one, we have
implicitly assumed that the difference is due to a systematic
offset in the mean distances of the HST Key Project Cepheid
galaxies and the groups in which they are thought to reside.
The alternative explanation is that the measured SBF magnitudes
are systematically too bright for the spiral bulges, either 
because of population effects or because of extra variance from
dust or other features.  Ferrarese \etal\ (1999) have likewise
chosen the former explanation and have derived a calibration
similar to that given here.

With this new calibration and depending on the flow model, we have 
derived values of $H_0$ in the range of 70--80\kmsMpc.  
This range correlates with the amount of
mass in the model attractors.  The value of $H_0 = 70$ corresponds to
a pure Hubble flow plus overall dipole model; at the other extreme
with $H_0 = 80$ is a dual isothermal ($\gamma=2$) attractor model with
$r_{\rm cut}$ set to infinity.  The former model is ruled out by the
extremely poor $\chi^2_N$; the latter model
overestimates the attractor masses, thus overestimating
$H_0$.  When we allow $r_{\rm cut}$ to vary and produce an $r^{-3}$
(and steeper) large radius falloff to the models, $H_0$ comes out near
74.  There is, however, a significant degree of anisotropy in the data
with respect to this model as evidenced by the dramatic improvement in
likelihood when a quadrupole component is introduced into the model.
As it happens, relieving this anisotropy with a quadrupole centered on
the Local Group causes our value for $H_0$ to
rise somewhat to $78$, whereas a quadrupole centered on Virgo yields
an $H_0$ of 76.  Despite our qualms about the ad hoc nature of the
latter model, the fact that it has a $\vec w$ consistent with zero 
causes us to prefer it slightly for estimating $H_0$.
Thus, for our best estimate of
a flow-corrected but unbiased value, we adopt
\begin{equation}\label{eq:Ho_val}
H_0 \;=\; 77 \pm 4 \pm 7 \hbox{ km\,s$^{-1}$\,Mpc$^{-1}$}\,.
\end{equation}
The first errorbar is the quadrature sum of the internal uncertainty
for a single flow model and an estimate of the likely range from
the different models.  The second errorbar includes the 0.10\,mag
uncertainty in the fitted calibration to Cepheids and an
estimated 0.17\,mag uncertainty in the Cepheid zero point itself
(see Ferrarese \etal); these systematic uncertainties also have 
been combined in quadrature.

Our value for $H_0$ differs from the value of 69\kmsMpc\ found by
the Ferrarese \etal\ (1999) analysis of SBF partially because of our
choice of zero point and partially because of the flow models used.
We have done extensive modeling to estimate and remove the effects
of large-scale flows from our data. 
Another difference is that we have used the full survey data set in
deriving $H_0$, rather than just four SBF measurements from \hst.  
Consequently, our best value is 11\% larger
than theirs, although marginally consistent within the random uncertainty.

As with the Key Project results, if the Cepheid calibration is changed
so that the distance modulus of the Large Magellanic Cloud is 0.2~mag
smaller than the standard 18.5~mag value, then $H_0$ would increase by
10\%; if the Kennicutt \etal\ (1998) metallicity dependence for the
Cepheid period-luminosity relation is correct, then $H_0$ would decrease
by 5\%.  However, our calibration would become quite discordant with the
theoretical calibration using Worthey's models if the zero point moved
by more than about 0.2\,mag, and those models are largely based on the
RR Lyrae and parallax distance scales, rather than Cepheids.

\section{Summary and Conclusions}

We have shown that the SBF technique is a powerful tool in mapping the
local and large-scale velocity field through accurate distance
measurements to early-type galaxies.  Judging by our best data, for
which the product of seeing and distance $PD < 1.0$, distance
measurements with an accuracy of $\pm5$\% are attainable for any
early-type galaxies with sufficiently good data.  Distance measurements
of this accuracy are essentially immune to Malmquist-like biases.

The parametric model we have developed is unique in the way it handles
both local peculiar motions and velocity dispersions, particularly for
galaxy clusters.  
Although it is important to follow this first analysis with 
non-parametric models, we find it remarkable that nearly all of
the deviation from a smooth Hubble flow can be accounted for by the
addition of two spherical, truncated isothermal attractors, one
centered on the Virgo cluster and the other centered just beyond
the most distant of the ellipticals associated with the Cen-30 
and Cen-45 clusters.  These two
attractors appear to center on enhancements in galaxy distribution,
consistent with the idea that galaxy light traces mass in at least
some approximate way.  The improvement in the model with the inclusion
of an additional dipole and quadrupole is statistically significant,
but our analysis indicates that these terms are more likely the result
of inadequacies in the spherical attractor model, rather than
manifestations of additional mass fluctuations beyond our survey
volume.  Further measurements, particularly beyond the $R \approx
40$\Mpc\ limit of most of our data, will be able to discriminate
between competing explanations, but it is clear from the size and
direction of these components that the influence of mass overdensities
in the Perseus-Pisces supercluster or Shapley concentration do not
play a substantial role in perturbing the velocity field within $3000
\kms$.

Not surprisingly, we find strong covariance between pairs of
parameters, for example, the amplitude of the GA infall and the
strength of the dipole component in this direction, or the cutoff
radius and the power-law exponent of mass distribution attributed to
the attractors.  New data are unlikely to reduce some of these
covariances, but additional measurements on
the backside of the Great Attractor, for example, will be able to
constrain much better the contribution from more distant concentrations
that produce a local dipole.  

Our standard flow model resolves most of the Local Group's motion with
respect to the CMB, except for the continued presence of a nearly 
$200 \kms$ component perpendicular to the supergalactic plane.  
This ``Local Anomaly'' has been seen before in other data sets. The
amplitude we find is generally smaller than in those previous studies,
and crude modeling of this ``anomalous'' motion as a ``push'' from 
a nearby void in the $+$SGZ direction can account for nearly all of it.

Our data are considerably more accurate than
those used in the analysis of Lynden-Bell et al. (1988).  Our best fit
model, while qualitatively similar, is significantly different, but
this is mainly a product of our inclusion of other components, for
example, the inclusion of a dipole in addition to the two attractors.
In particular, the thermal velocity of our sample is only $180 \kms$,
significantly lower than the Lynden-Bell et al. study. A more detailed
comparison of our study with previous results will appear in Dressler
et al. (1999).

We find zero or at most a very small dipole motion in the $R < 3000
\kms$ volume, which is consistent with the study of Giovanelli et al.
(1998), mildly inconsistent with the larger bulk flow reported by
Courteau et al.  (1994), and very inconsistent with the
large-amplitude, large-scale flows reported by Lauer \& Postman
(1994), Hudson et al.  (1999), and Willick (1999) unless the $R < 3000
\kms$ is at rest with respect to the CMB while the thick shell outside
$3000 \kms < R < 10,000 \kms$ in rapidly moving, a situation we think
is very unlikely.

Finally, we provide a new zero point for the SBF relation, based on SBF
distances to galaxies that have Cepheid distances determinedby the HST
Key Project.  The Hubble constant that results from this calibration,
$H_0 = 77 \pm 4 \pm 7$\kmsMpc\ is somewhat higher than
results of that study, partially because we choose a different zero
point and partially because we have a more extensive set of galaxies
and a more sophisticated model.  We
consider the zero point issue to be unresolved and in need of further
study. 

Our study shows the power of accurate SBF distance measures in
advancing our understanding of the local mass distribution and the
distortions in the velocity field that result from it.  Extending the
SBF technique, through bigger aperture telescopes with better
\PSF\ performance, through near-IR SBF measurements,
and through observations with HST and future space telescopes, promises
to provide decisive answers for the long-asked questions regarding
the large-scale distribution in the local universe.

\acknowledgements
We are grateful to all our friends who have helped us collect these
data over the years.  This
work was supported primarily by NSF grant  AST9401519.
Additional support for this work was provided by NASA through grant number
GO-06579.02-95A from the Space Telescope Science Institute, which is
operated by the Association of Universities for Research in Astronomy,
Inc., under NASA contract NAS5-26555.

\newpage

\appendix

\section{Appendix. The flow model}

Our standard flow model consists of 5 components: Hubble flow,
peculiar velocity, Virgo Attractor, Great Attractor, and quadrupole.
Alternative models are possible as discussed above which center the
quadrupole on Virgo or use different values of $\gamma$, but they give
very nearly the same model velocity as a function of position.  One
caveat is that our measure of the mean flow velocities around the VA
and GA are extremely uncertain for $r<6$~Mpc for the VA and $r<10$~Mpc
for the GA.

We work in supergalactic coordinates, using a transformation from
galactic to supergalactic coordinates by applying the rotation matrix
\begin{equation}
      \left(\matrix{-\sin\phi_S&+\cos\phi_S&0\cr
      -\sin\theta_S\,\cos\phi_S&-\sin\theta_S\,\sin\phi_S&\cos\theta_S\cr
      +\cos\theta_S\,\cos\phi_S&+\cos\theta_S\,\sin\phi_S&\sin\theta_S\cr
        }\right)
 \label{eq:rotmat}
\end{equation}
to galactic cartesian coordinates, 
where $\phi_S = 47.37^\circ$ and  $\theta_S = 6.32^\circ$.

Our value for the Sun's motion with respect to the CMB (hence
transformation from heliocentric velocities to CMB reference frame)
comes from Lineweaver et al. (1996), who find the sun moving at
369\kms\ towards SGL = $264.31^\circ$ and SGB = $48.05^\circ$.

Given a position $(x,y,z)$ in supergalactic coordinates with respect
to the Local Group, the first two terms of our model are
\begin{equation}
\vec v = H_0 \vec r + \vec w
\end{equation}
where $H_0 = 78.4$ and $\vec w = (-55,143,-8)$.

The two attractors have a radial infall around them.  Each causes a
deviation from the above velocity of the form (equation \ref{eq:upec})
\begin{equation}
u_{\rm infall} = {1\over3}\, H_0\, r_A \; \Omega_M^{0.6}\; 
        \delta(r_A)\, (1+\delta(r_A))^{-1/4}
\end{equation}
where $r_A$ is the distance from the point of interest $(x,y,z)$ to
the attractor, $\Omega_M = 0.2$, and $\delta(r_A)$ is the overdensity in
units of the background density at distance $r_A$ from the attractor.
The overdensity $\delta(r_A)$ is calculated from equation \ref{eq:rhobar}:
\begin{equation}
\delta(r_A) = {\delta_0\,e^{-r_A/r_{\rm cut}}\over1-\gamma/3} 
\left( r_A\over r_c\right)^{-3} \,
\left[(1+{r_A^3\over r_c^3})^{1-\gamma/3} - 1\right].
\end{equation}
$\delta_0$ is evaluated by plugging in the distance from the LG to the
attractor and matching to the model overdensities given below.
The contribution to $\vec v$ is obtained by resolving the radial
$u_{\rm infall}$ into components along $\vec r - \vec r_A$.

The Virgo Attractor is found at $(-3.1,16.6,-1.6)$ Mpc; it has
overdensity $\delta_V = 0.974$; its infall exponent is 
$\gamma_V = 1.5$; the core radius is $r_c = 2$; and the cutoff radius
is $r_{\rm cut} = 11.7$.

The Great Attractor is found at $(-36.7,14.6,-17.1)$ Mpc; it has
overdensity $\delta_G = 0.781$; its infall exponent is 
$\gamma_G = 2.0$; the core radius is $r_c = 2$; and the cutoff radius
is $r_{\rm cut} = 49.5$.

Finally, the quadrupole contribution is found by multiplying $\vec r$
by the matrix
\begin{equation}
      e^{-r^2 / 2r_{\rm quad}^2}
      \left(\matrix{
        ~2.57 &         \phn~2.04 & ~\llap{$-$}7.87\cr
        ~2.04 & ~\llap{$-$}11.12 & ~\llap{$-$}3.63\cr
 ~\llap{$-$}7.87 & \phn~\llap{$-$}3.63 &         ~8.55\cr
        }\right),
\end{equation}
where the quadrupole cutoff radius is $r_{\rm quad} = 50$~Mpc.

Our model for the (1-d) velocity dispersion as a function of position
has four components added in quadrature.  The first is an overall
thermal velocity of 187\kms.  The next two are components centered on
the VA and GA at the positions listed above, with amplitudes 650\kms\
and 500\kms\ respectively at the center, but falling off spatially as
a Gaussian with a sigma of 2\Mpc.  Finally, we incorporate a fourth
component for the Fornax cluster, centered at
$(-1.9,-15.0,-13.4)$\Mpc, with amplitude 235\kms\ and again a spatial
falloff with sigma of 2\Mpc.
Examples: 

\begin{flushleft}
\begin{tabular}{r@{,}r@{,}r@{ }r@{,}r@{,}r@{ }r@{  }r@{}}
 $x$  & $y$  & $z$   & $v_x$    & $v_y$    & $v_z$    & $v_r$  & $\sigma_v$ \\
 $-3$ & $18$ & $-2$  & $-590.0$ & $1101.7$ & $-298.6$ & $1209$ & $532$ \\
$-35$ & $15$ &$-15$ & $-3551.5$ &  $996.3$ & $-2118.3$& $4179$ & $272$ \\
\end{tabular}
\end{flushleft}
\newpage
\section{Appendix B. The zero point}

We find it difficult to choose a zero point for SBF using the various
Cepheid distances which are available because we are faced with two
options, each with virtues and faults.  SBF is
best measured in elliptical and S0 galaxies, and there are no direct
Cepheid distance to such galaxies.  Worse, there are presently no
distances to spiral galaxies which appear to be tidally interacting
with E or S0 galaxies (such as NGC~4647 and NGC~4649).  We therefore
are faced with the unpleasant choice of associating Cepheid distances
to SBF galaxies via group associations, or else measuring SBF directly
in spiral bulges where Cepheid distances exist.

\subsection{Option 1. Calibrating via Groups}

For a group comparison we refer to Ferrarese et al. (1999) for Cepheid
distances, but we reject the Coma groups because of the extreme
confusion in that region, the Cen~A group because it has a very large
depth inferred from its angular extent across the sky, and the
NGC~7331 group because it is really a galaxy-galaxy comparison with
all the inherent worries described below.  We also, for the sake of
argument, do not agree with Ferrarese et al. in the inclusion of
NGC~1425 as part of Fornax.  It is statistically inconsistent with the
other two Cepheid galaxies, NGC~1326A and NGC~1365; it is halfway
between the Fornax and the Eridanus clusters, and its Cepheid
distance is 0.40 mag more distant than the other two, very close to
the 0.46 mag that we find for the mean distance between Fornax and
Eridanus.  We also concur with Ferrarese et al. in rejecting NGC~4639
from the Virgo cluster because it is 0.77 mag more distant.

Table \ref{tab:grp} lists the mean \mi\ we find for these groups,
adjusted to a fiducial color of $\viz = 1.15$, the distance modulus
from Ferrarese et al., and the inferred SBF zero point, the absolute
\Mi\ at the fiducial color.

\insfig{
\begin{deluxetable}{lccc}
\tablewidth{0pt}
\tablecaption{Zero Points from Galaxy Groups}
\tablehead{
\colhead{Group}   & \colhead{$\langle \mi \rangle$} &
\colhead{\dmod}    & \colhead{$\langle \Mi \rangle$}
}
\startdata
M31    & $22.74 \pm 0.04$ & $24.44 \pm 0.10$ & $-1.70 \pm 0.11$ \nl
N1023  & $28.26 \pm 0.08$ & $29.84 \pm 0.08$ & $-1.58 \pm 0.11$ \nl
Fornax & $29.78 \pm 0.03$ & $31.41 \pm 0.06$ & $-1.63 \pm 0.07$ \nl
M81    & $26.27 \pm 0.09$ & $27.80 \pm 0.08$ & $-1.53 \pm 0.12$ \nl
LeoI   & $28.48 \pm 0.04$ & $30.08 \pm 0.06$ & $-1.60 \pm 0.07$ \nl
Virgo  & $29.41 \pm 0.02$ & $31.03 \pm 0.06$ & $-1.62 \pm 0.06$ \nl
\enddata
\label{tab:grp}
\end{deluxetable}
}

These groups paint a very consistent story: the zero point of the SBF
calibration is $\Mi = -1.61\pm0.03$.  The big advantage to
using a group calibration is that we have many, very good SBF
measurements in groups, as is reflected in the very small error bars.

The biggest disadvantage is that the association of spirals with
ellipticals is shaky at best.  As pointed out by Jacoby et al. (1992)
the spirals in groups seem to lie outside of an elliptical
core, as one might expect from the fragility of spirals and the harsh
environment in the center of a group.  However, in this case we are
finding a sheet of five Cepheid bearing spirals at precisely the
distance we assign to the core of the Virgo cluster, and two spirals
at precisely the same distance as the core of Fornax.  Had we included
NGC~1425 as part of Fornax and NGC~4639 as part of Virgo, we might have
expected the elliptical core to lie somewhere in between, leading to a
brighter zero point by 0.1--0.2 mag.

\subsection{Option 2. Calibrating via Galaxies}

We again refer to Ferrarese et al. for Cepheid distances to galaxies
where we have measured an SBF magnitude in the bulge.
Table \ref{tab:gxy} lists the mean \mi\ we find for these six
galaxies, adjusted to a  
fiducial color of $\viz = 1.15$, the distance modulus from
Ferrarese et al., and the inferred SBF zero point, the absolute \Mi\
at the fiducial color.

\insfig{
\begin{deluxetable}{lccc}
\tablewidth{0pt}
\tablecaption{Zero Points from Individual Spirals}
\tablehead{
\colhead{Galaxy}   & \colhead{$\langle \mi \rangle$} &
\colhead{\dmod}    & \colhead{$\langle \Mi \rangle$}
}
\startdata
N0224 & $22.67 \pm 0.06$ & $24.44 \pm 0.10$ & $-1.77 \pm 0.12$ \nl
N3031 & $26.21 \pm 0.25$ & $27.80 \pm 0.08$ & $-1.59 \pm 0.26$ \nl
N3368 & $28.34 \pm 0.21$ & $30.20 \pm 0.10$ & $-1.86 \pm 0.23$ \nl
N4548 & $29.68 \pm 0.54$ & $31.04 \pm 0.08$ & $-1.36 \pm 0.55$ \nl
N4725 & $28.87 \pm 0.34$ & $30.57 \pm 0.08$ & $-1.70 \pm 0.35$ \nl
N7331 & $28.85 \pm 0.16$ & $30.89 \pm 0.10$ & $-2.04 \pm 0.19$ \nl
\enddata
\label{tab:gxy}
\end{deluxetable}
}

These galaxies also are self consistent: the zero point of the SBF
calibration comes out as $\Mi = -1.80\pm0.08$.  Given the wide range
in the errors in the SBF measurement, it is also interesting to look
at the median: $\Mi = -1.735$.  Unfortunately the galaxy average is
inconsistent with the group calibration -- they differ by 2.2 sigma
which makes us very reluctant simply to take the group-galaxy
comparison average.

The advantages of using the very same galaxies to carry out the
calibration is that there is no doubt about the spatial relationship
of the stars contributing to the SBF mean and the Cepheids.  The
disadvantage is that the SBF measurement is very hard, as is evidenced
by the large error bars.  We can only work on the side of the bulge
which is in front of the plane of the disk; we must contend with dust
in the bulge; we are affected by a portion of the disk which underlays
the bulge we are analyzing; we encounter strong color gradients; and
we always worry that somehow SBF might be different in spiral bulges
than in ellipticals (despite the evidence to the contrary presented in
SBF-I).   We can try to improve our SBF measurements, but for some
galaxies (e.g., NGC~4414) even HST resolution only reveals more and more
dust threaded throughout the bulge.  We do our best to make an
accurate, unbiased measurement of SBF in spiral bulges, but we suspect
that if there is a bias, it is probably in the sense of assigning too
bright an SBF magnitude.

Given the probability that the ellipticals and spirals in these groups
may be systematically separated in space, we feel that we must use the
galaxy-galaxy calibration, although we are hedging our choice slightly
by using the galaxy median.  This is a situation which must be
improved by further observation.

\endinsfig  

\clearpage

\manufig{
\centerline{\bf FIGURE CAPTIONS}
\bigskip
}

\manufig{
\figcaption[f01.eps]{
The model velocity distribution function $P(v|r)$ is shown for a line
of sight which passes through the Virgo cluster.  The different
grayscale levels show the 2, 1, 0.5, and 0.2 sigma points on the
velocity distribution at a given distance.  A distance
observation is shown on the abscissa as a distance probability
function, and the resulting velocity probability function is shown on
the ordinate.  This is evaluated at the observed velocity and forms a
term in the likelihood product.
\label{fig:probs}}
}

\manufig{
\figcaption[f02.eps]{
Recession velocities in the CMB frame are plotted as a function of distance.
The line has a slope of 73\kmsMpc; 
the very deviant high points near 40~Mpc are Cen-45 galaxies.
\label{fig:hubble}}
}

\manufig{
\figcaption[f03.eps]{
Hubble ratios (in CMB frame) are plotted as a function of distance for
those points with small enough errors that the error in the Hubble
ratio is less than 10~\kmsMpc.
\label{fig:h0}}
}

\manufig{
\figcaption[f04.eps]{
Recession velocities (in Local Group frame) are plotted as a function of
distance in two directions.  The upper panel shows galaxies which lie
approximately in a $30^\circ$ (half angle) cone in the $\pm$SGX direction,
$90^\circ$ away from the direction of Virgo.  The lower panel shows
galaxies which lie within a $15^\circ$ cone towards Virgo (+SGY) and a
$45^\circ$ cone away from Virgo (which includes the Fornax cluster).  A
Hubble ratio of 73\kmsMpc\ is also drawn.
\label{fig:aniso}}
}

\manufig{
\figcaption[f05.eps]{
Residual velocities after removal of a model Hubble
flow.  Residual velocities which are less than $1{-}\sigma$ are shown as
gray; greater than $1{-}\sigma$ are shown as black (approaching) or white
(receding).  For clarity only the group residual is shown for galaxies
in groups.
\label{fig:h0fit}}
}

\manufig{
\figcaption[f06.eps]{
Observed peculiar velocities are compared to model peculiar velocities.
Only galaxies with model velocity uncertainty less than 250\kms\
(essentially $[(H_0\,\delta r)^2+\delta v_{\rm virial}^2]^{1/2}$) are
shown, thereby excluding galaxies where a significant virial velocity is
expected and showing the quality of match between the model
and observed large scale flow field.
\label{fig:h0resi}}
}

\manufig{
\figcaption[f07.eps]{
Residual velocities after removal of a model
consisting of Hubble flow, Virgo Attractor, and constant velocity vector.
\label{fig:virgofit}}
}

\manufig{
\figcaption[f08.eps]{
Comparison between model and observed residual velocities
consisting of Hubble flow, Virgo Attractor, and constant velocity vector.
\label{fig:virgoresi}}
}

\manufig{
\figcaption[f09.eps]{
Residual velocities after removal of a model
consisting of Hubble flow, Virgo and Great Attractors,
and constant velocity vector.
\label{fig:gafit}}
}

\manufig{
\figcaption[f10.eps]{
Comparison of velocity residuals after removal of a model
consisting of Hubble flow, Virgo Attractor, GA,
and constant velocity vector.
\label{fig:garesi}}
}

\manufig{
\figcaption[f11.eps]{
Residual velocities after removal of a model
consisting of Hubble flow, Virgo and Great Attractors,
constant velocity vector, and quadrupole.
\label{fig:qfit}}
}

\manufig{
\figcaption[f12.eps]{
Comparison of velocity residuals after removal of a model
consisting of Hubble flow, Virgo and Great Attractors,
constant velocity vector, and quadrupole.
\label{fig:qresi}}
}

\manufig{
\figcaption{
Joint confidence contours of the fitted infall velocity,
distance, and location of the sky of the Virgo Attractor.
The points show the location of the galaxies in the RC3 with $v_{h} < 2800$.
\label{fig:locvir}}
}

\manufig{
\figcaption{
Joint confidence contours of the fitted infall velocity,
distance, and location of the sky of the GA attractor.
The points are from the SPS survey; Centaurus lies at $(156,-12)$,
Abell~1060 is the cluster at $(139,-37)$, and the Galactic plane is
evident on the left side.
\label{fig:locga}}
}

\manufig{
\figcaption{
Joint confidence contours of the SGX component $w_x$ of the dipole
velocity with GA infall amplitude (left) and $H_0$ (right).
\label{fig:wcov}}
}

\manufig{
\figcaption{
Joint confidence contours of the cutoff radius $r_{\rm cut}$ and power
law slope $\gamma$ (left) and infall amplitude (right) for
the Virgo Attractor.
\label{fig:rcutvir}}
}

\manufig{
\figcaption{
Joint confidence contours of the cutoff radius $r_{\rm cut}$ and power
law slope $\gamma$ (left) and infall amplitude (right) for
the GA attractor.
\label{fig:rcutga}}
}

\manufig{
\figcaption{
The net model velocity is shown as a function of position along vectors
running through the Virgo and Great Attractors using various values of
$\gamma$ and $r_{\rm cut}$ which leave $\cal N$ constant.  The dashed
line is just $H_0\,r$.  The points are those galaxies within
$25^\circ$ of the vectors ($|\cos\theta| > 0.9$).
\label{fig:flows}}
}

\manufig{
\figcaption{
The model infall velocity is shown as a function of distance from the
Virgo and Great Attractors for the four sets of $\gamma$ values.
A solid line delineates the range where there is substantial agreement
and where we believe the models are well constrained by the data.
\label{fig:infalls}}
}

\manufig{
\figcaption{
Contours are shown of our model of the radial component of the
flow field in a plane which cuts through the Local Group, Virgo,
and Great Attractors.  The model velocity of the Local Group with
respect to the CMB frame is apparent in the discontinuity at the
origin.  The Virgo Attractor is found at ``SGX,SGY'' = $(-3,+17)$
(quotes because the plane doesn't correspond perfectly to
supergalactic coordinates); the Great Attractor is at $(-42,+16)$,
and radial lines are drawn at $\pm10^\circ$ and $\pm20^\circ$ from it.
\label{fig:vcont}}
}

\manufig{
\figcaption{
This is the same figure as the previous one except that our survey
galaxies are overplotted.  Note that the Centaurus galaxies which
appear to be going through the Great Attractor actually pass above it
by about $15^\circ$, hence lie in the stall zone.
\label{fig:vcontgxy}}
}

\manufig{
\figcaption{
The CMB velocities of
galaxies from the RC3 are plotted as a function of angular separation
from the GA (left), and as a function of supergalactic longitude SGL
(right).  The concentration at $v_{\rm CMB} = 3150$
(left) and the trend for $v_{\rm CMB}$ to change from 1100 to
1500\kms\ (right) inherent in our model is apparent in these redshift
data.
\label{fig:gauma}}
}

\manufig{
\figcaption{
Joint confidence contours of the Hubble constant and relative
quadrupole amplitude when the quadrupole is centered on the Local
Group are shown on the left. Confidence contours of Hubble constant
and quadrupole cutoff radius when the quadrupole is centered on Virgo
instead are displayed on the right.
\label{fig:hqsig}}
}

\begin{figure}[H]
\plotone{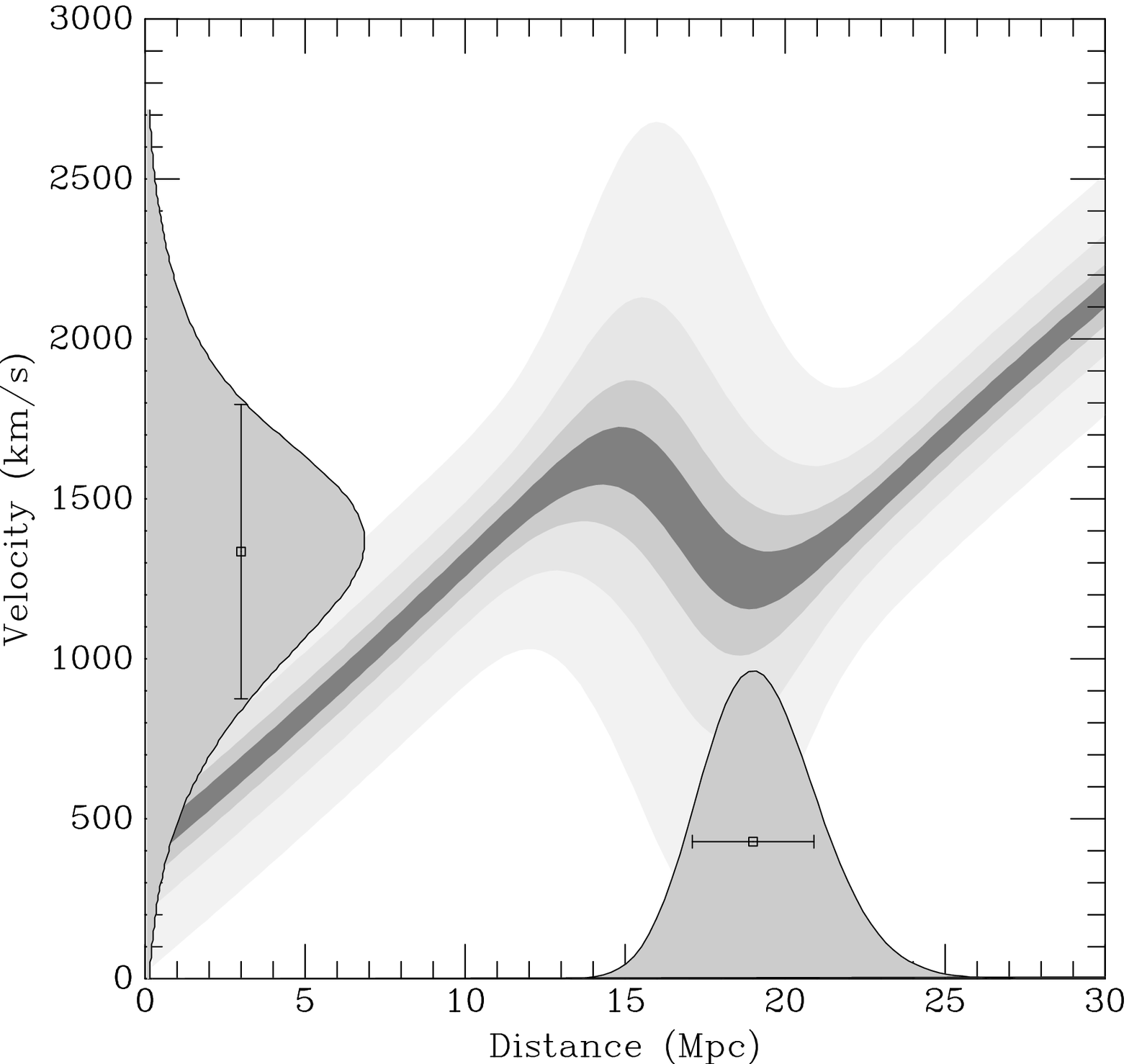}
\capfig{
\caption[f01.eps]{
The model velocity distribution function $P(v|r)$ is shown for a line
of sight which passes through the Virgo cluster.  The different
grayscale levels show the 2, 1, 0.5, and 0.2 sigma points on the
velocity distribution at a given distance.  A distance
observation is shown on the abscissa as a distance probability
function, and the resulting velocity probability function is shown on
the ordinate.  This is evaluated at the observed velocity and forms a
term in the likelihood product.
\label{fig:probs}}
}
\end{figure}

\begin{figure}[H]
\plotone{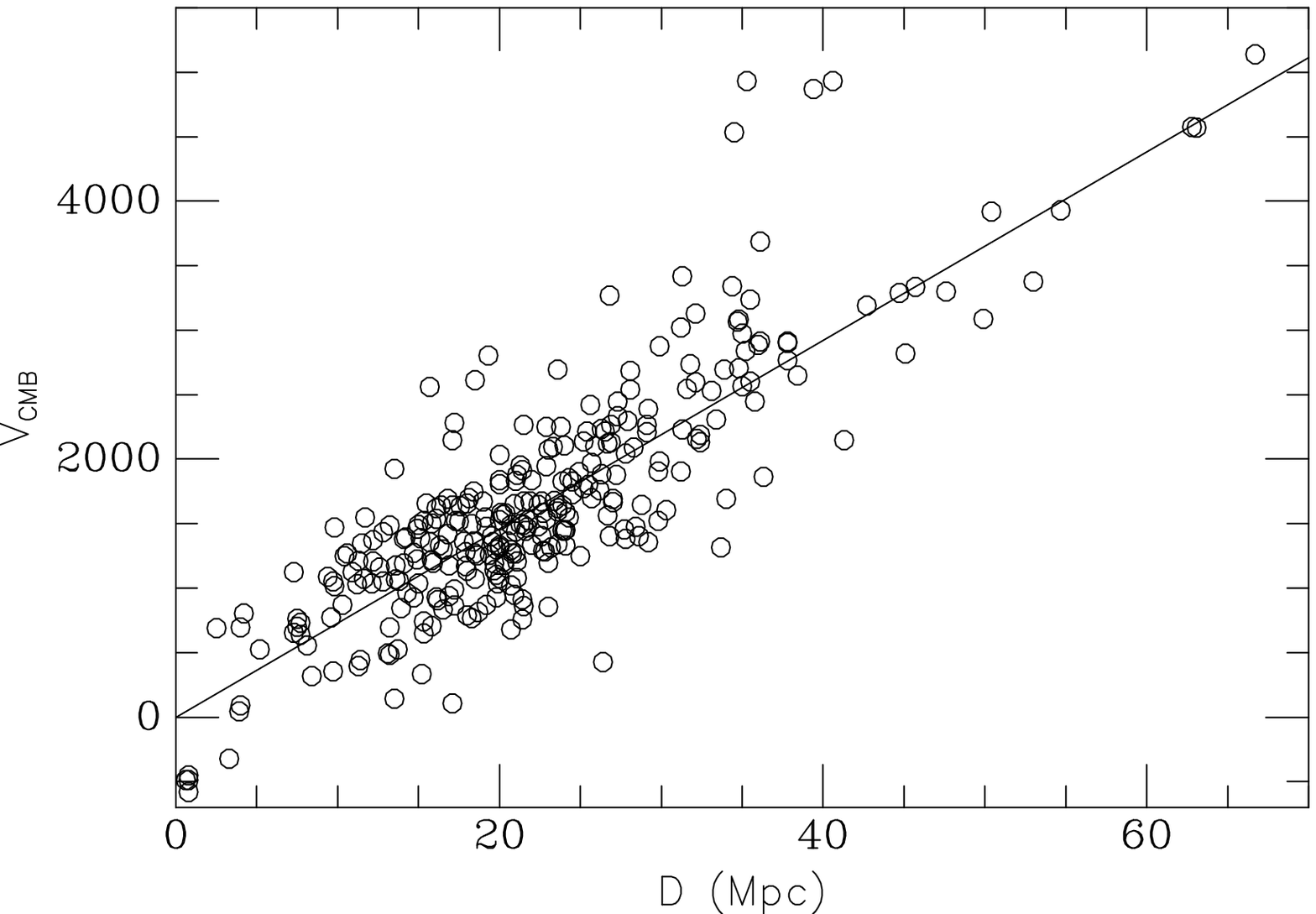}
\capfig{
\caption[f02.eps]{
Recession velocities in the CMB frame are plotted as a function of distance.
The line has a slope of 73\kmsMpc; 
the very deviant high points near 40~Mpc are Cen-45 galaxies.
\label{fig:hubble}}
}
\end{figure}

\begin{figure}[H]
\plotone{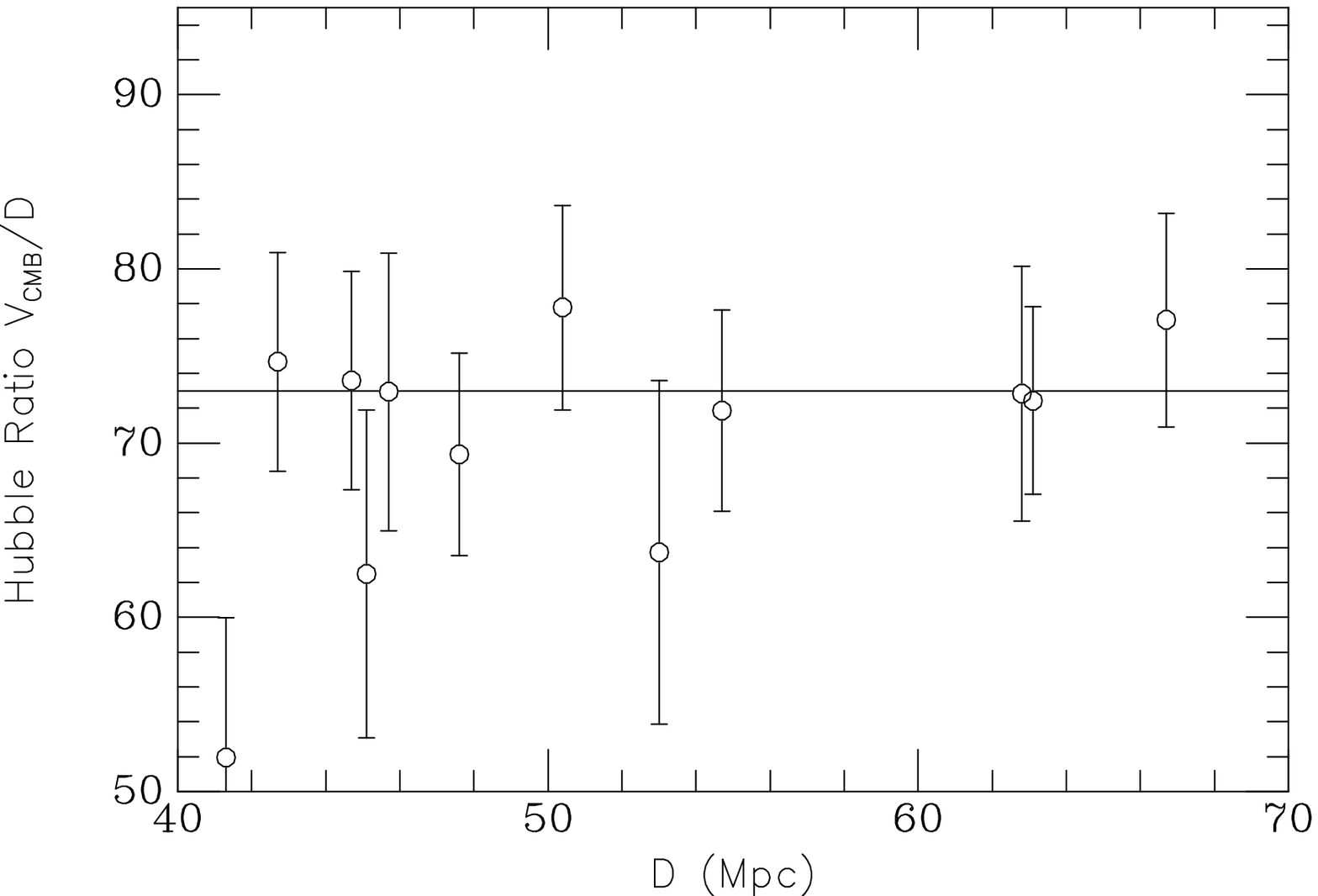}
\capfig{
\caption[f03.eps]{
Hubble ratios (in CMB frame) are plotted as a function of distance for
those points with small enough errors that the error in the Hubble
ratio is less than 10~\kmsMpc.
\label{fig:h0}}
}
\end{figure}

\clearpage

\begin{figure}[H]
\epsscale{0.8}
\plotone{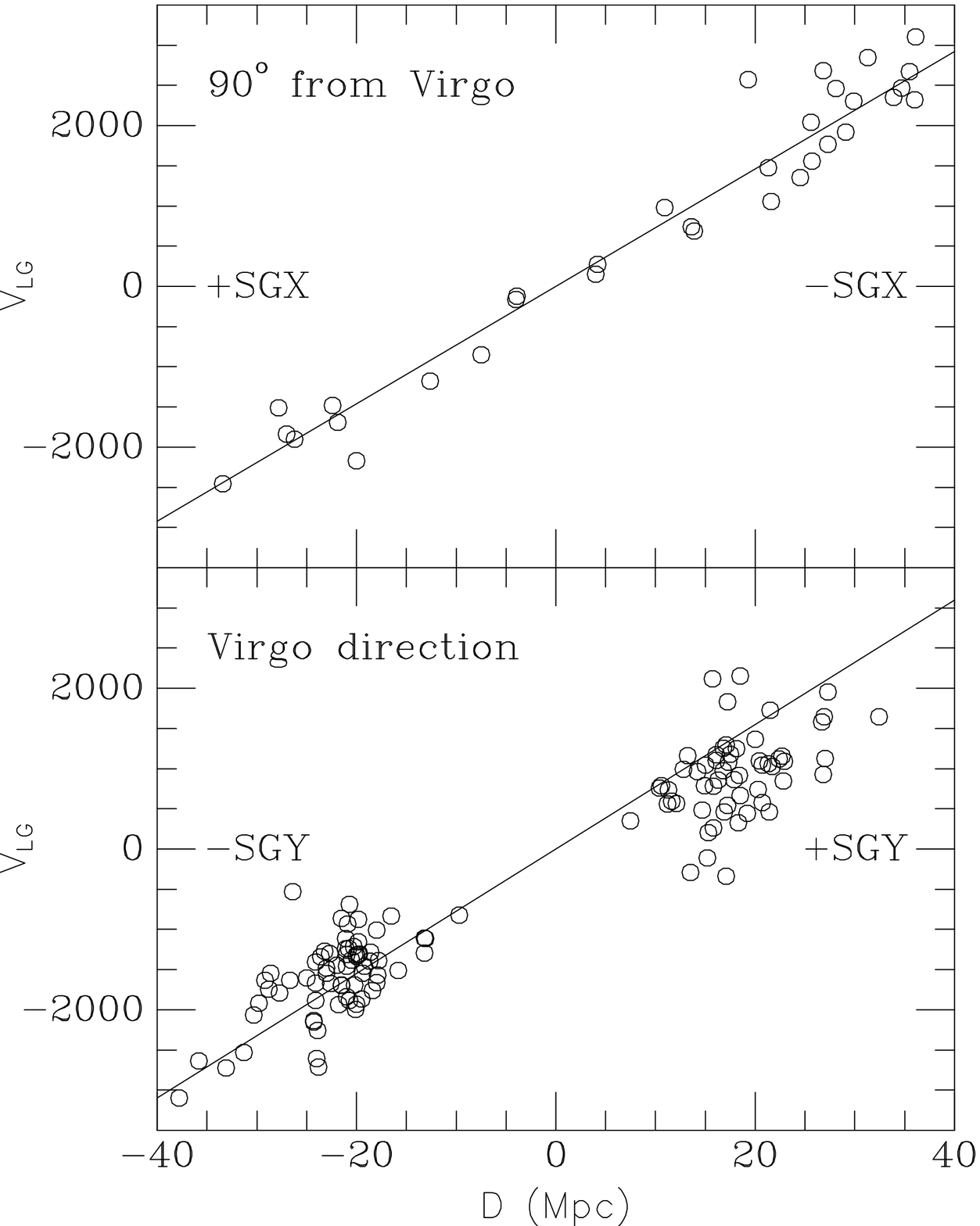}
\capfig{
\caption[f04.eps]{
Recession velocities (in Local Group frame) are plotted as a function of
distance in two directions.  The upper panel shows galaxies which lie
approximately in a $30^\circ$ (half angle) cone in the $\pm$SGX direction,
$90^\circ$ away from the direction of Virgo.  The lower panel shows
galaxies which lie within a $15^\circ$ cone towards Virgo (+SGY) and a
$45^\circ$ cone away from Virgo (which includes the Fornax cluster).  A
Hubble ratio of 73\kmsMpc\ is also drawn.
\label{fig:aniso}}
}
\end{figure}

\begin{figure}[H]
\plotone{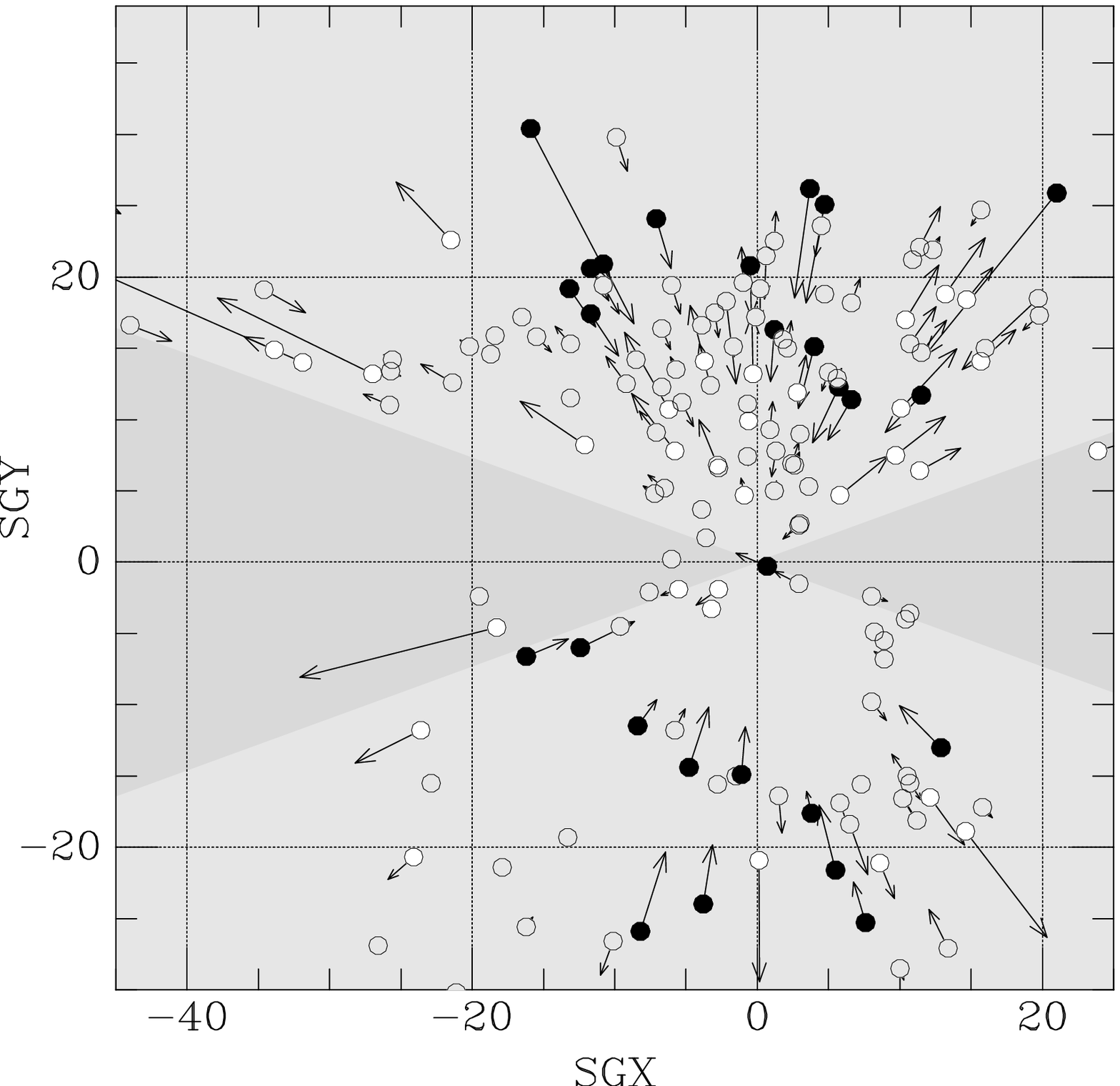}
\capfig{
\caption[f05.eps]{
Residual velocities after removal of a model Hubble
flow.  Residual velocities which are less than $1{-}\sigma$ are shown as
gray; greater than $1{-}\sigma$ are shown as black (approaching) or white
(receding).  For clarity only the group residual is shown for galaxies
in groups.
\label{fig:h0fit}}
}
\end{figure}

\begin{figure}[H]
\plotone{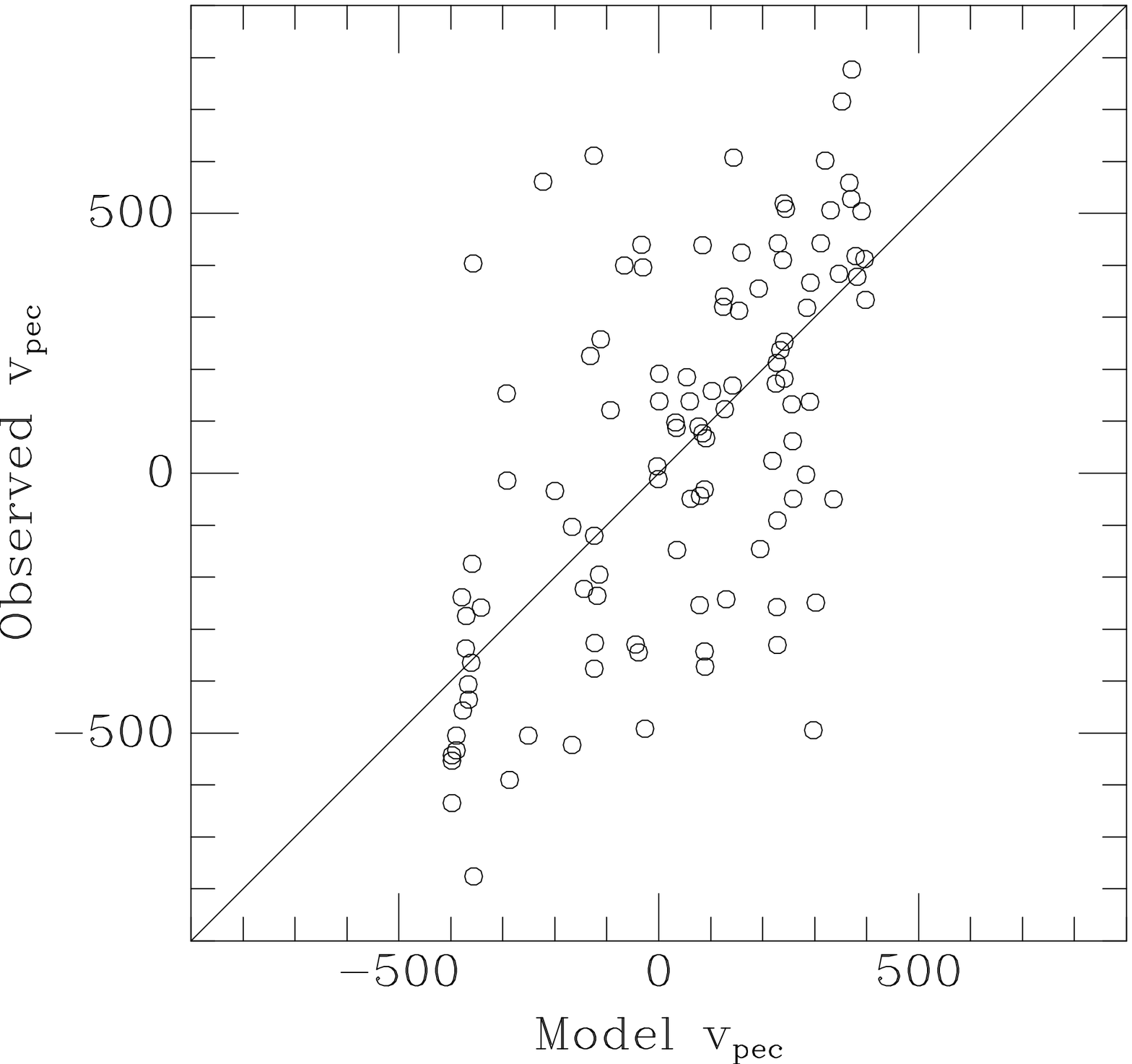}
\capfig{
\caption[f06.eps]{
Observed peculiar velocities are compared to model peculiar velocities.
Only galaxies with model velocity uncertainty less than 250\kms\
(essentially $[(H_0\,\delta r)^2+\delta v_{\rm virial}^2]^{1/2}$) are
shown, thereby excluding galaxies where a significant virial velocity is
expected and showing the quality of match between the model
and observed large scale flow field.
\label{fig:h0resi}}
}
\end{figure}

\begin{figure}[H]
\plotone{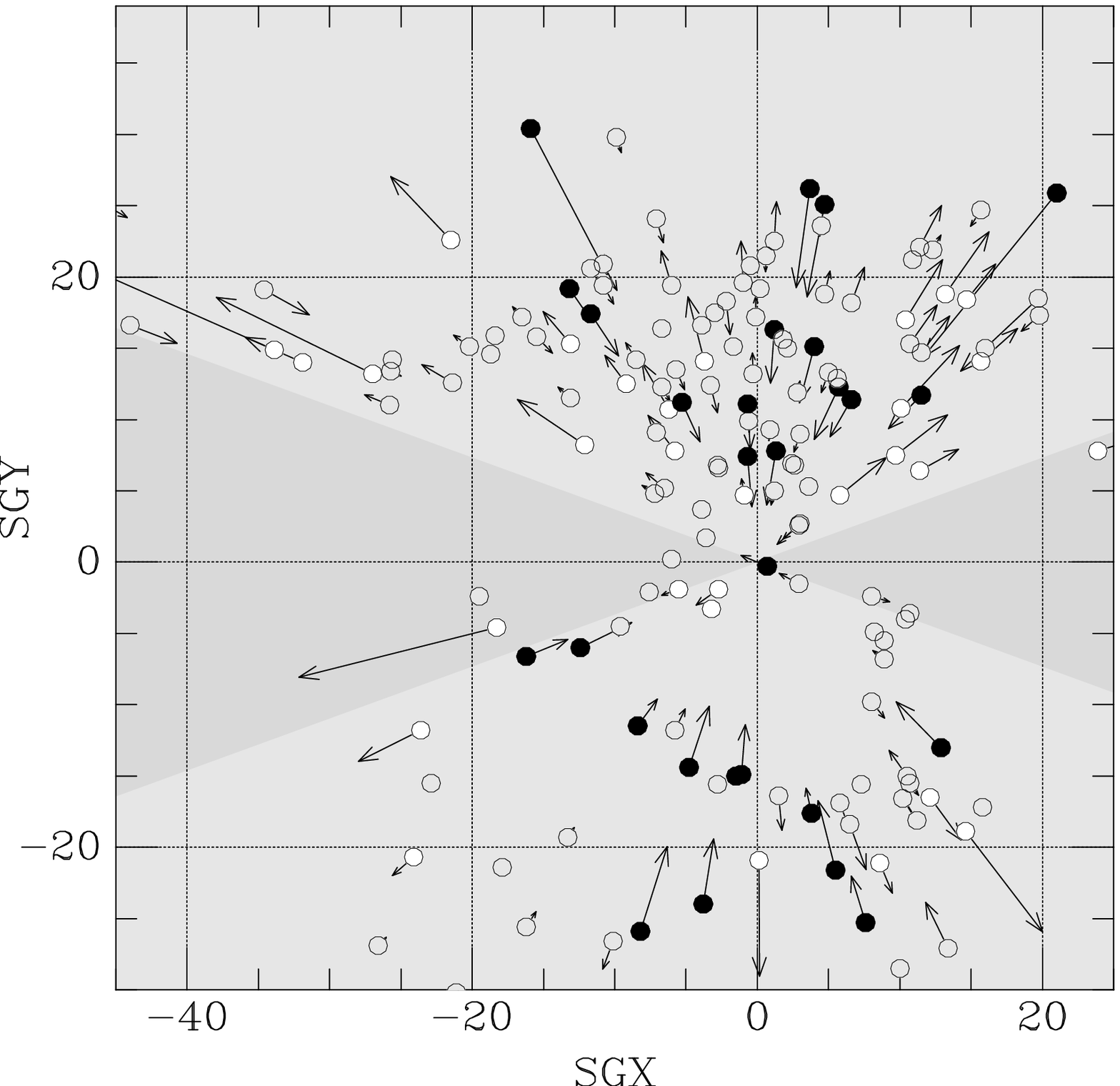}
\capfig{
\caption[f07.eps]{
Residual velocities after removal of a model
consisting of Hubble flow, Virgo Attractor, and constant velocity vector.
\label{fig:virgofit}}
}
\end{figure}

\begin{figure}[H]
\plotone{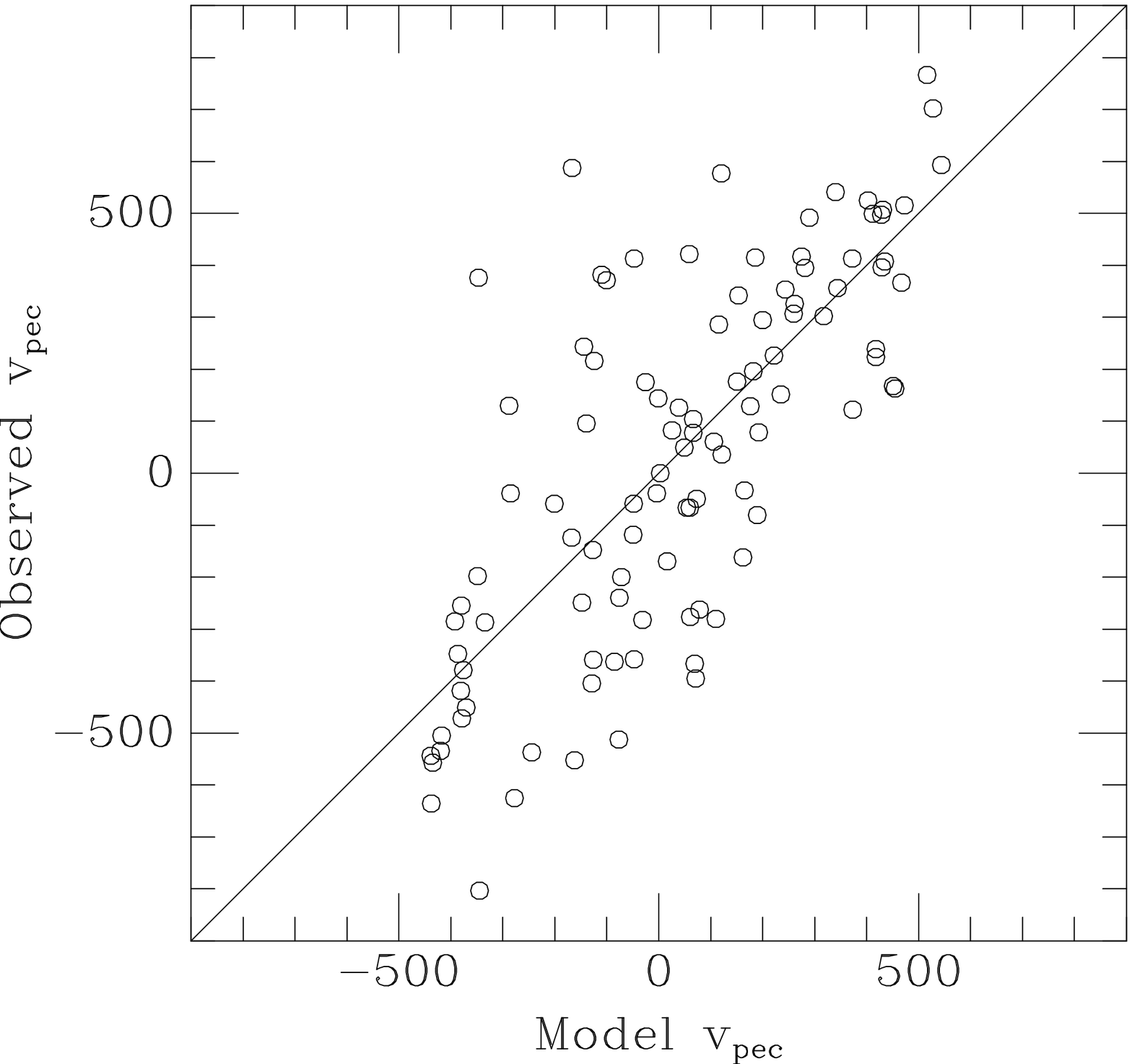}
\capfig{
\caption[f08.eps]{
Comparison between model and observed residual velocities
consisting of Hubble flow, Virgo Attractor, and constant velocity vector.
\label{fig:virgoresi}}
}
\end{figure}

\begin{figure}[H]
\plotone{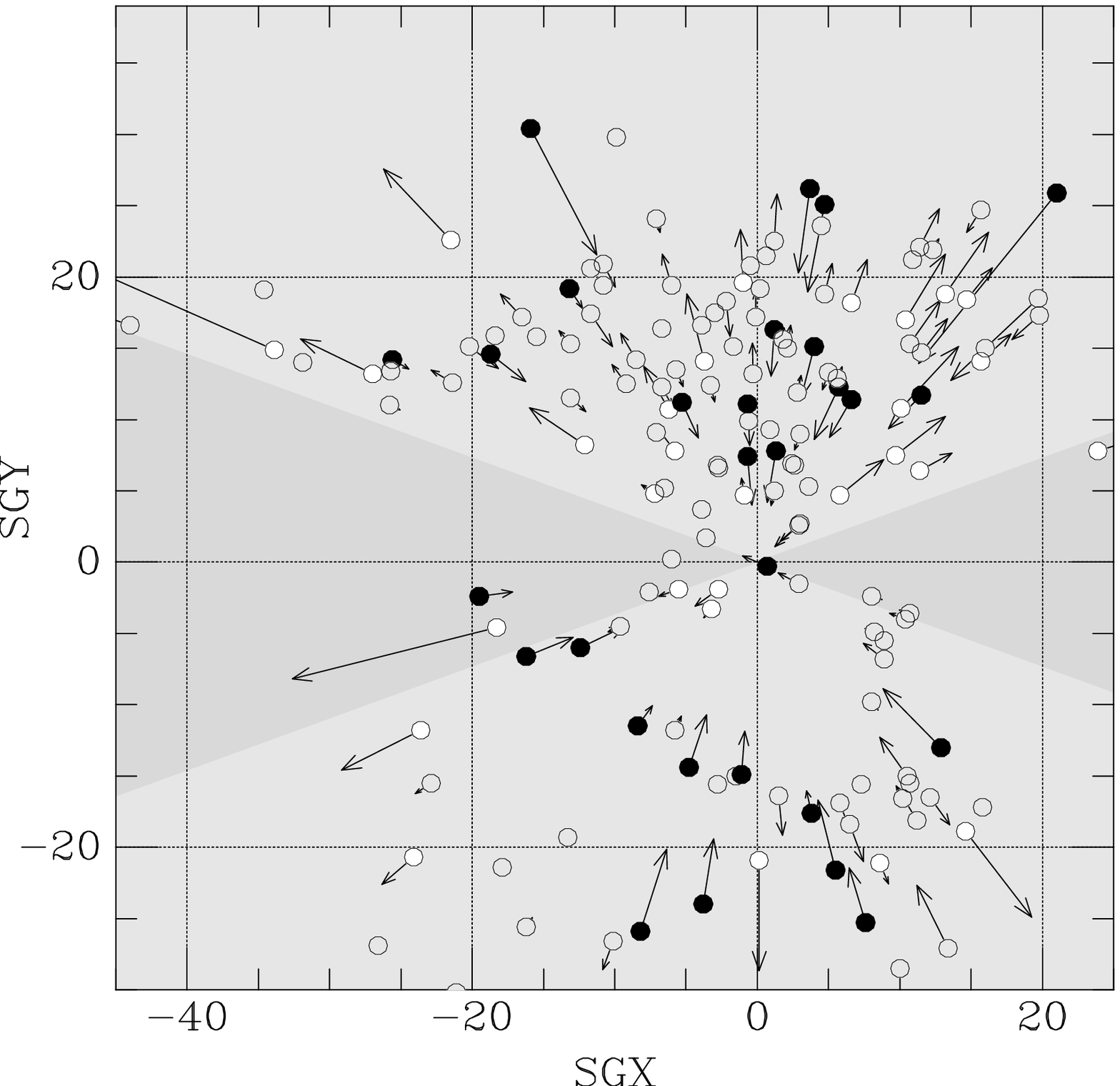}
\capfig{
\caption[f09.eps]{
Residual velocities after removal of a model
consisting of Hubble flow, Virgo and Great Attractors,
and constant velocity vector.
\label{fig:gafit}}
}
\end{figure}

\begin{figure}[H]
\plotone{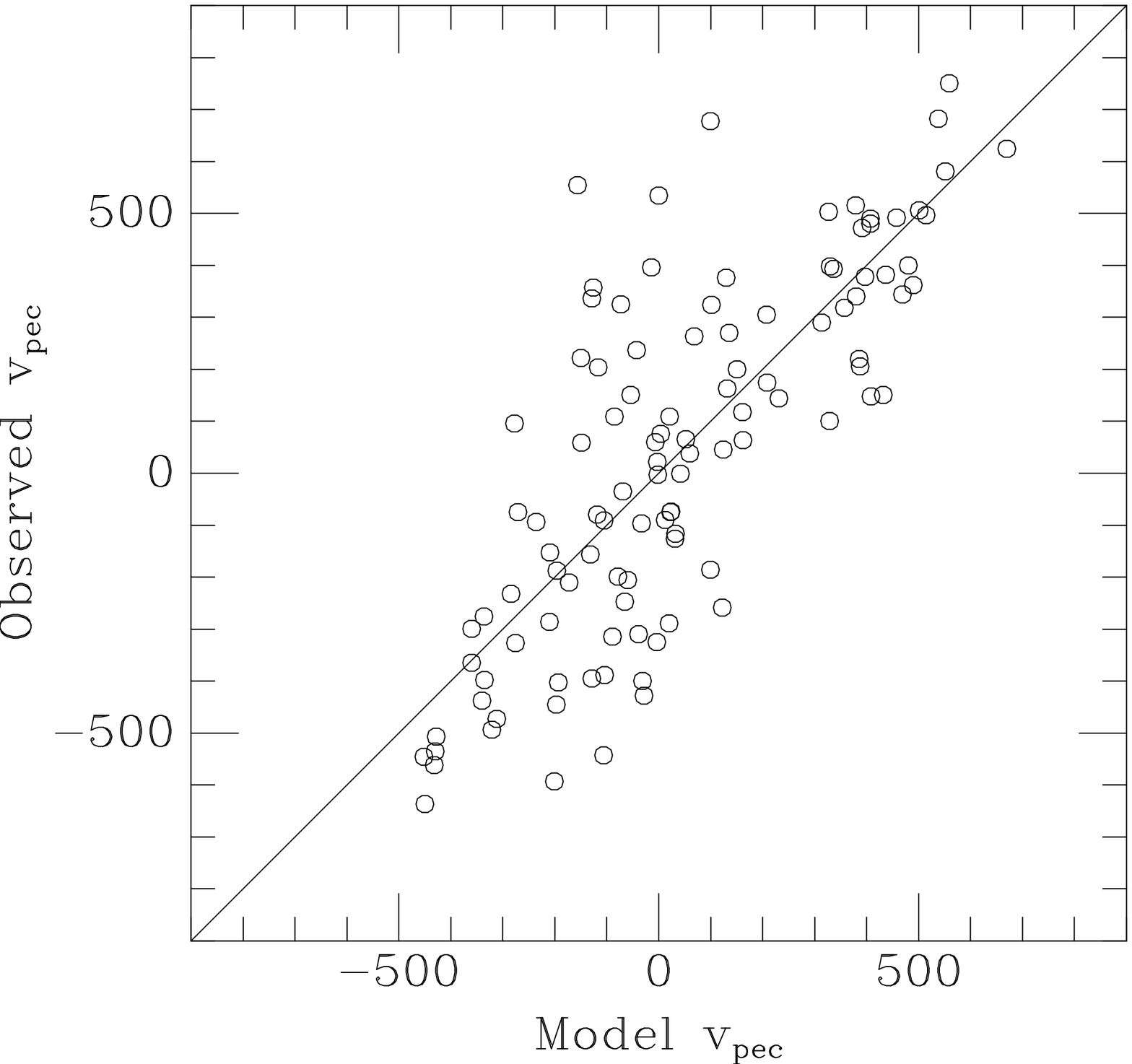}
\capfig{
\caption[f10.eps]{
Comparison of velocity residuals after removal of a model
consisting of Hubble flow, Virgo Attractor, GA,
and constant velocity vector.
\label{fig:garesi}}
}
\end{figure}

\begin{figure}[H]
\plotone{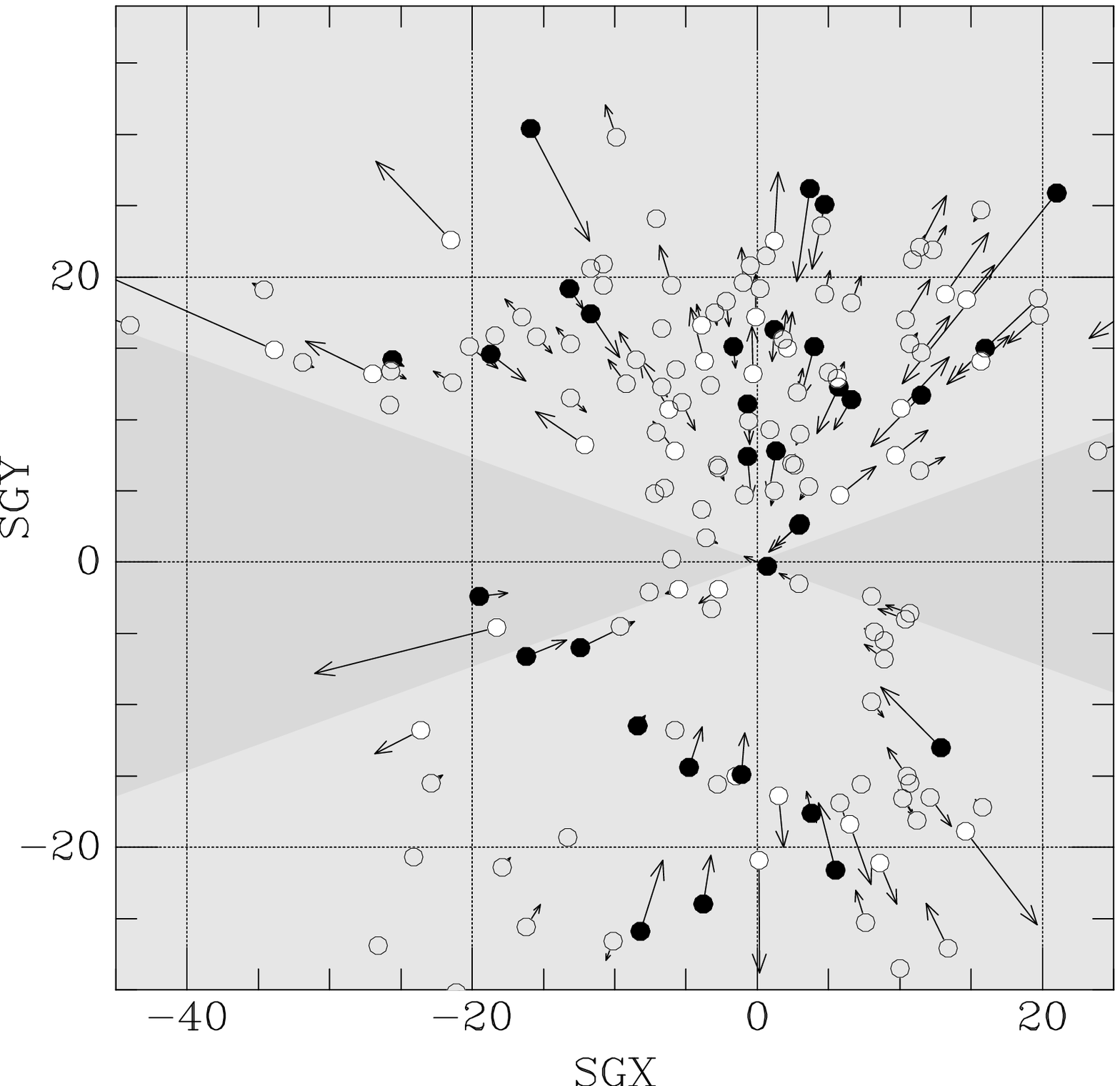}
\capfig{
\caption[f11.eps]{
Residual velocities after removal of a model
consisting of Hubble flow, Virgo and Great Attractors,
constant velocity vector, and quadrupole.
\label{fig:qfit}}
}
\end{figure}

\begin{figure}[H]
\plotone{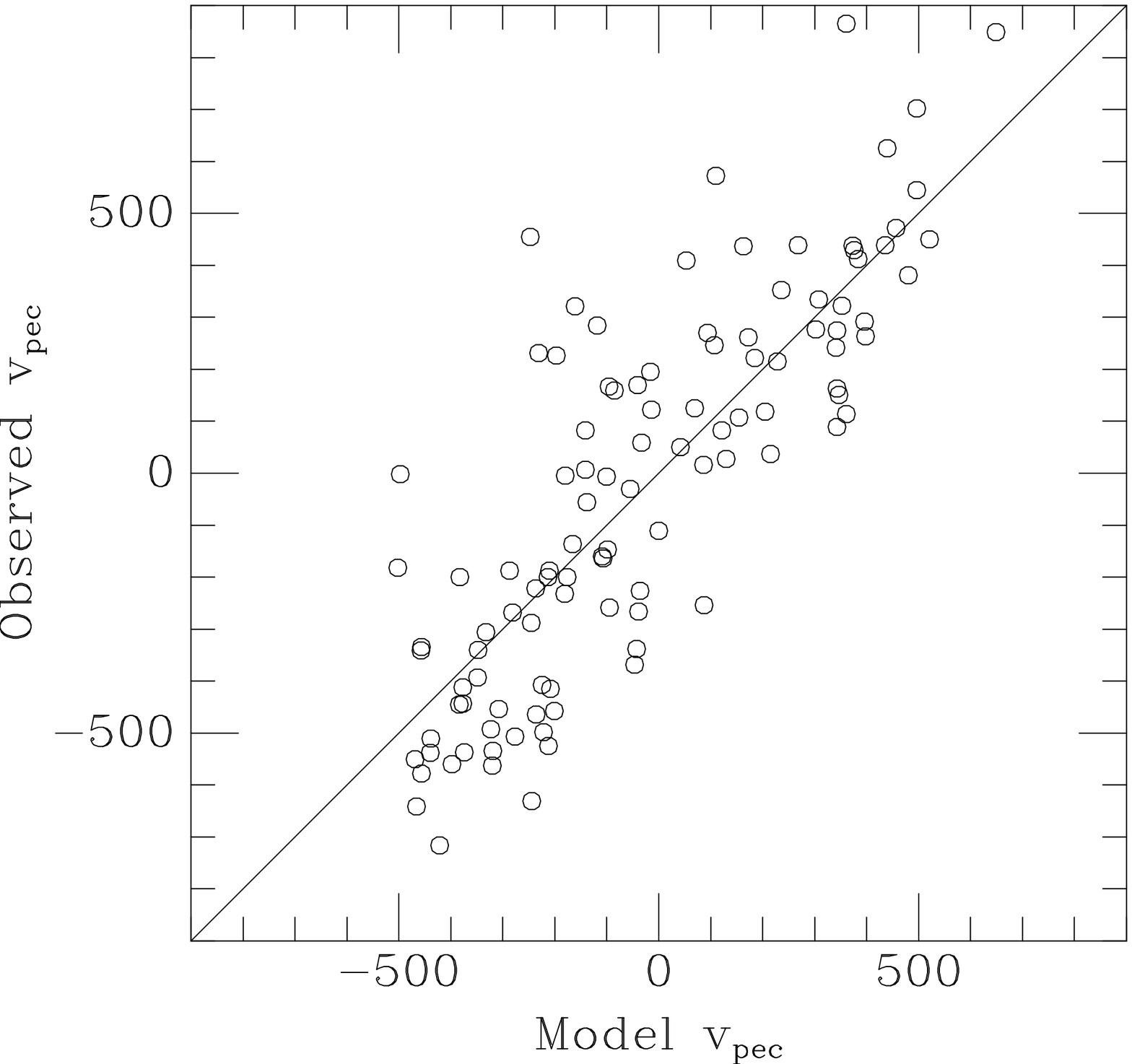}
\capfig{
\caption[f12.eps]{
Comparison of velocity residuals after removal of a model
consisting of Hubble flow, Virgo and Great Attractors,
constant velocity vector, and quadrupole.
\label{fig:qresi}}
}
\end{figure}

\begin{figure}[H]
\plottwo{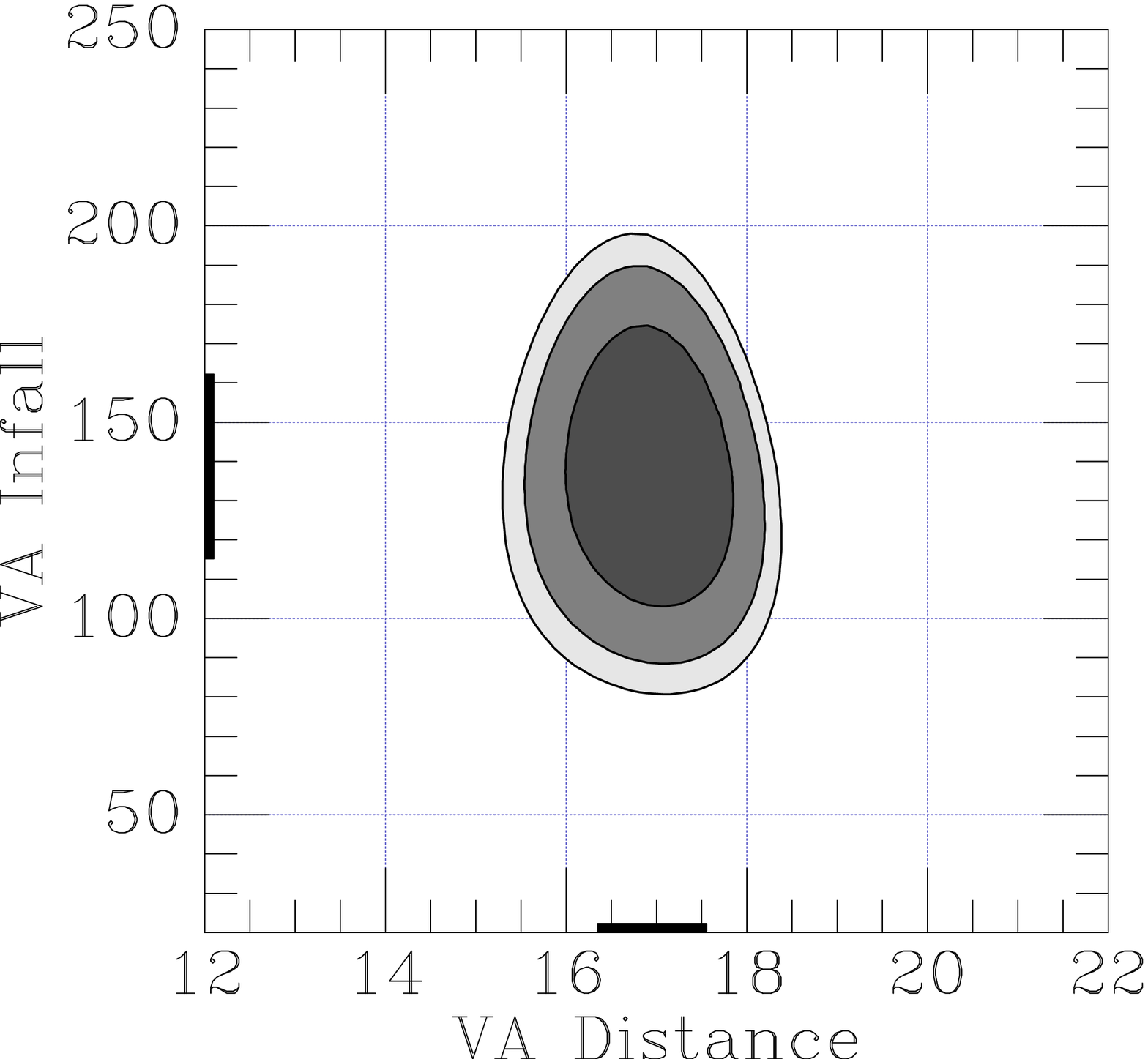}{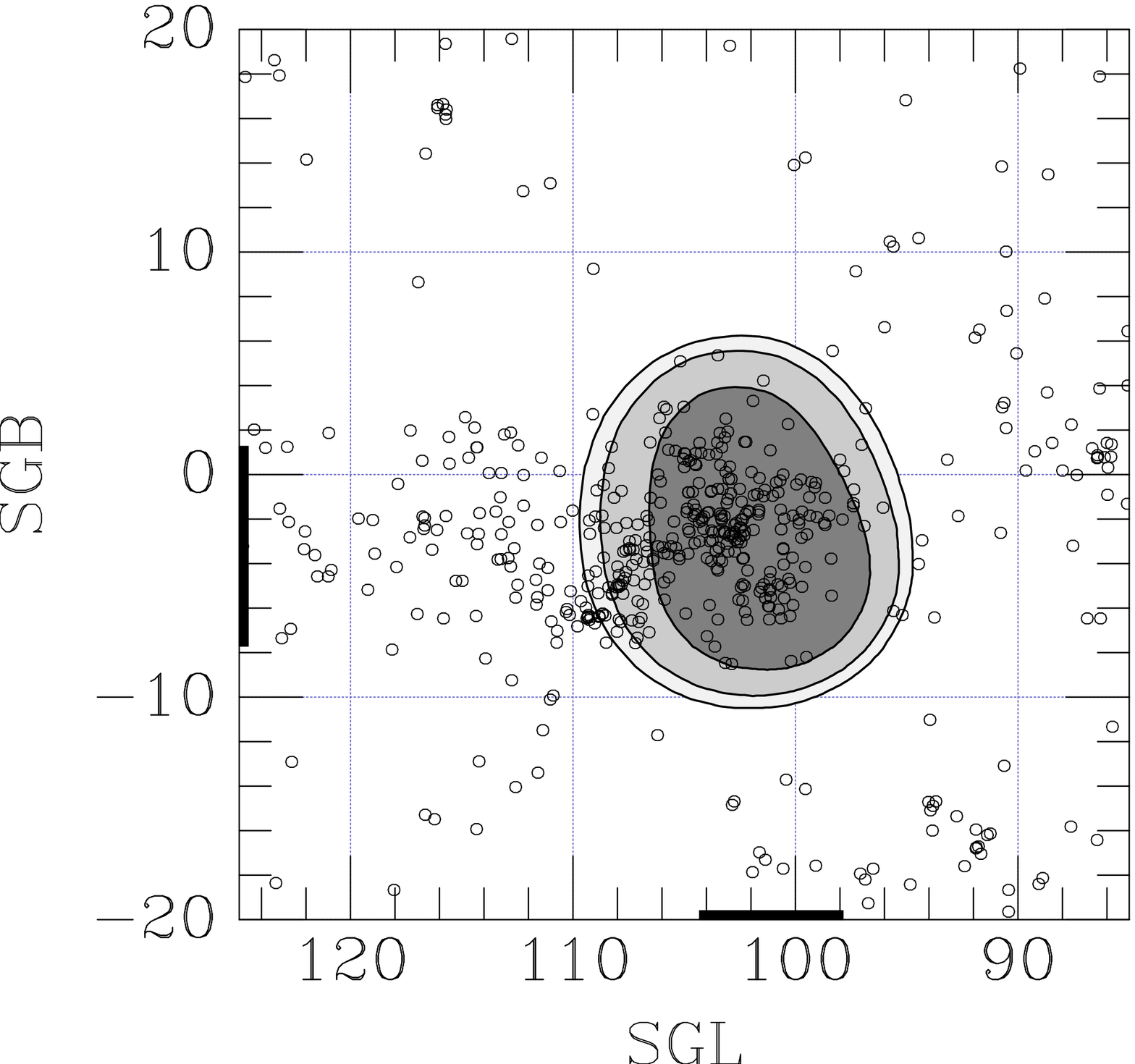}
\capfig{
\caption{
Joint confidence contours of the fitted infall velocity,
distance, and location of the sky of the Virgo Attractor.
The points show the location of the galaxies in the RC3 with $v_{h} < 2800$.
\label{fig:locvir}}
}
\end{figure}

\begin{figure}[H]
\plottwo{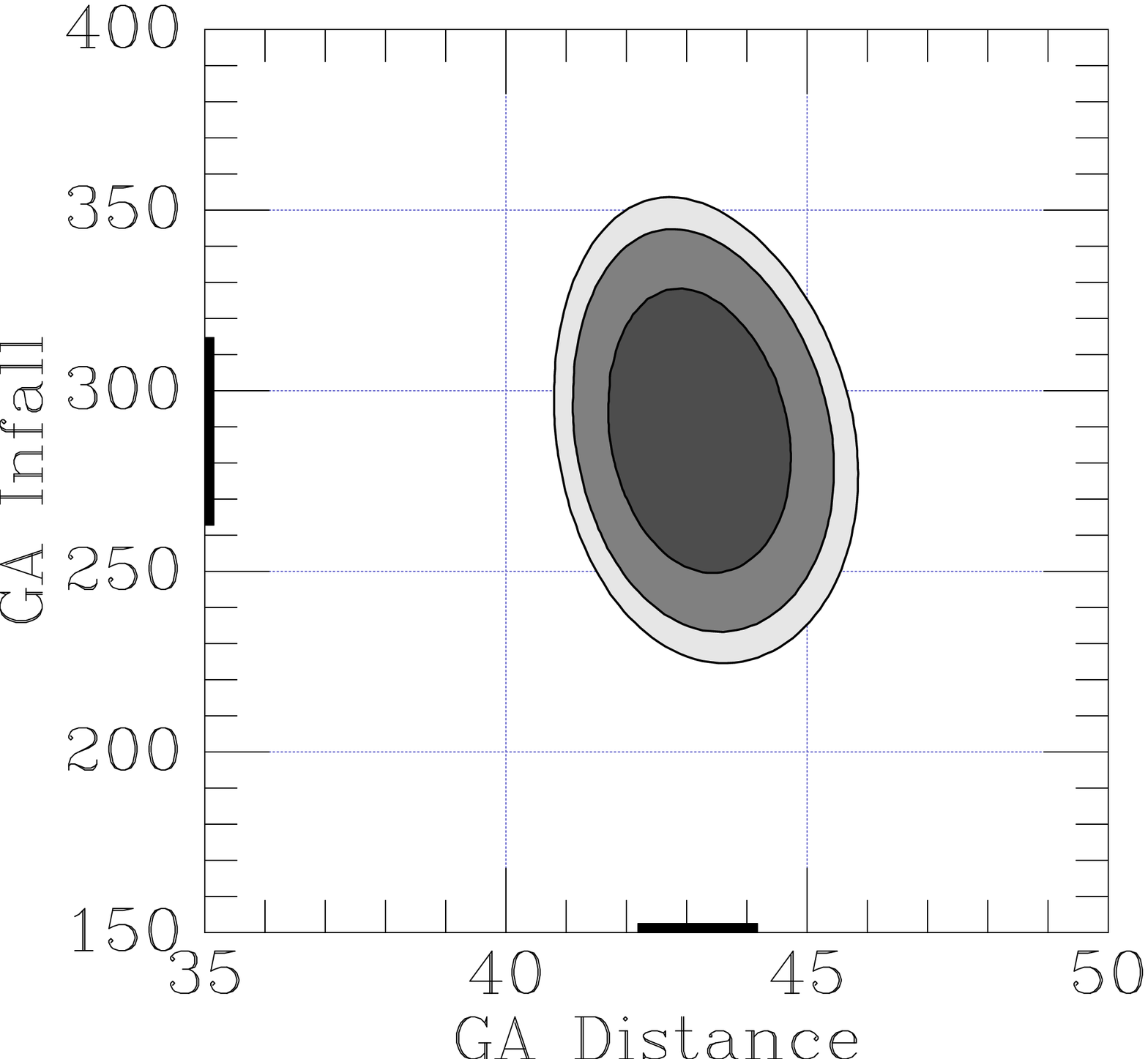}{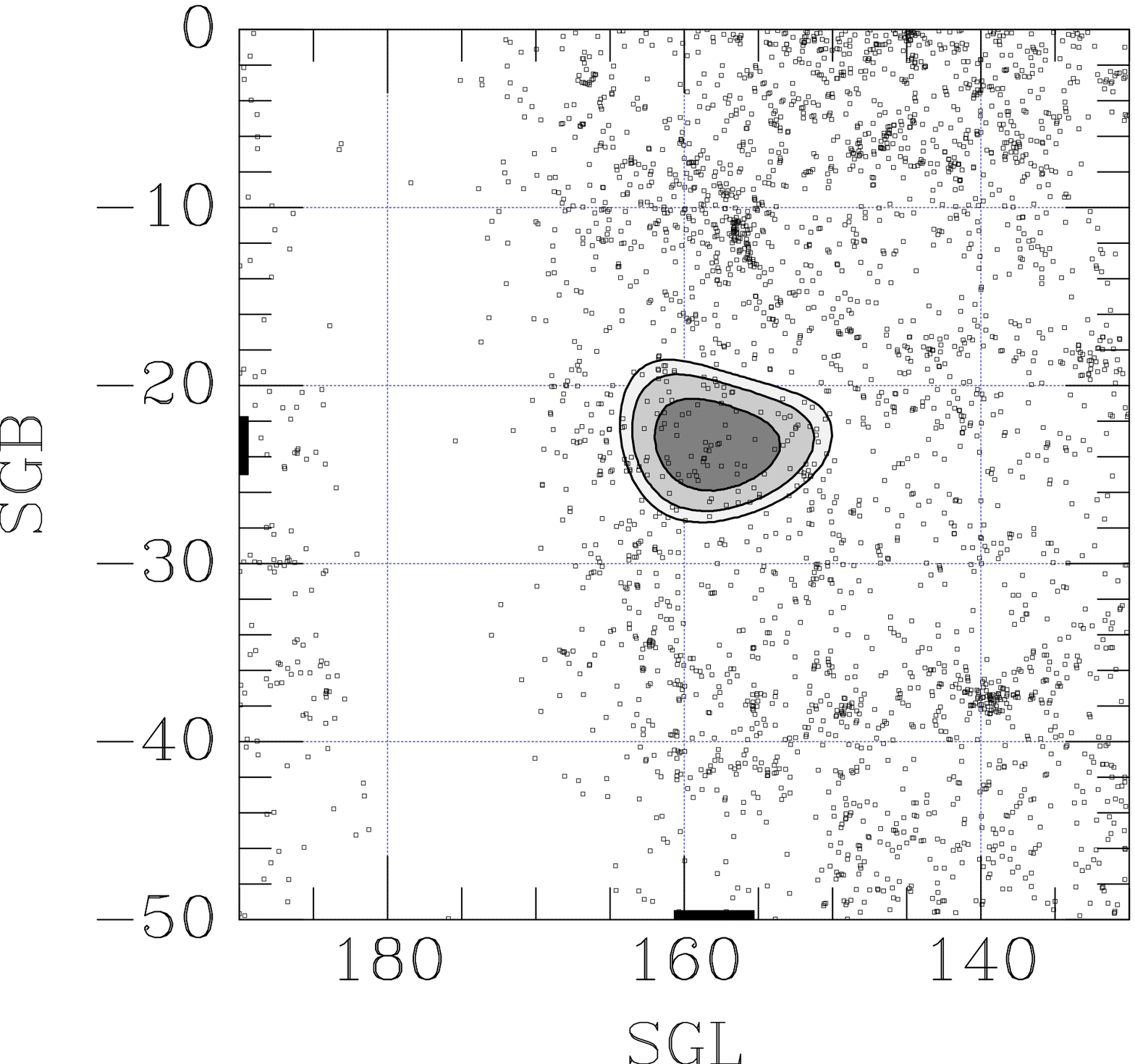}
\capfig{
\caption{
Joint confidence contours of the fitted infall velocity,
distance, and location of the sky of the GA attractor.
The points are from the SPS survey; Centaurus lies at $(156,-12)$,
Abell~1060 is the cluster at $(139,-37)$, and the Galactic plane is
evident on the left side.
\label{fig:locga}}
}
\end{figure}

\begin{figure}[H]
\plottwo{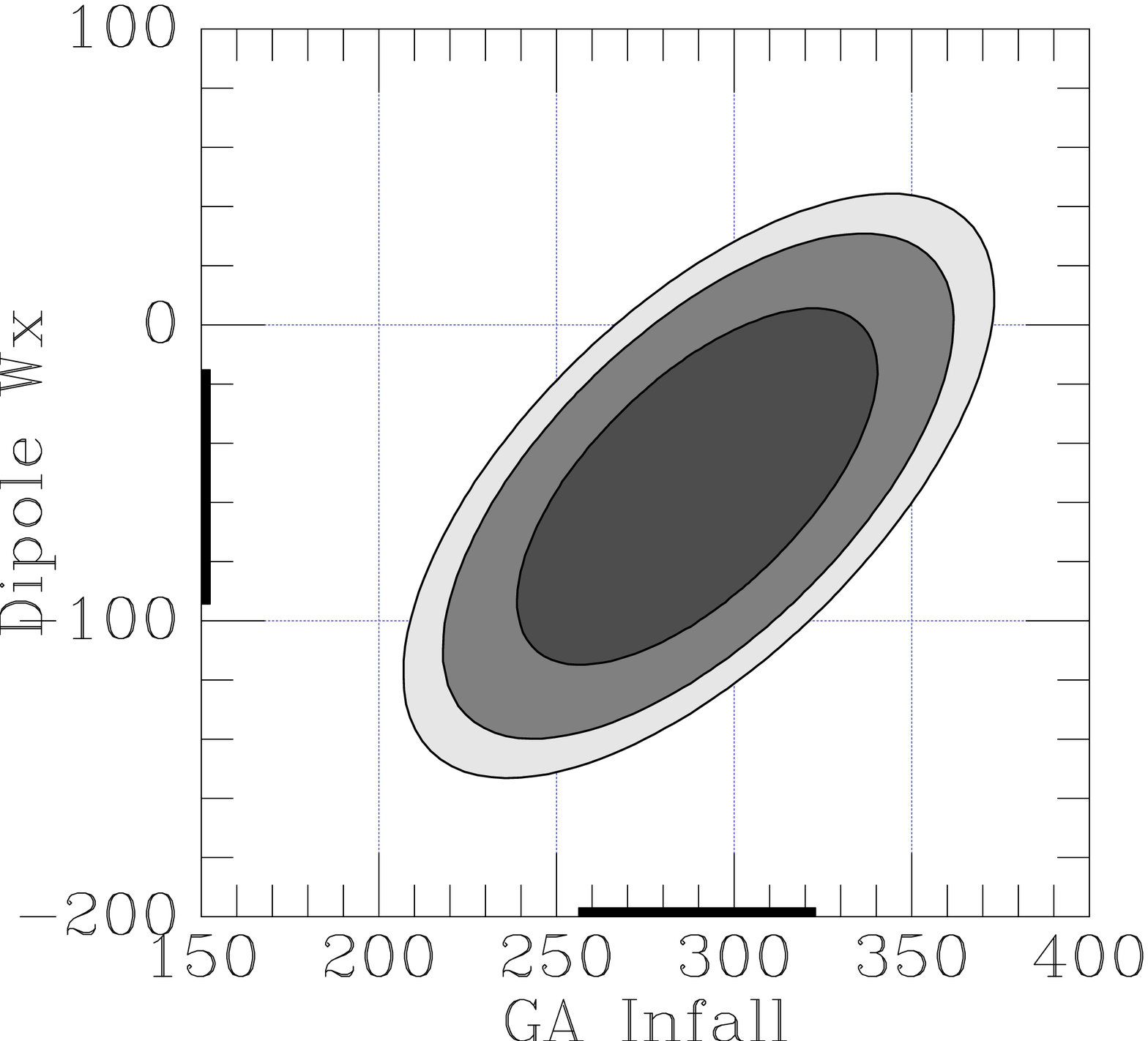}{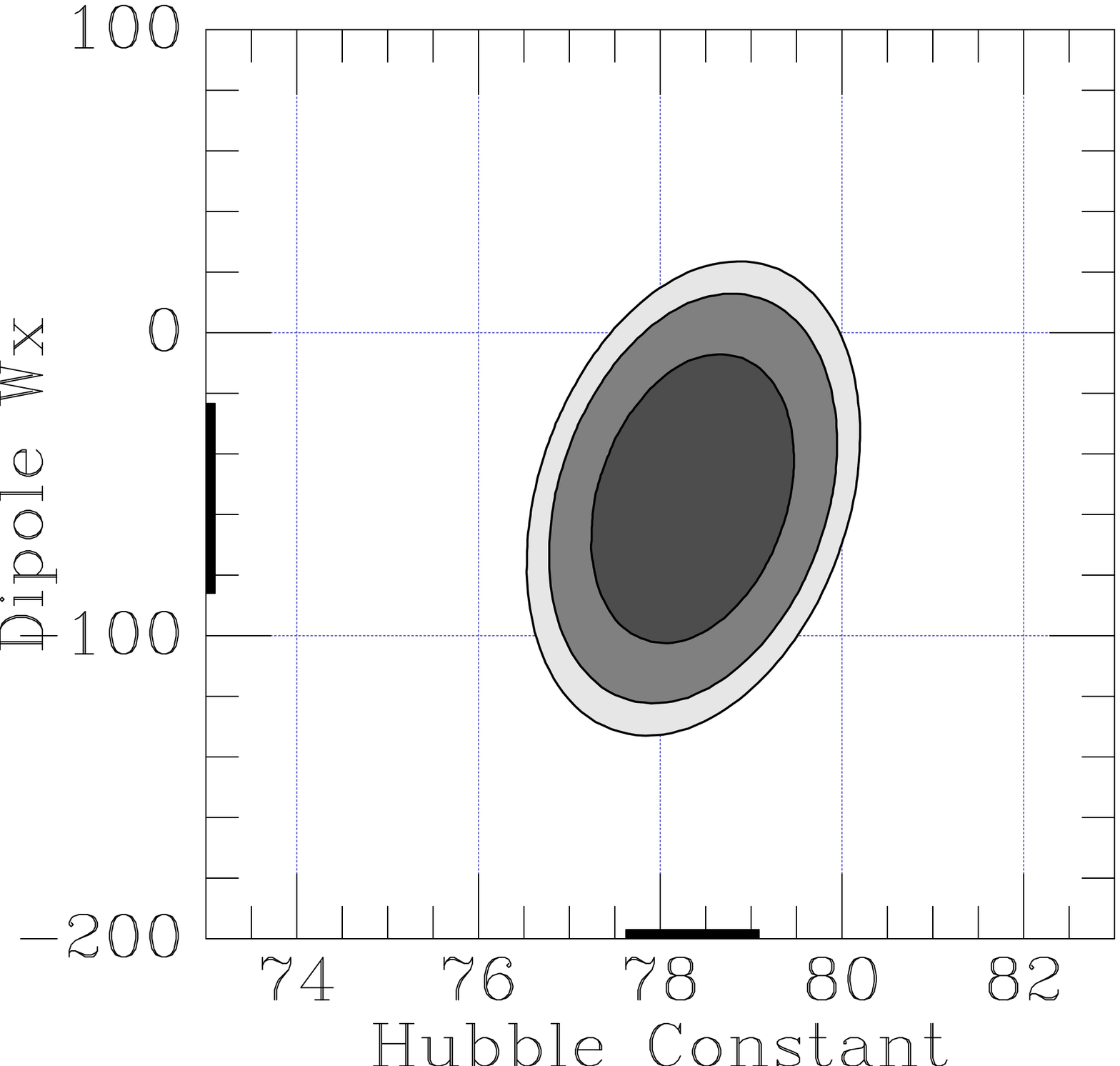}
\capfig{
\caption{
Joint confidence contours of the SGX component $w_x$ of the dipole
velocity with GA infall amplitude (left) and $H_0$ (right).
\label{fig:wcov}}
}
\end{figure}

\clearpage

\begin{figure}[H]
\plottwo{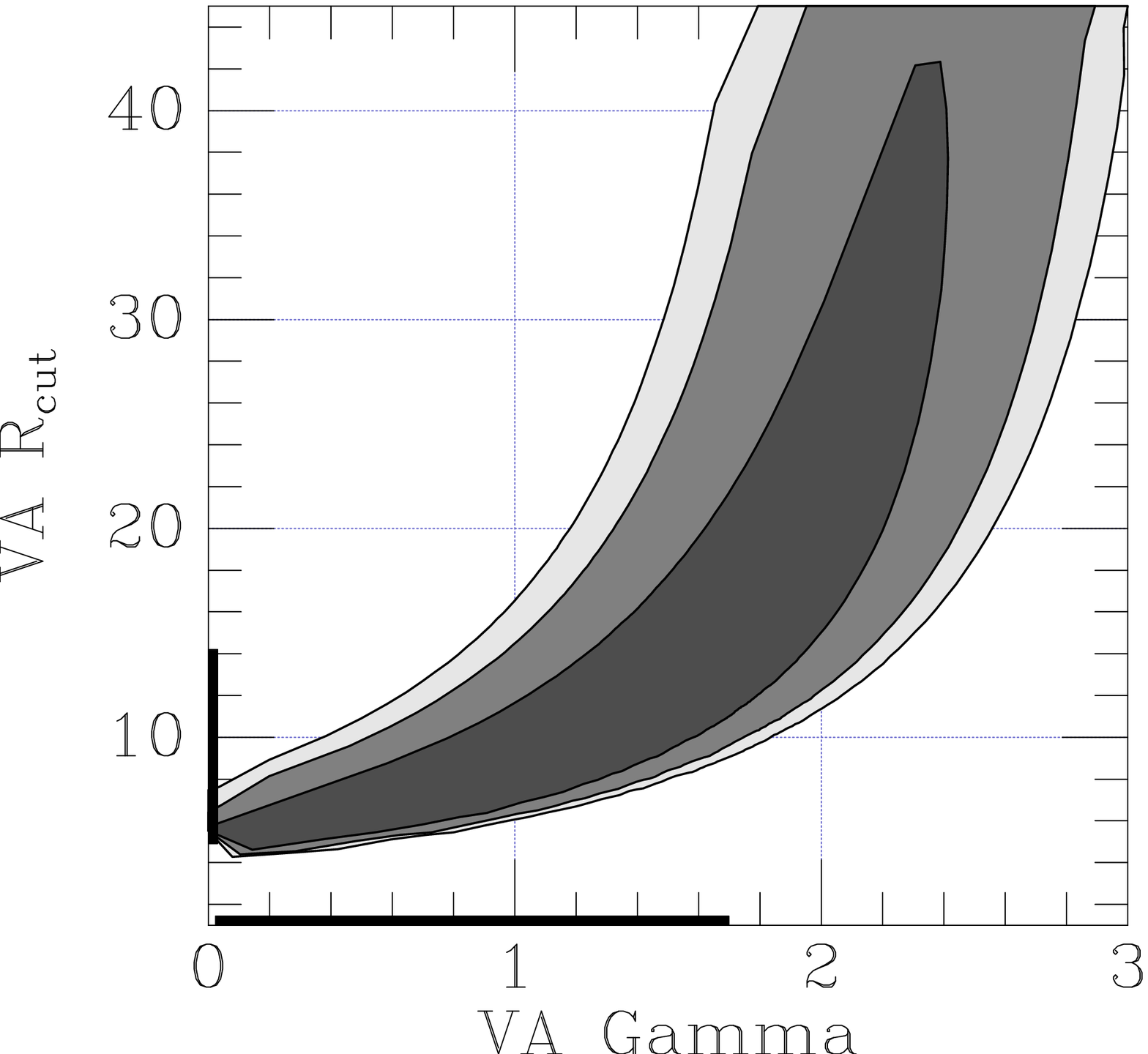}{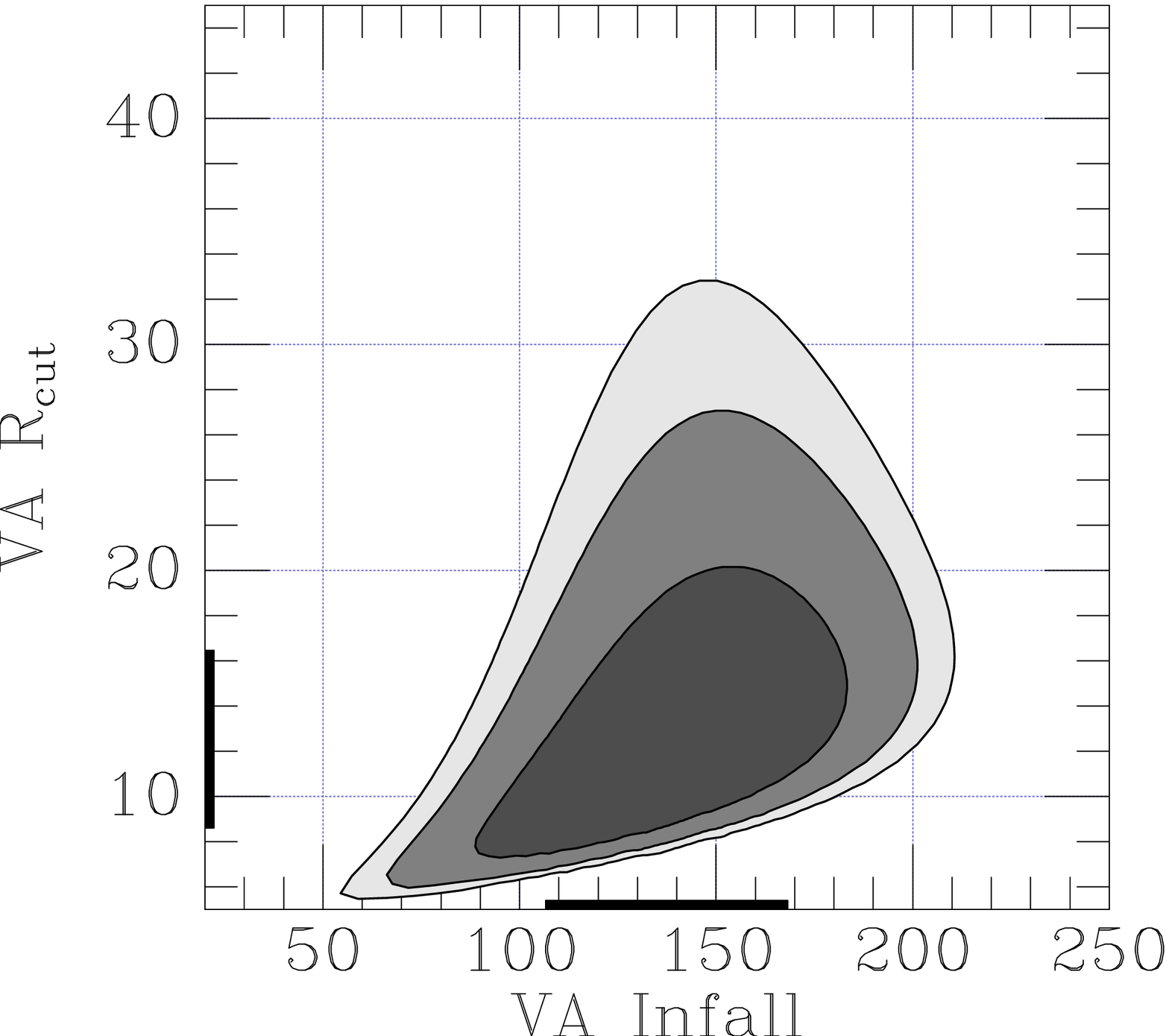}
\capfig{
\caption{
Joint confidence contours of the cutoff radius $r_{\rm cut}$ and power
law slope $\gamma$ (left) and infall amplitude (right) for
the Virgo Attractor.
\label{fig:rcutvir}}
}
\end{figure}

\begin{figure}[H]
\plottwo{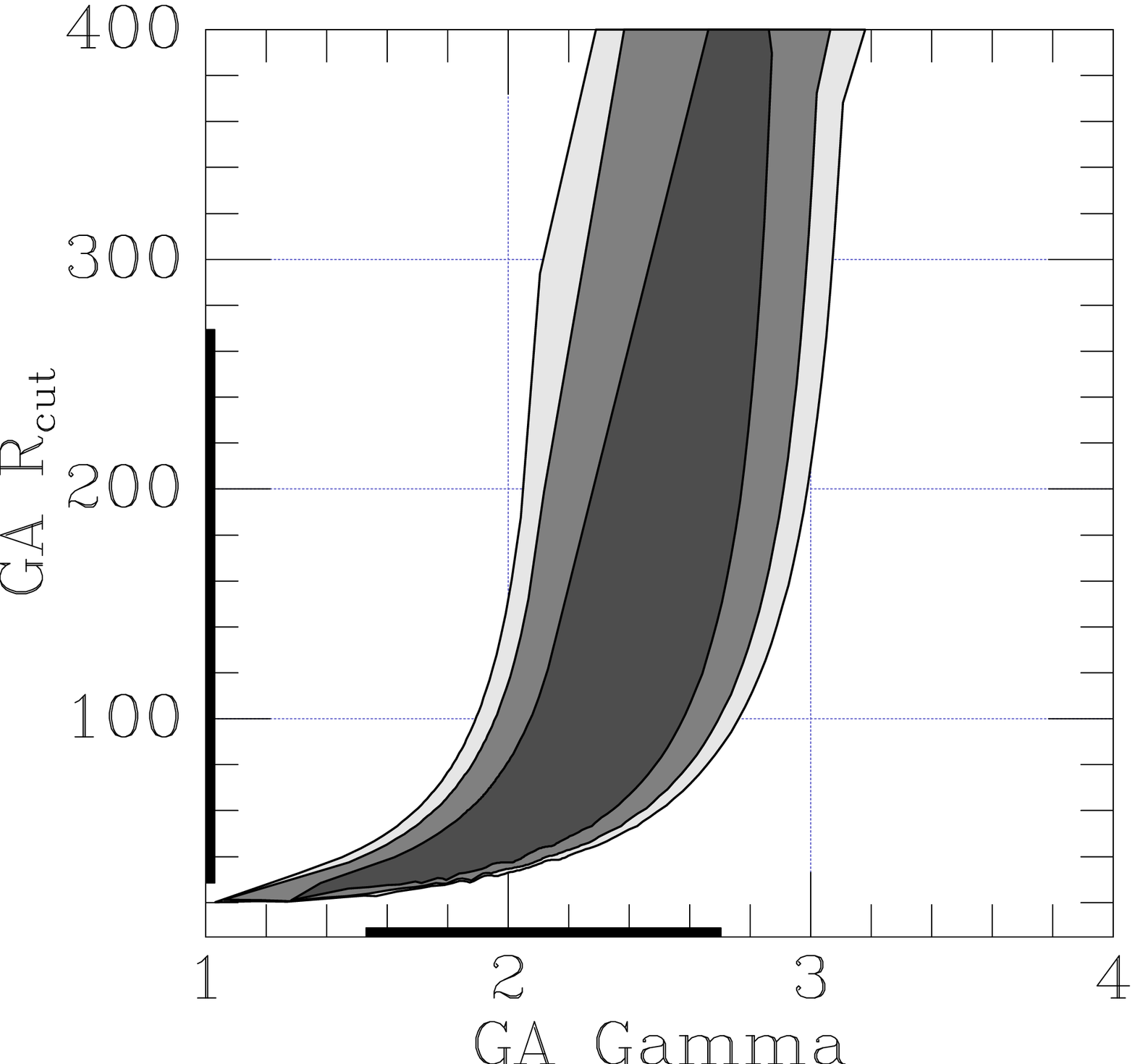}{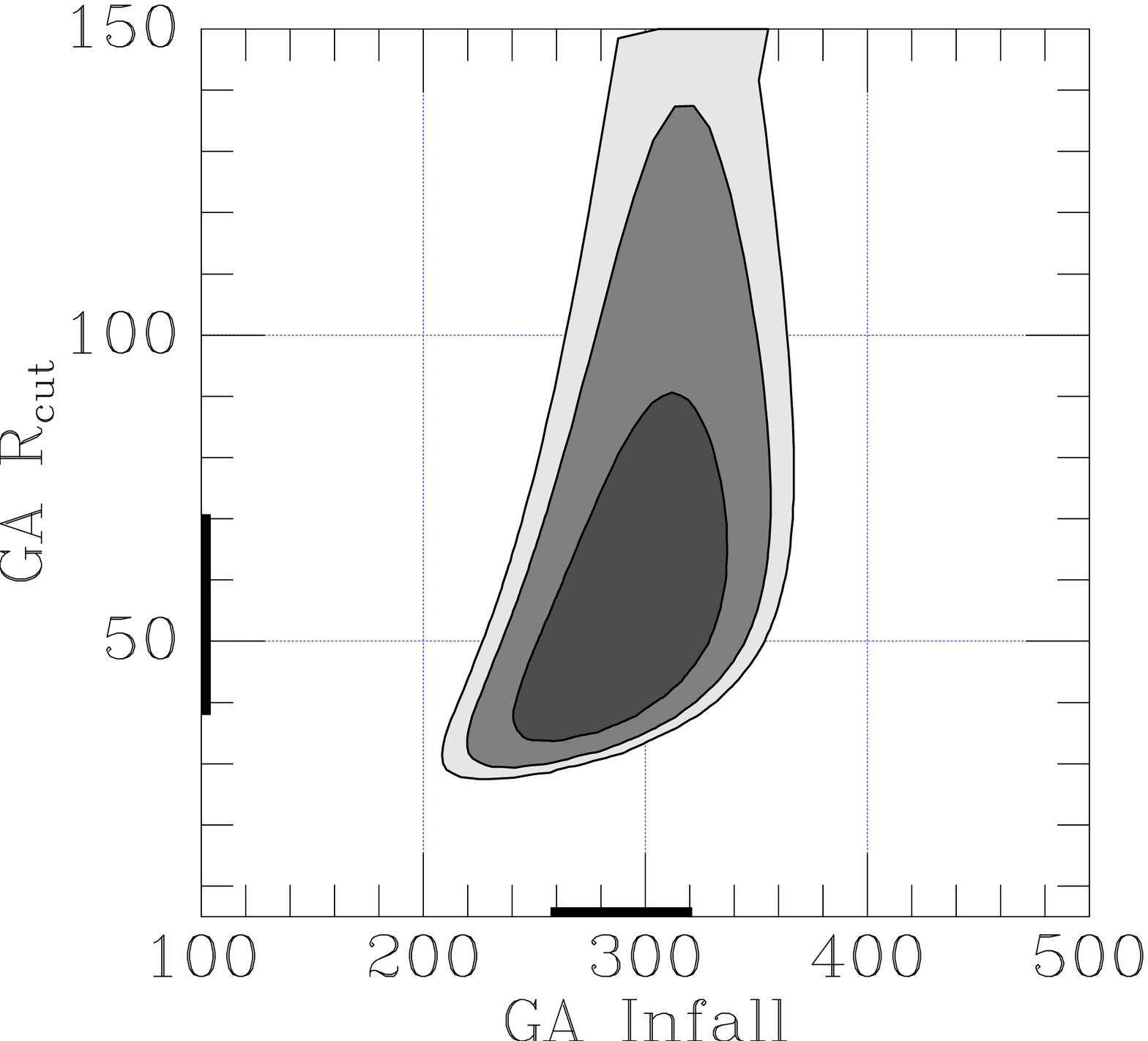}
\capfig{
\caption{
Joint confidence contours of the cutoff radius $r_{\rm cut}$ and power
law slope $\gamma$ (left) and infall amplitude (right) for
the GA attractor.
\label{fig:rcutga}}
}
\end{figure}

\begin{figure}[H]
\plotone{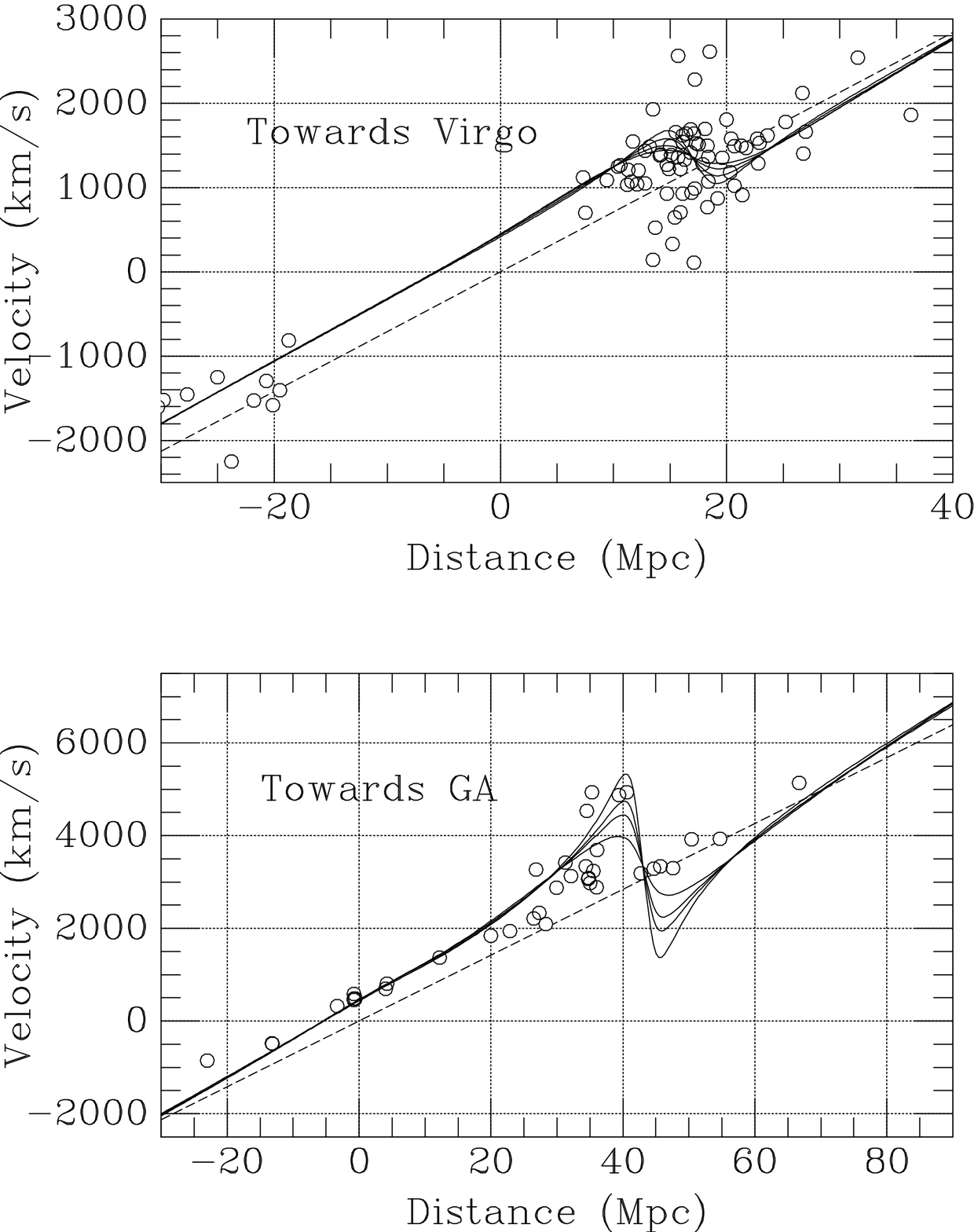}
\capfig{
\caption{
The net model velocity is shown as a function of position along vectors
running through the Virgo and Great Attractors using various values of
$\gamma$ and $r_{\rm cut}$ which leave $\cal N$ constant.  The dashed
line is just $H_0\,r$.  The points are those galaxies within
$25^\circ$ of the vectors ($|\cos\theta| > 0.9$).
\label{fig:flows}}
}
\end{figure}

\begin{figure}[H]
\plotone{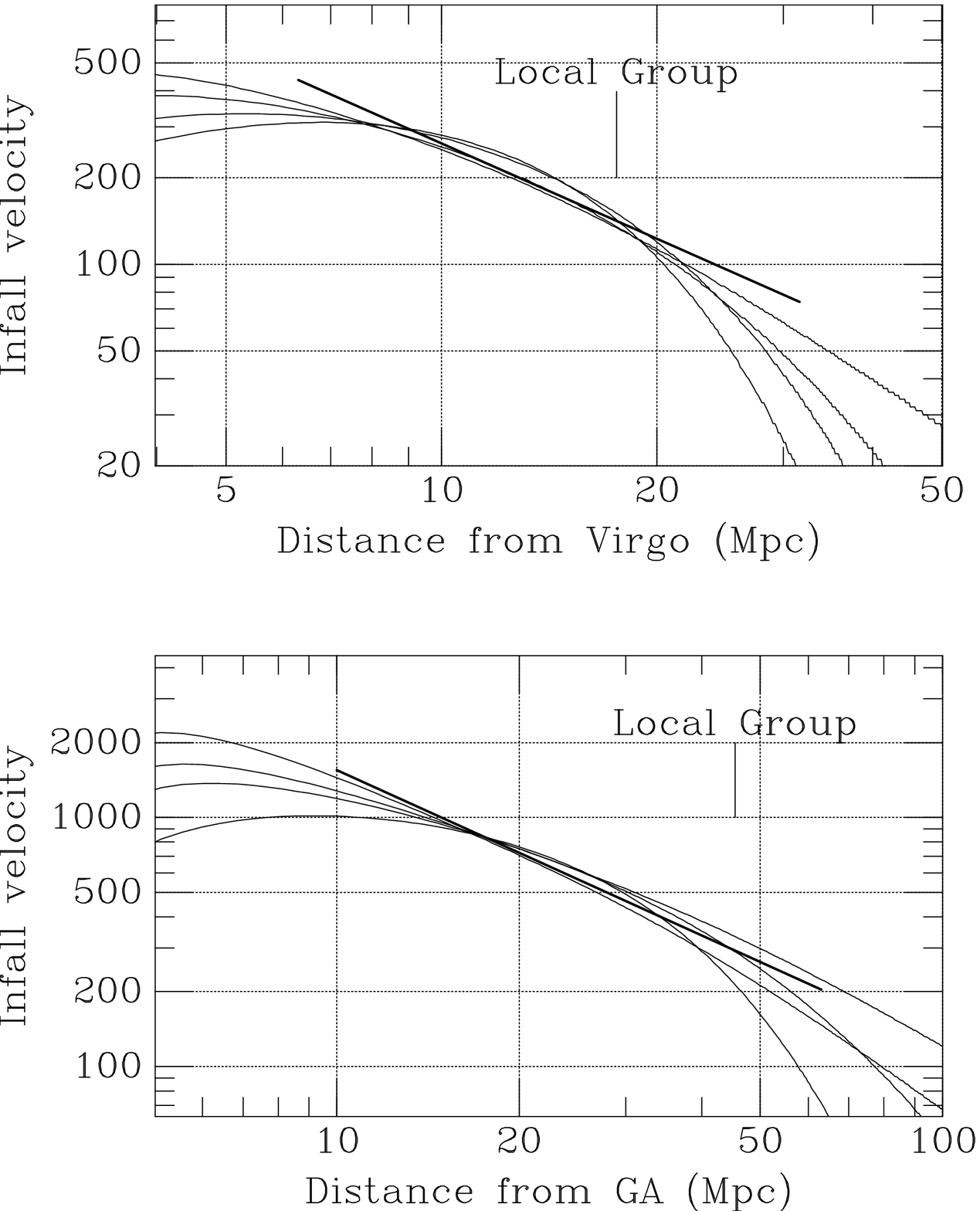}
\capfig{
\caption{
The model infall velocity is shown as a function of distance from the
Virgo and Great Attractors for the four sets of $\gamma$ values.
A solid line delineates the range where there is substantial agreement
and where we believe the models are well constrained by the data.
\label{fig:infalls}}
}
\end{figure}

\begin{figure}[H]
\plotone{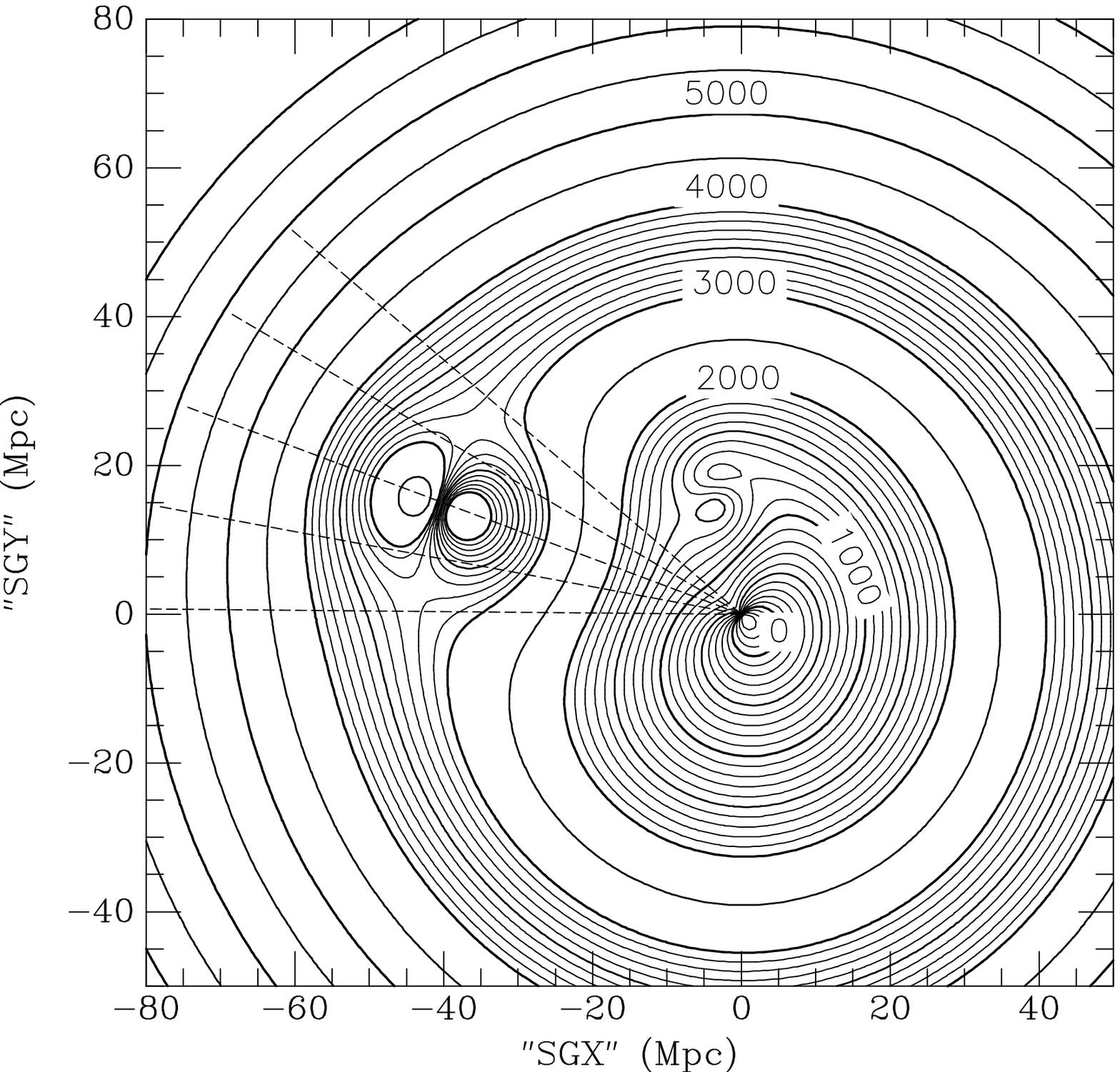}
\capfig{
\caption{
Contours are shown of our model of the radial component of the
flow field in a plane which cuts through the Local Group, Virgo,
and Great Attractors.  The model velocity of the Local Group with
respect to the CMB frame is apparent in the discontinuity at the
origin.  The Virgo Attractor is found at ``SGX,SGY'' = $(-3,+17)$
(quotes because the plane doesn't correspond perfectly to
supergalactic coordinates); the Great Attractor is at $(-42,+16)$,
and radial lines are drawn at $\pm10^\circ$ and $\pm20^\circ$ from it.
\label{fig:vcont}}
}
\end{figure}

\clearpage

\begin{figure}[H]
\plotone{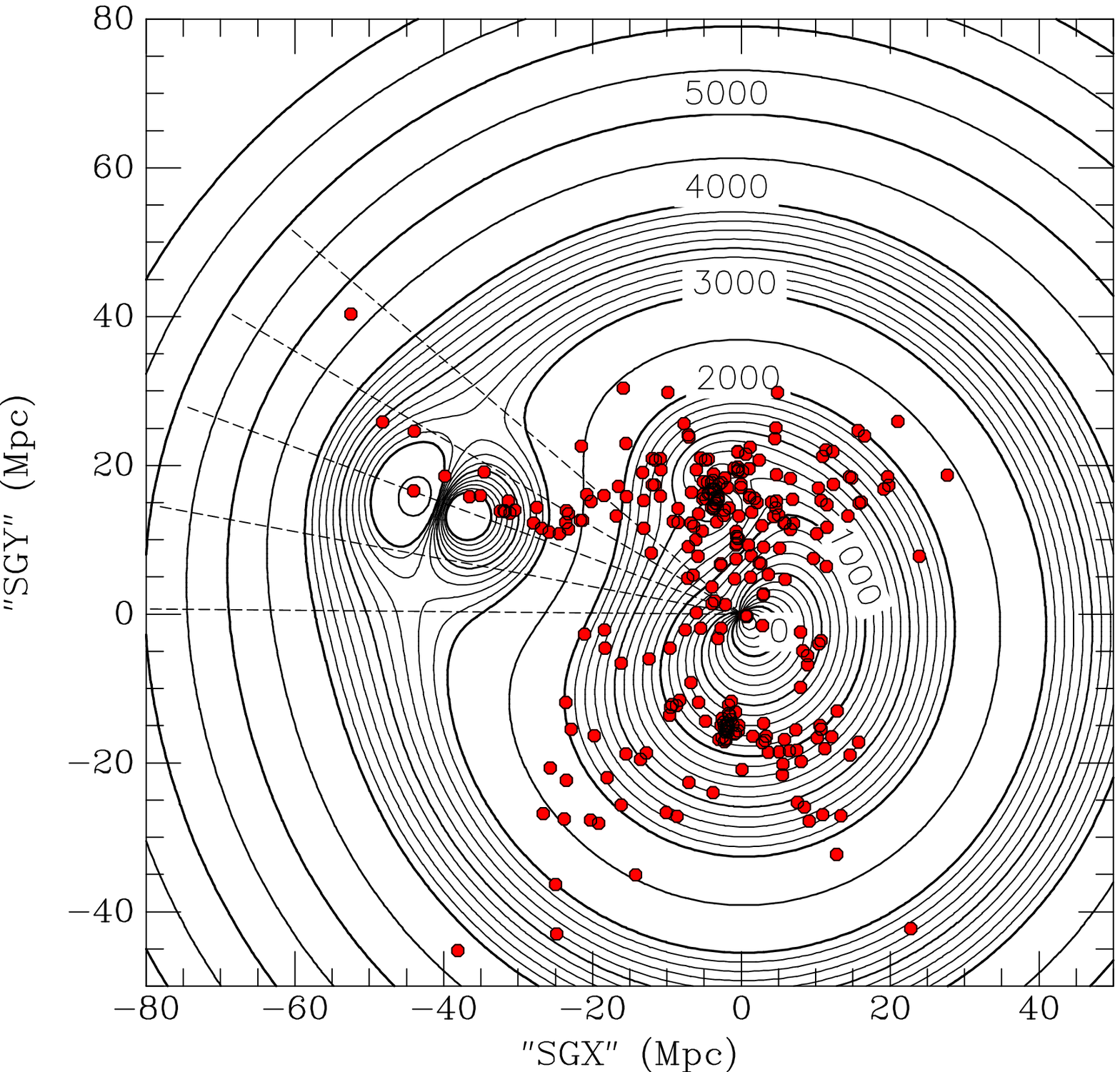}
\capfig{
\caption{
This is the same figure as the previous one except that our survey
galaxies are overplotted.  Note that the Centaurus galaxies which
appear to be going through the Great Attractor actually pass above it
by about $15^\circ$, hence lie in the stall zone.
\label{fig:vcontgxy}}
}
\end{figure}

\begin{figure}[H]
\plottwo{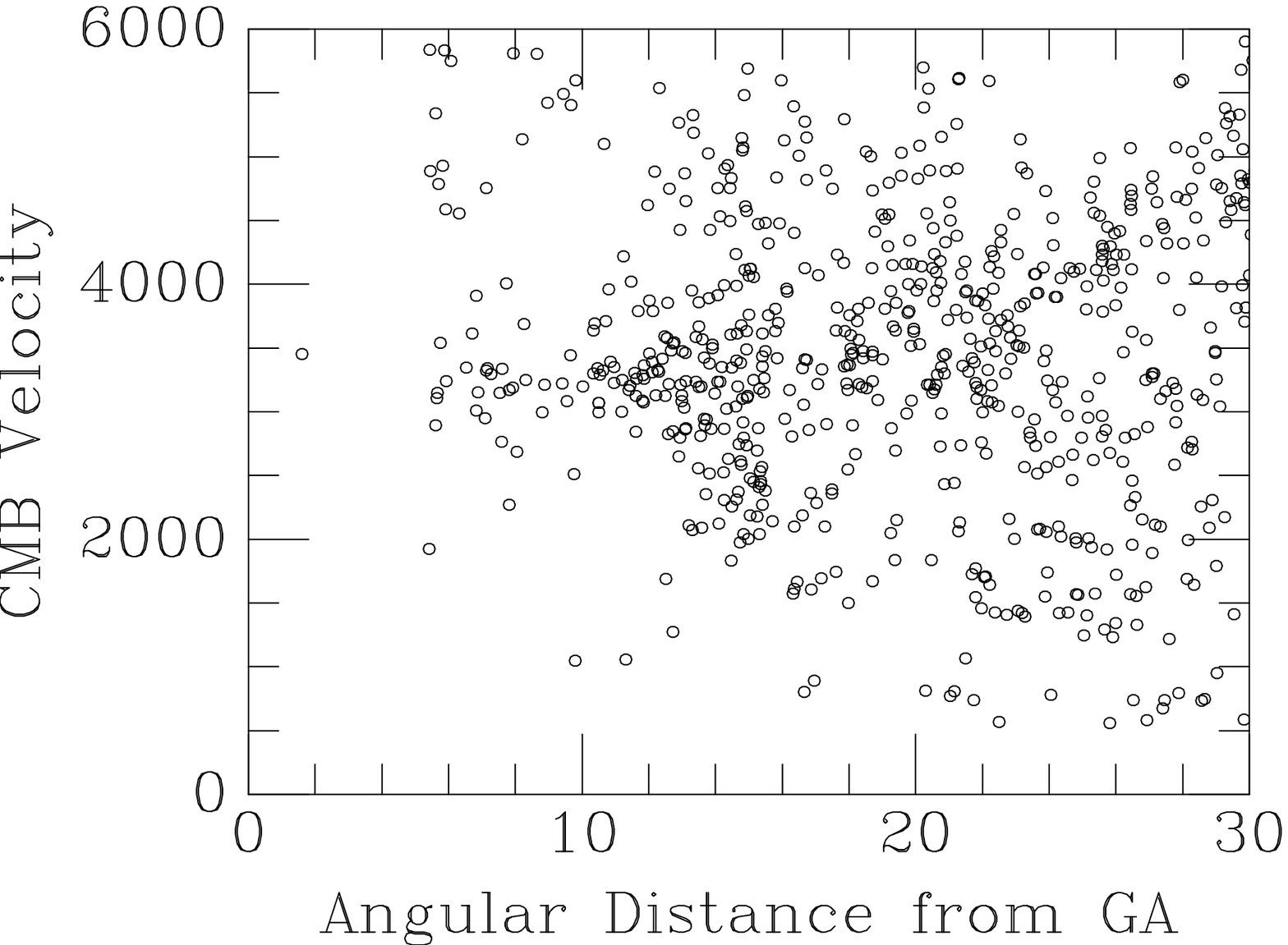}{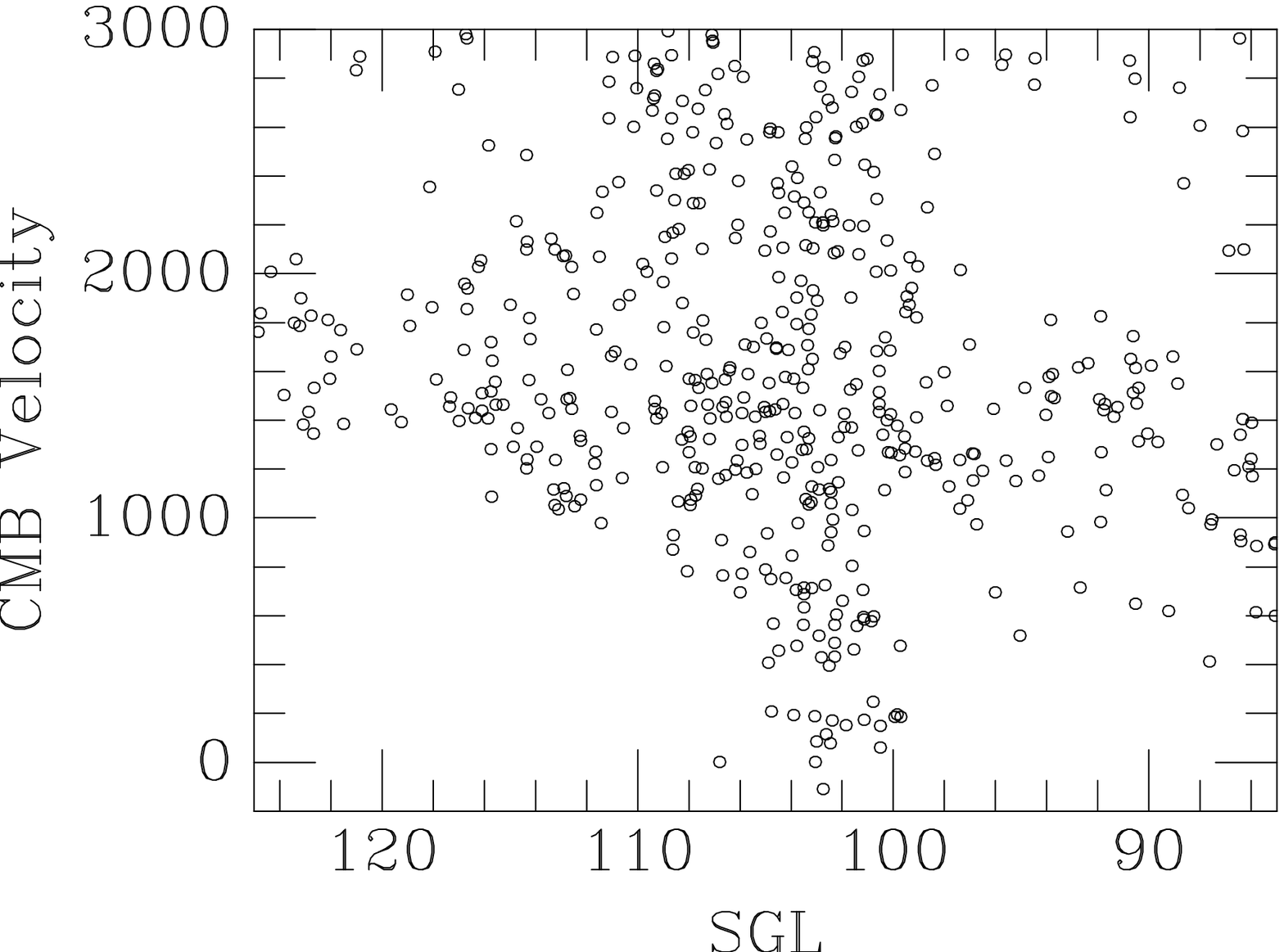}
\capfig{
\caption{
The CMB velocities of
galaxies from the RC3 are plotted as a function of angular separation
from the GA (left), and as a function of supergalactic longitude SGL
(right).  The concentration at $v_{\rm CMB} = 3150$
(left) and the trend for $v_{\rm CMB}$ to change from 1100 to
1500\kms\ (right) inherent in our model is apparent in these redshift
data.
\label{fig:gauma}}
}
\end{figure}

\begin{figure}[H]
\plottwo{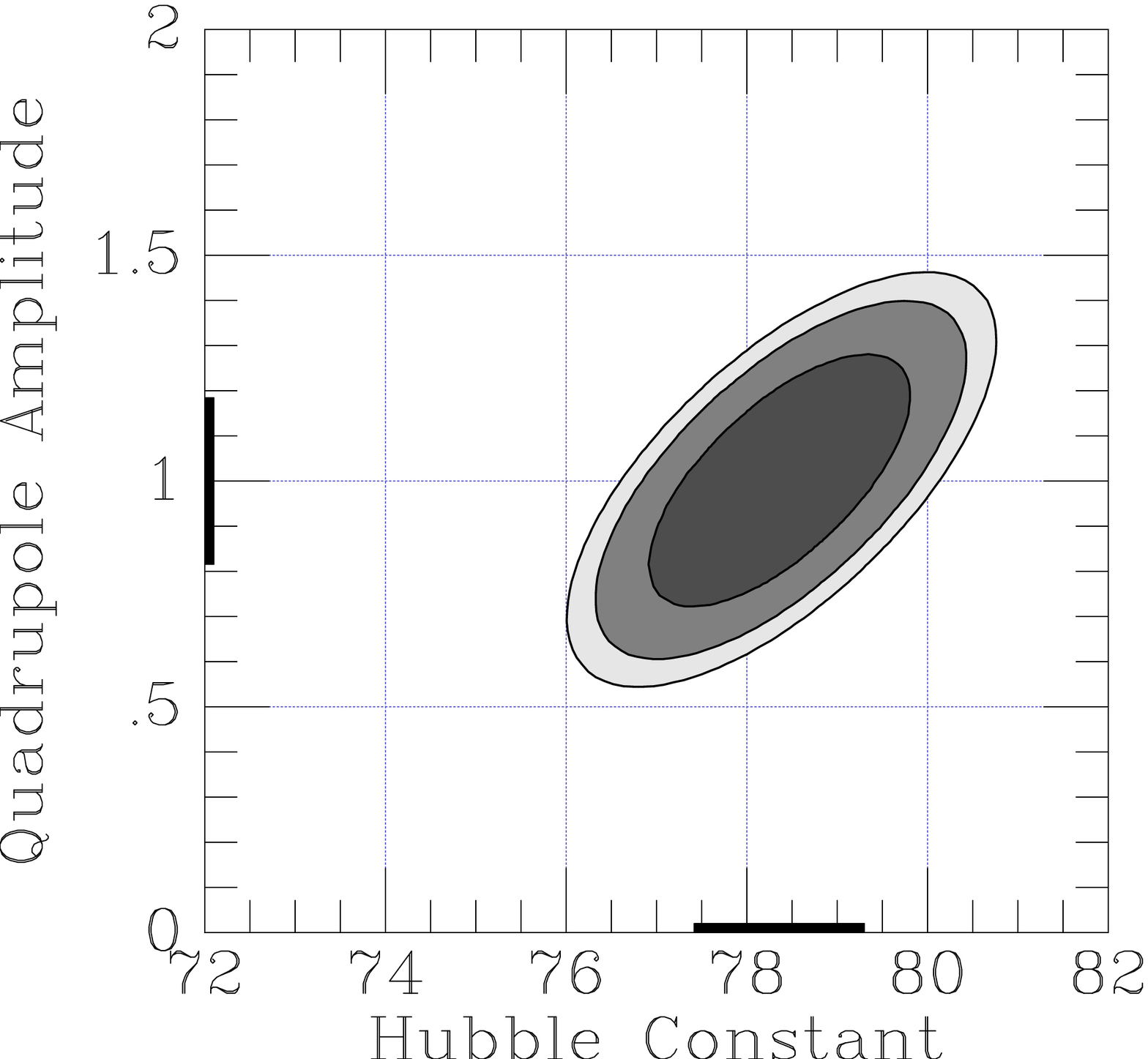}{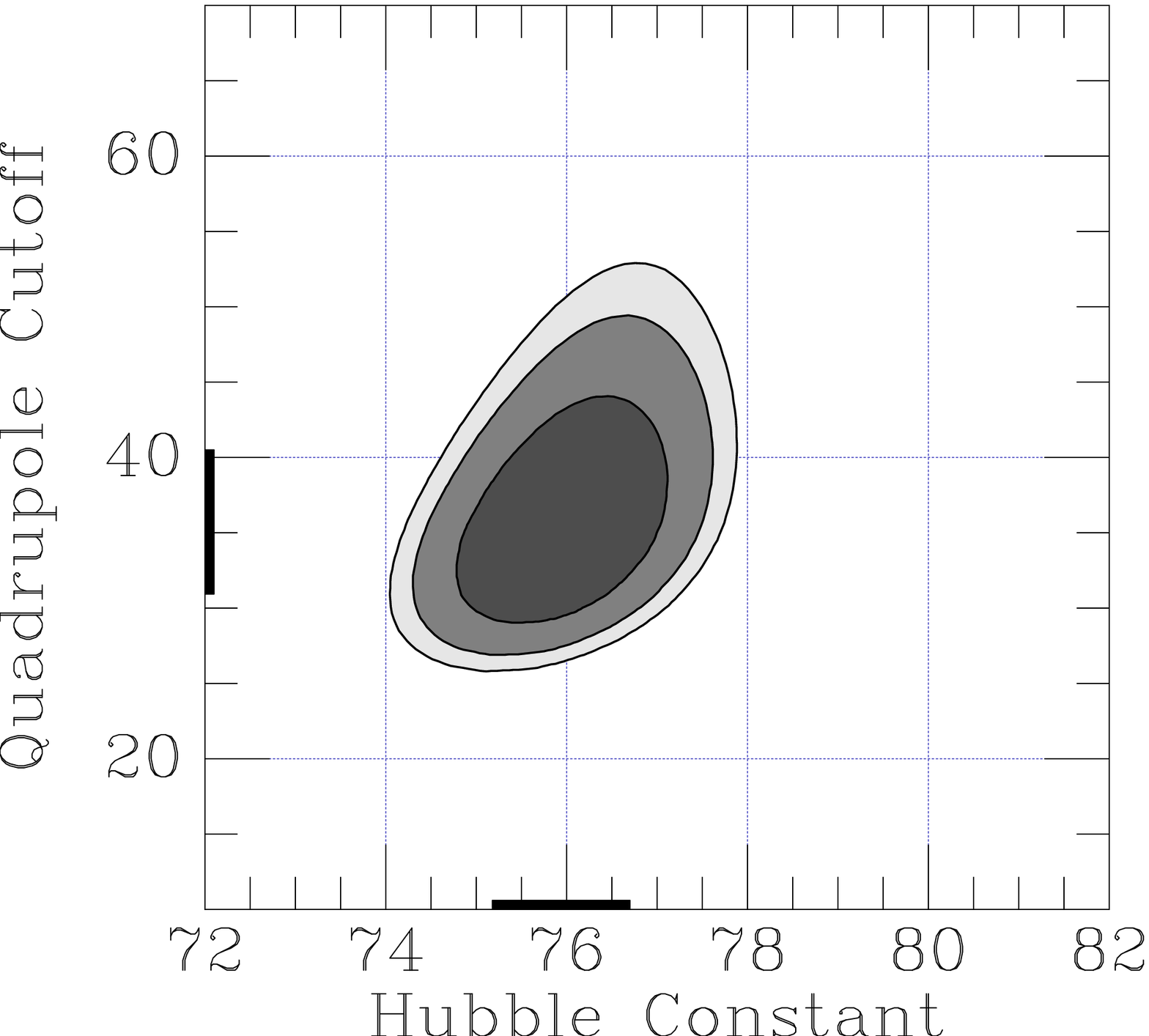}
\capfig{
\caption{
Joint confidence contours of the Hubble constant and relative
quadrupole amplitude when the quadrupole is centered on the Local
Group are shown on the left. Confidence contours of Hubble constant
and quadrupole cutoff radius when the quadrupole is centered on Virgo
instead are displayed on the right.
\label{fig:hqsig}}
}
\end{figure}
\clearpage
\begin{deluxetable}{lccc}
\tablewidth{0pt}
\tablecaption{Zero Points from Galaxy Groups}
\tablehead{
\colhead{Group}   & \colhead{$\langle \mi \rangle$} &
\colhead{\dmod}    & \colhead{$\langle \Mi \rangle$}
}
\startdata
M31    & $22.74 \pm 0.04$ & $24.44 \pm 0.10$ & $-1.70 \pm 0.11$ \nl
N1023  & $28.26 \pm 0.08$ & $29.84 \pm 0.08$ & $-1.58 \pm 0.11$ \nl
Fornax & $29.78 \pm 0.03$ & $31.41 \pm 0.06$ & $-1.63 \pm 0.07$ \nl
M81    & $26.27 \pm 0.09$ & $27.80 \pm 0.08$ & $-1.53 \pm 0.12$ \nl
LeoI   & $28.48 \pm 0.04$ & $30.08 \pm 0.06$ & $-1.60 \pm 0.07$ \nl
Virgo  & $29.41 \pm 0.02$ & $31.03 \pm 0.06$ & $-1.62 \pm 0.06$ \nl
\enddata
\label{tab:grp}
\end{deluxetable}

\clearpage
\begin{deluxetable}{lccc}
\tablewidth{0pt}
\tablecaption{Zero Points from Individual Spirals}
\tablehead{
\colhead{Galaxy}   & \colhead{$\langle \mi \rangle$} &
\colhead{\dmod}    & \colhead{$\langle \Mi \rangle$}
}
\startdata
N0224 & $22.67 \pm 0.06$ & $24.44 \pm 0.10$ & $-1.77 \pm 0.12$ \nl
N3031 & $26.21 \pm 0.25$ & $27.80 \pm 0.08$ & $-1.59 \pm 0.26$ \nl
N3368 & $28.34 \pm 0.21$ & $30.20 \pm 0.10$ & $-1.86 \pm 0.23$ \nl
N4548 & $29.68 \pm 0.54$ & $31.04 \pm 0.08$ & $-1.36 \pm 0.55$ \nl
N4725 & $28.87 \pm 0.34$ & $30.57 \pm 0.08$ & $-1.70 \pm 0.35$ \nl
N7331 & $28.85 \pm 0.16$ & $30.89 \pm 0.10$ & $-2.04 \pm 0.19$ \nl
\enddata
\label{tab:gxy}
\end{deluxetable}

\end{document}